\documentclass[useAMS,usenatbib,usegraphicx]{mn2e}

\usepackage{graphicx} 

\title[UFOs in radio-loud AGNs]{Ultra-fast outflows in radio-loud active galactic nuclei}
\author[F. Tombesi et al.]{F. Tombesi$^{1,2}$\thanks{E-mail: ftombesi@astro.umd.edu}, F. Tazaki$^3$, R.~F. Mushotzky$^{2}$, Y. Ueda$^{3}$, M. Cappi$^{4}$, J. Gofford$^{5}$, \newauthor J.~N. Reeves$^{5,6}$ and M. Guainazzi$^7$ \\  
$^{1}$X-ray Astrophysics Laboratory, NASA/Goddard Space Flight Center, Greenbelt, MD 20771, USA\\
$^{2}$Department of Astronomy, University of Maryland, College Park, MD 20742, USA\\
$^{3}$Department of Astronomy, Kyoto University, Kyoto 606-8502, Japan\\
$^{4}$INAF-IASF Bologna, Via Gobetti 101, I-40129 Bologna, Italy\\
$^{5}$Astrophysics Group, School of Physical and Geographical Sciences, Keele University, Keele, Staffordshire, ST5 5BG, UK\\
$^{6}$Center for Space Science and Technology, University of Maryland Baltimore County, 1000 Hilltop Circle, Baltimore, MD 21250, USA\\
$^{7}$European Space Astronomy Centre of ESA, PO Box 78, Villanueva de la Ca\~{n}ada, E-28691, Madrid, Spain\\
}
\begin{document}

\date{Accepted ???. Received ???; in original form ???}


\maketitle

\label{firstpage}

\begin{abstract}





Recent X-ray observations show absorbing winds with velocities up to mildly-relativistic values of the order of $\sim$0.1c in a limited sample of 6 broad-line radio galaxies. They are observed as blue-shifted Fe XXV--XXVI K-shell absorption lines, similarly to the ultra-fast outflows (UFOs) reported in Seyferts and quasars. In this work we extend the search for such Fe K absorption lines to a larger sample of 26 radio-loud AGNs observed with \emph{XMM-Newton} and \emph{Suzaku}. The sample is drawn from the \emph{Swift} BAT 58-month catalog and blazars are excluded. X-ray bright FR II radio galaxies constitute the majority of the sources. Combining the results of this analysis with those in the literature we find that UFOs are detected in $>$27\% of the sources. However, correcting for the number of spectra with insufficient signal-to-noise, we can estimate that the incidence of UFOs is this sample of radio-loud AGNs is likely in the range $f$$\simeq$$(50\pm20)$\%.  A photo-ionization modeling of the absorption lines with XSTAR allows to estimate the distribution of their main parameters. The observed outflow velocities are broadly distributed between $v_\mathrm{out}$$\la$1,000~km~s$^{-1}$ and $v_\mathrm{out}$$\simeq$0.4c, with mean and median values of $v_{out}$$\simeq$0.133c and $v_\mathrm{out}$$\simeq$0.117c, respectively. The material is highly ionized, with an average ionization parameter of log$\xi$$\simeq$4.5~erg~s$^{-1}$~cm, and the column densities are larger than $N_\mathrm{H}$$>$$10^{22}$~cm$^{-2}$. Overall, these characteristics are consistent with the presence of complex accretion disk winds in a significant fraction of radio-loud AGNs and demonstrate that the presence of relativistic jets does not preclude the existence of winds, in accordance with several theoretical models.










\end{abstract}


\section{Introduction}

Increasing evidences show that most galaxies harbor a Super-massive Black Hole (SMBH) at their center (e.g., Ferrarese \& Ford 2005). The accretion of material from these SMBHs provide the basic power source for Active Galactic Nuclei (AGNs). Depending on the ratio of their luminosity in radio and optical/X-rays, AGNs are classified as radio-loud or radio-quiet (e.g., Terashima \& Wilson 2003). Physically, this is related to the presence or absence of strong relativistic jets. The origin of this dichotomy is still debated and it is intimately linked to the mechanisms producing the jets, for instance the black hole spin (e.g., Sikora et al.~2007; Garofalo et al.~2010; Tchekhovskoy \& McKinney 2012) . 

In radio-loud AGNs the highly collimated relativistic jets are frequently observed at radio, optical and X-rays. They can travel for very long distances, impacting areas far away from the center of these galaxies. Until very recently, this was the main known mechanism for the deposition of mechanical energy from the SMBH into the host galaxy environment in radio-loud sources, contributing to the AGN cosmological feedback (e.g., Fabian 2012). Deep X-ray observations changed this view, showing an increasing evidence for accretion disc winds/outflows in this class of sources as well. 

For instance, the systematic analysis of the \emph{Suzaku} spectra of a small sample of five bright Broad-Line Radio Galaxies (BLRGs) performed by Tombesi et al.~(2010a) showed significant blue-shifted Fe XXV/XXVI K-shell absorption lines at E$>$7~keV at least in three of them, namely 3C~111, 3C~120 and 3C~390.3. They imply an origin from highly ionized and high column density gas outflowing with mildly relativistic velocities of the order of 10\% of the speed of light. Successive studies confirmed these findings and found similar outflows also in 3C~445 and 4C$+$74.26 (Reeves et al.~2010; Tombesi et al.~2011a, 2012a, 2013a; Ballo et al.~2011; Braito et al.~2011; Gofford et al.~2013). The slower ($v_\mathrm{out}$$\sim$100--1,000~km~s$^{-1}$) and less ionized warm absorbers have also been detected in the soft X-ray spectra of some bright BLRGs (Reynolds 1997; Ballantyne 2005; Reeves et al.~2009a, 2010; Torresi et al.~2010, 2012; Braito et al.~2011). These observations indicate that, at least for some radio-loud AGNs, the presence of relativistic jets does not preclude the existence of winds and the two might possibly be linked (e.g., Tombesi et al.~2012a, 2013a; Fukumura et al.~2014).

The characteristics of these winds are similar to the Ultra-Fast Outflows (UFOs) with velocities $v_{out} \ge$10,000~km/s observed in $\sim$40\% of radio-quiet AGNs (e.g., Chartas et al. 2002, 2003; Pounds et al.~2003; Dadina et al.~2005; Markowitz et al.~2006; Papadakis et al.~2007; Braito et al.~2007; Cappi et al.~2009; Reeves et al.~2009b; Chartas et al.~2009; Tombesi et al.~2010b, 2011b; Giustini et al.~2011; Gofford et al.~2011; Lobban et al.~2011; Patrick et al.~2012; Lanzuisi et al.~2012; Dauser et al.~2012; Gofford et al.~2013; Gupta et al.~2013). The mechanical power of these UFOs seems to be high enough to influence the host galaxy environment (Tombesi et al.~2012b; 2013b), in accordance with simulations of SMBH-galaxy feedback driven by AGN winds/outflows (e.g. King \& Pounds 2003, 2014; Di Matteo et al.~2005; King 2010; Ostriker et al.~2010; Hopkins \& Elvis 2010; Gaspari et al.~2011a, 2011b, 2012; Zubovas \& King 2012, 2013; Wagner et al.~2013). 

The typical location of the UFOs is estimated to be in the range between $\sim$10$^2$--$10^4$~$r_s$ ($r_s$$=$$GM_{BH}/c^2$) from the central SMBH, indicating a direct association with winds originating from the accretion disk (e.g., Tombesi et al.~2012b, 2013b). Their high ionization levels and velocities are, to date, consistent with both radiation pressure and magnetohydrodynamic (MHD) processes contributing to the outflow acceleration (e.g., King \& Pounds 2003; Everett \& Ballantyne 2004; Everett 2005; Ohsuga et al.~2009; King 2010; Fukumura et al.~2010, 2014; Kazanas et al.~2012; Ramirez \& Tombesi 2012; Reynolds 2012; Tombesi et al.~2013b). The study of UFOs in radio-loud AGNs can provide important insights into the characteristics of the inner regions of these sources and can provide an important test for theoretical models trying to explain the disk-jet connection (e.g., McKinney~2006; Tchekhovskoy et al.~2011; Yuan, Bu \& Wu 2012; Sadowski et al.~2013, 2014). 

The discovery of mildly-relativistic UFOs in a few BLRGs provided us a potential new avenue of investigation into the central engines of radio-loud AGNs. However, this sample is too small to draw any general conclusions and many questions are still open, for instance: what is the actual incidence of UFOs in radio-loud AGNs? Are the UFOs in radio-loud and radio-quiet AGNs the same phenomena? In other words, does the jet related radio-quiet/radio-loud dichotomy holds also for AGN winds or not? What are the main ingredients for jet/winds formation: the black hole spin, the accretion disk or the strength of the magnetic field? The main goal of this paper is to answer the first question by conducting a systematic and uniform search for blue-shifted Fe K absorption lines in a larger sample of 26 radio-loud AGNs observed with \emph{XMM-Newton} and \emph{Suzaku}. 

A detailed comparison with respect to the UFOs detected in radio-quiet AGNs will be presented in a subsequent paper (Tombesi et al. in prep.). This will allow to perform a more detailed discussion of the validity of the jet related radio-quiet/radio-loud dichotomy for winds and possibly to better constrain the winds acceleration mechanisms and energetics.

\section{Sample selection and data reduction}

The initial sample of radio-loud AGNs was chosen from the \emph{Swift} BAT 58-month hard X-ray survey catalog \footnote{http://heasarc.gsfc.nasa.gov/docs/swift/results/bs58mon/}. This survey reaches a flux level of $1.48 \times 10^{-11}$~erg~s$^{-1}$~cm$^{-2}$ in the energy interval E$=$14--195~keV over 90\% of the sky (Baumgartner et al.~2010). This allows us to obtain an almost complete and homogeneous sample of hard X-ray selected bright sources. First, we select the radio sources identified in the third and the fourth Cambridge (3C and 4C) catalogs (Edge et al.~1959; Pilkington \& Scott 1965) and The Parkes catalogue (Bolton et al.~1964). The sample comprises also other four radio galaxies observed with \emph{Swift} BAT, namely Centaurus~A, IC~5063, NGC~612 and Pictor~A, that were studied in X-rays in previous works (e.g., Fukazawa et al.~2011; Tazaki et al.~2011; Eguchi et al.~2011; Eracleous et al.~2000). We do not include the sources classified as blazars, because the strong contribution of the jet in the X-ray band in this case might complicate the study of emission and absorption features. The analysis of blazars is deferred to a future work. 

This initial sample was cross-correlated with the \emph{XMM-Newton} and \emph{Suzaku} catalogs. These archives are ideal for this project as they provide high-sensitivity hard X-ray spectra for a large number of AGNs. We then downloaded and reduced all the observations using the standard procedures and the latest calibration database. We use all the public data as of December 2012. After the data reduction and screening, described in the subsequent section \S2, we are left with a total of 26 sources and 61 observations. The number of sources (observations) optically classified as type 1 (from 1 to 1.5) and type 2 (from 1.6 to 2) is 12 (40) and 12 (19), respectively.
We numbered each source and included letters to identify the different observations. The details of the observations are listed in Table~B1 in Appendix B. The same sample, with the exclusion of PKS~0558$-$504 and Centaurus~A, is also used in a complementary work by Tazaki et al.~(in prep.), which is focused on the X-ray study of the accretion disk and torus reflection features and also in quantifying the jet contribution in the X-ray band.

Following the classification of Fanaroff \& Riley (1974), radio galaxies can be divided depending on their radio jet morphology and radio power in FR~I and FR~II. The former have low radio brightness regions further from the central galaxy and an overall lower radio luminosity than the latter. From Table~B1 in Appendix B we note that the great majority of the sources are classified as FR~II. In particular, of the 23 sources with an FR classification, 20 are identified as FR~II and only 3 as FR~I, namely 3C~120, IC~5063 and Centaurus~A. It is known that the X-ray luminosity of FR~II galaxies is systematically higher than that of FR~Is (e.g., Fabbiano et al.~1984; Hardcastle et al.~2009). Therefore, given that this is an X-ray selected flux-limited sample, the higher number of FR~IIs compared to FR~Is can be simply due to a selection effect. In fact, even if their X-ray luminosity is relatively low, IC~5063 and Centaurus~A are included in this flux-limited sample because of their very low cosmological redshifts.  

As shown in Figure~1, the sources are local and their cosmological redshifts are broadly distributed between $z \simeq 0$ and $z \simeq 0.2$. As we can see in Figure~2, this sample consists of X-ray bright sources, with the majority of them having observed E$=$4--10~keV fluxes higher than $\sim$$10^{-12}$~erg~s$^{-1}$~cm$^{-2}$ and an average value of $\simeq$$2\times 10^{-11}$~erg~s$^{-1}$~cm$^{-2}$.

This sample of radio-loud AGNs is directly complementary to that of Tombesi et al.~(2010b), which focused on a search for UFOs in 42 local radio-quiet AGNs (i.e., Seyferts) observed with \emph{XMM-Newton}. A detailed comparison with respect to the sample of radio-quiet AGNs of Tombesi et al.~(2010b) will be presented in a subsequent paper (Tombesi et al. in prep.). Another recent systematic search for UFOs using \emph{Suzaku} observations was reported by Gofford et al.~(2013). In this case the authors selected an heterogeneous flux-based sample, considering spectra having $>50.000$ counts in the 2--10 keV band and both low/high redshift radio-quiet and radio-loud AGNs, in particular 34 Seyferts, 5 quasars and 6 BLRGs. As discussed in Gofford et al.~(2013), the results from the two \emph{XMM-Newton} and \emph{Suzaku} samples are formally consistent, with the \emph{Suzaku} results suggesting that strong fast outflows also appear to be present at high luminosities/redshift (e.g., PDS~456 and APM~08279$+$5255), indicating that such winds may play an important role also near the peak of quasar growth/feedback at $z\sim2$.

   \begin{figure}
   \centering
    \includegraphics[width=7cm,height=7cm,angle=0]{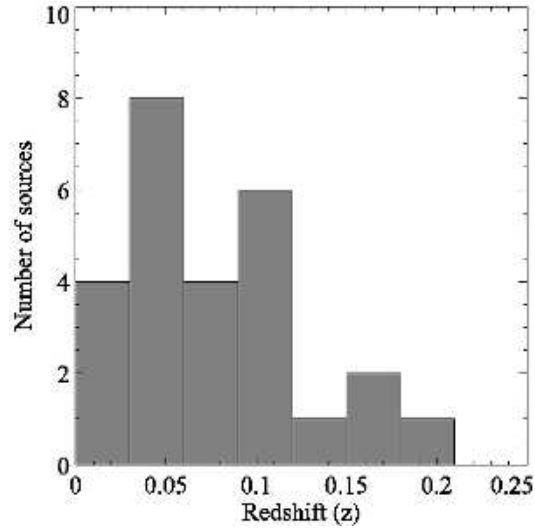}
   \caption{Distribution of the cosmological redshifts of the sources in the radio-loud AGN sample.}
    \end{figure}

   \begin{figure}
   \centering
    \includegraphics[width=7cm,height=7cm,angle=0]{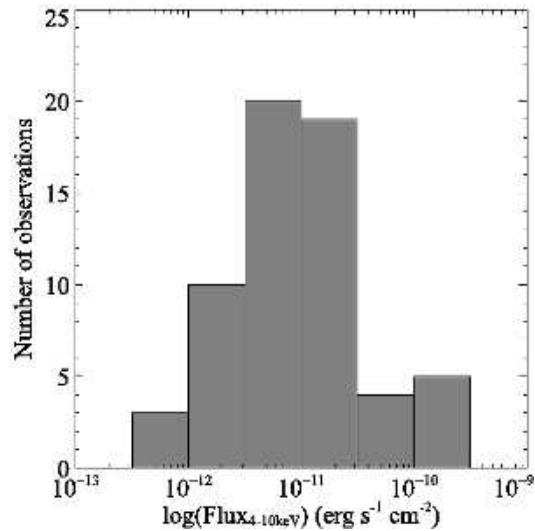}
   \caption{Distribution of the observed flux in the E$=$4--10~keV energy band for the observations of the radio-loud AGN sample.}
    \end{figure}

\subsection{XMM-Newton}

We reduced the \emph{XMM-Newton} observations using the standard procedures and the latest calibration database. The data reduction was performed using the SAS version 12.0.1 and the \emph{heasoft} package version 6.12. We consider only the EPIC-pn instrument, which offers the highest sensitivity in the Fe K band between E$=$3.5--10.5~keV. We checked for high background flare intervals and excluded these periods from the successive spectral analysis. The exposure time for the EPIC-pn spectra reported in Table~B1 in Appendix B is net of the high background flare periods and detector dead-time.

The majority of the observations did not show pile-up problems and the source and background regions were extracted from circles with 40$\arcsec$ radii. However, there are some observations with significant pile-up, which required the source extraction region to be an annulus in order to exclude the inner part of the PSF, which is more affected by this problem. The outer radius of the source and background regions in these cases was of 50$\arcsec$. Instead, the inner radius of the source annulus which was excluded from the analysis was of 5$\arcsec$ for Pictor~A (observation number 11 in Table~B1 in Appendix B), 10$\arcsec$ for PKS~0558$-$504 (observations 2b and 2c) and 20$\arcsec$ for Centaurus~A (observations 26a and 26b), respectively. The parameters of the \emph{XMM-Newton} observations are listed in Table~B1 in Appendix B.

\subsection{Suzaku}

The \emph{Suzaku} data were reduced starting from the unfiltered event files and then screened applying the standard selection criteria described in the \emph{Suzaku} ABC guide\footnote{http://heasarc.gsfc.nasa.gov/docs/suzaku/analysis/abc/}. We follow the same method outlined in Tazaki et al.~(2013). The source spectra were extracted from circular regions of 3$\arcmin$ radius centered on the source, whereas background spectra were extracted from a region of same size offset from the main target and avoiding the calibration sources. We generate the RMF (redistribution matrix file) and ARF (ancillary response file) of the XIS with the \emph{xisrmfgen} and \emph{xissimarfgen} ftools (Ishisaki et al. 2007), respectively. The spectra were inspected for possible pile-up contamination and this was corrected when present. The spectra of the front illuminated XIS-FI instruments (XIS~0, XIS~3 and also XIS~2 when available) were merged after checking that their fluxes were consistent. Instead, the data of the back illuminated XIS-BI instrument, the XIS~1, are not used due to the much lower sensitivity in the Fe K band and the possible cross-calibration uncertainties with the XIS-FI.

For the purpose of the broad-band spectral analysis, we also reduced the \emph{Suzaku} PIN data for those sources showing possible absorption features in the Fe K band. The PIN data reduction and analysis was performed following the standard procedures.  The source and background spectra were extracted within the common good time intervals and the source spectrum was corrected for the detector dead-time. The ``tuned'' non X-ray background (NXB) event files provided by the HXD team are utilized for the background subtraction of the PIN data. The simulated cosmic X-ray background (CXB) is added to the NXB spectrum based on the formula given by Gruber et al.~(1999). The PIN response file relative to each observation is used. We do not consider the \emph{Suzaku} GSO data given the limited sensitivity of this instrument for the study of these sources. The parameters of the \emph{Suzaku} observations are listed in Table~B1 in Appendix B.

\section{Initial Fe K band spectral analysis}

In this section we describe the initial spectra fitting of the data focused in the energy range E$=$3.5--10.5~keV, i.e. the Fe K band. The different \emph{XMM-Newton} EPIC-pn and \emph{Suzaku} XIS spectra were grouped to a minimum of 25 counts for each energy bin in order to enable the use of the $\chi^2$ statistics when performing the spectral fitting. As already discussed in Tombesi et al.~(2011b) and Gofford et al.~(2013), equivalent results are obtained grouping to a slightly higher number of counts per bin and using of the $\chi^2$ statistic, not grouping the data at all and using the C-statistic (Cash 1979), or alternatively grouping the data to 1/5 of the FWHM of the instrument to sample the intrinsic energy resolution of the detector. The spectral analysis was carried out using the XSPEC v.~12.7.1 software. As pointed out in Table~B1 in Appendix B, we do not analyze again those observations of 3C~111 (6), 3C~120 (4), 3C~382 (1), 3C~390.3 (1), 3C~445 (1) and 4C$+$74.26 (1) for which a detailed spectral analysis and search for Fe K absorbers was already reported in the literature using methods equivalent to the one employed here (e.g., Tombesi et al.~2010a, b; 2011a, b; Ballo et al.~2011; Braito et al.~2011; Sambruna et al.~2009, 2011; Gofford et al.~2013). 

The analysis of the remaining 47 spectra was carried out in a homogeneous manner following a series of steps. Each spectrum was analyzed in the E$=$3.5--10.5~keV band using a baseline model similar to Tombesi et al.~(2010a, b), i.e., a Galactic absorbed power-law continuum with possible narrow Gaussian Fe K emission lines. When required, we considered also additional neutral absorption intrinsic to the source. At this stage of the spectral analysis we did not include any Gaussian absorption lines but only emission lines when required. Throughout the analysis we consider only spectral components required at $>$99\% with the F-test. This baseline model provides a good representation of all the spectra in this restricted energy band. The best-fit parameters for each observation are listed in Table~A2 in Appendix A. Throughout this work the error bars refer to the 1$\sigma$ level, if not otherwise stated.  

Energy-intensity contour plots were calculated following the procedure described in Tombesi et al.~(2010a, b). This consists in the fitting the spectrum with the previously derived baseline model without the inclusion of any absorption line and then stepping an additional narrow Gaussian line at intervals of 50~eV in the rest-frame energy interval between 5--10~keV with normalization free to be positive or negative. Each time the new $\chi^2$ is stored. In this way we derive a grid of $\chi^2$ values for different line energies and normalizations and then calculated the contours with the same $\Delta\chi^2$ level relative to the baseline model fit. We use $\Delta\chi^2$ levels of $-2.3$, $-4.61$ and $-9.21$, which correspond to F-test confidence levels for the addition of two more parameters of 68\%, 90\% and 99\%, respectively. Due to the relatively high cosmological redshift ($z = 0.1522$) and the possible presence of absorption lines at high energies, the contour plots of PKS~1549$-$79 were calculated in the interval E$=$5--12~keV for clarity. The ratio of the spectra with respect to the absorbed power-law continuum (without the inclusion of any Gaussian emission lines) and the contour plots with respect to the best-fit baseline models (including Gaussian emission lines) are reported in Figure~C1 in Appendix C.

The contour plots were inspected for the possible presence of absorption features. The data analysis of a particular observation was interrupted if the contour plots showed no evidence for absorption features at rest-frame energies E$>$6~keV with a significance at least $>$99\%, indicated by the blue contours in Figure~C1 in Appendix C. 


The background counts are on average of the order or less than 2\% of the source ones in the 4--10~keV. However, it is important to note that both the EPIC-pn onboard \emph{XMM-Newton} and the XIS cameras onboard \emph{Suzaku} have an intense instrumental background emission line due to Cu K$\alpha$ at E$=$8.05~keV and Ni K$\alpha$ at E$=$7.48~keV, respectively (Katayama et al.~2004; Yamaguchi et al.~2006; Carter \& Read 2007). These lines originate from the interaction of the cosmic rays with the sensor housing and electronics and this also causes their intensity to be slightly dependent on the location on the detector. Therefore, the selection of the background on a region of the CCD where the intensity of the lines is higher than that on the source extraction region can possibly induce spurious absorption lines in the background subtracted spectrum. If an indication of possible absorption lines was present in the contour plots, we then performed several checks to exclude the possibility that they were induced by an inadequate subtraction of instrumental background emission lines.

First, since the background emission lines are present at specific energies, we checked that the observed energies of the absorption lines are not consistent with those values. Secondly, we checked if the absorption lines were still present in the contour plots even without background subtraction. Only in two cases we find that the absorption lines could be induced by the background emission lines, namely 3C~206 and PKS~0707$-$35 (observations number 8 and 14 in Table~A2 in Appendix A, respectively). Therefore, these two cases are excluded from the successive analysis.


From this initial study focused in the E$=$3.5--10.5~keV band we find possible significant Fe K absorption lines at E$>$6~keV in the contour plots of 8 observations of 5 sources, namely 4C$+$74.26 (observations 1a and 1c), 3C~120 (observation 9f), PKS~1549$-$79 (observation 12a), 3C~105 (observation 16) and Centaurus~A (observations 26c, 26d and 26e). In the next section and in Appendix A we investigate these observations in more detail. In particular, we derive a more robust detection of the possible absorption lines and reduce the uncertainties on the continuum level extending the analysis to the relative \emph{XMM-Newton} and \emph{Suzaku} broad-band spectra.

\section{Broad-band spectral analysis}

In order to derive a more robust detection of the possible Fe K absorption lines and reduce the uncertainties on the continuum level, we then exploited the full broad-band capabilities of \emph{XMM-Newton} EPIC-pn in the E$=$0.5--10.5~keV and \emph{Suzaku} in the E$=$0.6--70~keV combining the XIS-FI (E$=$0.6-10.5~keV, but excluding the interval between E$=$1.5--2~keV due to calibration uncertainties) and PIN (E$=$14--70~keV). We perform a broad-band spectral analysis only of the 8 observations of 5 sources showing possible absorption lines in the contour plots in Figure~C1 in Appendix C, namely 4C$+$74.26 (observations 1a and 1c), 3C~120 (observation 9f), PKS~1549$-$79 (observation 12a), 3C~105 (observation 16) and Centaurus~A (observations 26c, 26d and 26e).

The data analysis has been performed following an homogeneous standard procedure for all the observations. We start the broad-band analysis using a simple baseline model composed of a Galactic and intrinsically absorbed power-law continuum. Then, we inspect the residuals and include additional components that are required at more than the 99\% level. If required, for type 2 sources we include also a scattered continuum component in the soft X-rays with slope tied to the main continuum. The neutral Fe K emission lines previously modeled with Gaussians in the E$=$3.5--10.5~keV analysis are now parameterized with a cold reflection component \emph{pexmon} in XSPEC (Nandra et al.~2007), which takes simultaneously into account the neutral E$=$6.4~keV Fe K$\alpha$ fluorescent emission line, the relative Fe K$\beta$ at E$=$7.06~keV, the E$=$7.45~keV Ni K$\alpha$ and the high energy reflection hump. In this model we assume standard Solar abundances, a high energy cut-off at $E$$=$300~keV (e.g., Dadina 2007) and an inclination of 30$^{\circ}$ or 60$^{\circ}$ for type 1 and type 2 sources, respectively. 

In a few cases we find that the spectra show some deviations from this standard baseline model. In particular, for 4C$+$74.26 we included two lowly ionized, soft X-ray warm absorber components modeled with two XSTAR tables with turbulent velocity of 100~km/s. For the first observation (1a) of this source, the Fe K emission line detected at E$\simeq$6.5~keV required also a mildly ionized reflection component modeled with a \emph{xillver} table (Garc{\'{\i}}a et al.~2010, 2013, 2014) instead of \emph{pexmon}. The broad-band spectrum of Centaurus~A requires a hot plasma emission component, \emph{apec} in XSPEC, with temperature $kT\simeq 0.65$~keV to model the cluster/lobe emission in the soft X-rays and the neutral reflection component \emph{pexmon} requires an iron abundance that is about 40\% of the Solar value. The details of the fits for each observation are reported in the next sections. 

Once a best-fit was found, this broad-band model was used as a baseline to calculate the contour plots as discussed in \S3 for the initial analysis focused in the Fe K band. The E$=$5--10~keV contour plots were inspected again to check the presence of absorption feature(s) at rest-frame energies $>$6~keV with F-test significance $>$99\%. If the absorption line was confirmed, we modeled it with an inverted Gaussian and estimated the relative parameters, such as energy, width, intensity, EW and $\Delta\chi^2$ improvement to the fit.

In order to derive a more robust estimate of the detection significance for the absorption lines detected at E$>$7~keV, we also performed 1000 Monte Carlo spectral simulations following the method outlined in Tombesi et al.~(2010a, b). This allows to better estimate the incidence of random fluctuations when considering lines in the relatively wide energy interval E$=$7--10~keV. This usually provides a slightly lower significance than the standard F-test, which is applicable for a line observed at a certain expected energy, such as in the range E$=$6.4--7~keV for K shell lines from neutral to hydrogen-like iron. Therefore, this test was required for those lines detected at rest-frame energies higher than 7~keV.    

The Monte Carlo simulations allow to test the null hypothesis that the spectra were adequately fitted by a model that did not include the absorption line. The analysis has been carried out as follows: 1) we simulated a spectrum (with the \emph{fakeit} command in XSPEC) using the broad-band baseline model without any absorption line and considered the same exposure as the real data. We subtracted the appropriate background and grouped the spectrum to a minimum of 25 counts per energy bin; 2) this simulated spectrum was fitted with the baseline model and the resultant $\chi^2$ was stored; 3) then a narrow Gaussian line was added to the model, with its normalization initially set to zero and let free to vary between positive and negative values. In order to account for the possible range of energies in which the line could be detected, we then stepped its centroid energy between 7~keV and 10~keV at intervals of 100~eV, each time making a fit and stored only the maximum $\Delta\chi^2$; 4) this procedure was repeated 1000 times and consequently a distribution of simulated $\Delta\chi^2$ values was generated. This indicates the fraction of expected random generated emission/absorption features in the E$=$7--10~keV energy band with a certain $\Delta\chi^2$ value. Thus, the Monte Carlo significance level of the observed line is represented by the ratio between the number of random lines with $\Delta\chi^2$ greater or equal to the real one with respect to the total number of simulated spectra. Following Tombesi et al.~(2010a, b) and Gofford et al.~(2013), we then select only the absorption features with a Monte Carlo confidence level higher than 95\%. 

Finally, we perform a more physically self-consistent modeling of the absorption lines using the XSTAR tables and method described in Tombesi et al.~(2011b). The nuclear X-ray ionizing continuum in the input energy range of XSTAR (E$=$0.1~eV to E$=$1~MeV) consists of a typical power-law continuum with photon index $\Gamma=2$ and high energy cut-off at E$>$100~keV. This is consistent with the average spectrum of local X-ray bright radio-quiet/radio-loud AGNs (e.g., Dadina 2008; Tombesi et al.~2011b, 2013a; Kataoka et al.~2007; Ballo et al.~2011; Sambruna et al.~2007, 2009, 2011; Lohfink et al~2013; Tazaki et al.~2013). In fact, the jet contribution to the X-ray emission of the sources in our sample is found by Tazaki et al.~(in prep.) to be negligible. We use standard Solar abundances (Asplund et al.~2009) and tested three turbulent velocities of 1,000~km~s$^{-1}$, 3,000~km~s$^{-1}$ and 5,000~km~s$^{-1}$.

The broad-band analysis of the 8 observations of 5 sources (namely, 4C$+$74.26, 3C~120, PKS~1549$-$79, 3C~105 and Centaurus~A) with detected Fe K absorption lines are described in detail in Appendix A. The best-fit XSTAR parameters of their Fe K absorbers, along with those already reported in the literature for other radio-loud AGNs of this sample, are listed in Table~1. In Appendix A we discuss the robustness of these results for each case, checking that they are not affected by the possible inclusion of rest-frame neutral or ionized partial covering absorption or disk reflection features. For the latter, this was also expected due to the already reported fact that reflection features are generally weak in radio galaxies (e.g., Ballantyne et al.~2004; Kataoka et al.~2007; Larsson et al.~2008; Sambruna et al.~2009, 2011; Tombesi et al.~2011a, 2013a; Lohfink et al.~2013; Tazaki et al.~2013). We point to Tazaki et al.~(in prep.) for a detailed description of the accretion disk and torus reflection features for most of the sources listed in Table~B1.    

The observation number 1c of 4C$+$74.26 is the only case in which we find complexities in the Fe K band in the form of broad emission and absorption which could be modeled as disk reflection and a fast ionized outflow or as a rest-frame partial covering ionized absorber. This is discussed in Appendix A in \S~A1.3. Given that we can not strongly distinguish between these two models, we conservatively exclude the discussion of this particular observation from Table~1 and \S~5. This does not change the overall conclusions on the whole sample.

\begin{table*}
\centering
\begin{minipage}{153mm}
\caption{Best-fit XSTAR parameters of the Fe K absorbers from the broad-band spectral analysis and radio jet inclination for each source.}
\begin{center}
\begin{tabular}{l@{\hspace{0.3cm}}l@{\hspace{0.3cm}}c@{\hspace{0.3cm}}c@{\hspace{0.3cm}}c@{\hspace{0.3cm}}c@{\hspace{0.3cm}}c@{\hspace{0.3cm}}c@{\hspace{0.3cm}}c}
\hline\hline       
Source & Num & log$\xi$ & $N_H$ & $v_\mathrm{out}$ & $\Delta\chi^2/\Delta\nu$ & $\chi^2/\nu$ & $P_\mathrm{F}$ & $\alpha$\\ 
 \scriptsize{(1)} & \scriptsize{(2)} & \scriptsize{(3)} & \scriptsize{(4)} & \scriptsize{(5)} & \scriptsize{(6)} & \scriptsize{(7)} & \scriptsize{(8)} & \scriptsize{(9)}\\
\hline
\multicolumn{9}{c}{This work}\\
\hline
4C$+$74.26 & 1a & $4.62\pm0.25$ & $> 4^*$ & $0.045\pm0.008$ & 13.7/3 & 1451.1/1427 & 99.6\% & $<$49$^{\circ}$$^{f}$\\[+2pt]
3C~120 & 9f & $4.91\pm1.03$ & $> 2^*$ & $0.161\pm0.006$ & 15.3/3 & 2933.0/2540 & 99.5\% & 20.5$^{\circ}$$\pm$1.8$^{\circ}$$^g$\\[+2pt]
PKS~1549-79 & 12a & $4.91\pm0.49$ & $> 14$ & $0.276\pm0.006$ & 25.0/4 & 850.0/930 & 99.99\% & $<$55$^{\circ}$$^h$\\[+2pt]
          &                   &     tied      & tied   & $0.427\pm0.005$ & & & & \\[+2pt]
3C~105 & 16 & $3.81\pm1.30$ & $> 2^*$ & $0.227\pm0.033$ & 10.5/3 & 226.4/253 & 99\% & $>$60$^{\circ}$\\[+2pt]
Centaurus~A & 26c & $4.33\pm0.03$ & $4.2\pm0.4$ & $<$0.004$^*$ & 126.5/2 & 2830.9/2682 & $\gg$99.99\% & 50$^{\circ}$--80$^{\circ}$$^i$ \\[+2pt]
            & 26d & $4.39\pm0.06$ & $4.1\pm0.9$ & $<$0.005$^*$ & 54.7/2 & 2852.2/2661 & $\gg$99.99\% & \\[+2pt]
            & 26e & $4.31\pm0.10$ & $4.0\pm1.1$ & $<$0.003$^*$ & 105.3/2 & 2855.8/2698 & $\gg$99.99\% & \\[+2pt]
\hline
\multicolumn{9}{c}{From the literature}\\
\hline
4C$+$74.26 & 1b$^c$ & $4.06\pm0.45$ & $>0.6^{*}$ & $0.185\pm0.026$ & & & & \\[+2pt]
3C~390.3 & 6c$^a$ & $5.60^{+0.20}_{-0.80}$ & $>3^*$ & $0.146\pm0.004$ & & & & 19$^{\circ}$--33$^{\circ}$$^j$ \\[+2pt]
3C~111 & 7b$^a$ & $5.00\pm0.30$ & $>20^*$ & $0.041\pm0.003$ & & & & 18.1$^{\circ}$$\pm$5.1$^{\circ}$$^g$ \\[+2pt]
       & 7d$^b$ & $4.32\pm0.12$ & $7.7\pm2.9$ & $0.106\pm0.006$ & & & & \\[+2pt]
3C~120 & 9c,d,e$^a$ & $3.80\pm0.20$ & $1.1^{+0.5}_{-0.4}$ & $0.076\pm0.003$ & & & & \\[+2pt]
3C~445 & 18b$^{d, e}$ & $1.42^{+0.13}_{-0.08}$ & $18.5^{+0.6}_{-0.7}$ & $0.034\pm0.001$ & & & & 60$^{\circ}$--70$^{\circ}$$^k$ \\[+2pt]
\hline   
\end{tabular}
\end{center}
{\em Note.} (1) Object name. (2) Observation number. number. (3) Logarithm of the ionization parameter in units of erg~s$^{-1}$~cm. (4) Column density in units of $10^{22}$~cm$^{-2}$. (5) Outflow velocity in units of the speed of light, c. (6) $\chi^2$ improvements with respect to the additional parameters with respect to the baseline model. (7) Final best-fit $\chi^2$ and degrees of freedom $\nu$. (8) F-test confidence level. (9) Radio jet inclination angle. Given the lack of a reported value in the literature, we assume an inclination angle of $\alpha$$>$60$^{\circ}$ for 3C~105 given its optical classification as type 2. $^*$ upper or lower limit at the 90\%. $^a$ Tombesi et al.~(2010a), $^b$ Tombesi et al.~(2011a), $^c$ Gofford et al.~(2013), $^d$ Braito et al.~(2011), $^e$ Reeves et al.~(2010), $^f$ Pearson et al.~(1992), $^g$  Jorstad et al.~(2005), $^h$ Holt et al.~(2006), $^i$ Tingay et al.~(1998), $^j$Eracleous \& Halpern (1998), $^k$ Sambruna et al.~(2007).
\end{minipage}
\end{table*}

\section{Results}

\subsection{Incidence of UFOs in radio-loud AGNs}

Following the definition of UFOs as those highly ionized Fe K absorbers with outflow velocities higher than 10,000~km~s$^{-1}$ (Tombesi et al.~2010a,b), here we derive an estimate of the incidence of such absorbers in the sample of radio-loud AGNs. If we consider only the spectra analyzed in this work, we find that UFOs are detected in only 4/48 ($\simeq$8\%) of the observations or 4/25 ($\simeq$16\%) of the sources, respectively. Instead, if we include also all the observations of the sources in the sample that were already studied in detail in the literature, we find that UFOs are detected in a total of 12/61 ($\simeq$20\%) of the observations or 7/26 ($\simeq$27\%) of the sources, respectively.    


It is important to note that these detection fractions can significantly underestimate the real values because they do not take into account the number of observations that have low counts levels and therefore do not have enough statistics for the detection of Fe K absorption lines even if present. 

Following Tombesi et al.~(2010b), the net 4--10 keV counts of each observation can be regarded as an indication of their statistics and we can estimate the number of counts required to obtain a reliable line detection. We can estimate the effect of the limited statistics available in the spectra of the whole sample considering a mean EW of 50~eV and a line energy of 8~keV, corresponding to a blue-shifted velocity of about 0.1c for Fe~XXVI Ly$\alpha$ (Tombesi et al.~2010b). These values are equivalent to the average EW and energy of the blue-shifted Fe K absorption lines in radio-loud AGNs reported here and in the literature of $<$EW$>$$\simeq$45~eV and $<$E$>$$\simeq$8.2~keV. The 4−-10~keV counts level needed for a 3$\sigma$ detection of such a line is $\simeq$$10^4$ and $\simeq$$2\times 10^{4}$ counts for the EPIC-pn instrument onboard \emph{XMM-Newton} and the combined XIS front illuminated CCDs of \emph{Suzaku}, respectively. 
This criterion is met by 15 \emph{XMM-Newton} and 19 \emph{Suzaku} observations, respectively. 
Therefore, as we can see from the cumulative distributions of counts in Figure~3, only about 56\% of the available \emph{XMM-Newton} and \emph{Suzaku} observations have enough counts for a proper detection of highly ionized UFOs with a typical velocity of $\sim$0.1c. 

This test demonstrates that the comparison of UFO detections with respect to the total number of observations or sources can drastically underestimate the actual fraction because this does not take into account the fact that several observations can have insufficient counts. Therefore, considering only the observations with enough counts we have a UFO detection fraction of 12/34 ($\simeq$32\%), which corresponds to an incidence of 6/12 ($\simeq$50\%) for this sub-sample of sources. 

However, f we want to extrapolate this fraction to the whole sample, we have to take into account our ignorance due to the number of sources with low S/N observations. Assuming that a UFO is present in either none or all of the sources with low S/N we can estimate the lower and upper limits corresponding to 7/26 ($\simeq$30\%) and 20/26 ($\simeq$70\%), respectively. Therefore, the incidence of UFOs in our sample of radio-loud AGNs can be estimated to be in the range $f$$\simeq$$(50\pm20)$\%. Interestingly, this is consistent with what reported for similar studies of large samples of radio-quiet AGNs (Tombesi et al. 2010; Patrik et al.~2012; Gofford et al.~2013). 

For the sources optically classified as type 1 or type 2 we have a detection fraction of 5/12 ($\simeq$42\%) and 1/12 ($\simeq$8\%), respectively. However, taking into account that the sources having spectra with enough counts are only 10 for type 1s and 2 for type 2s, the most likely fractions can be estimated to be 5/10 and 1/2, respectively. This translates in an incidence of UFOs of $\sim$50\% for both type 1s and type 2s. However, these values are only indicative, given the small number statistics, especially for type 2s. 

The Monte Carlo simulations discussed in \S4 ensure that the reported blue-shifted absorption lines indicative of UFOs are detected at a confidence level $>$95\% against random positive and negative fluctuations in the wide energy band between E$=$7--10~keV. Therefore, we would naively expect to have a random emission or absorption line in the interval E$=$7--10~keV in one spectrum out of 20. Considering all the 61 observations, the maximum fraction of random detections would be $\sim$3. Moreover, considering that there are only 34 observations with enough S/N to detect a typical UFO if present, this would results in a maximum of $\sim$1--2 random lines. These estimates are much lower than 12, the number of actual spectra with at least one highly blue-shifted absorption line detected.

As already noted by Tombesi et al.~(2010b) and Gofford et al.~(2013), a more quantitative estimate of the global probability that the observed absorption features are purely due to statistical noise can be derived using the binomial distribution. Given that there are $n = 12$ spectra with at least one blue-shifted absorption line detected with a Monte Carlo probability $\ge$95\% out of a total of $N = 61$, the probability of one of these lines to be due to random noise is $p < 0.05$. Therefore, their global random probability is $P < 3\times 10^{-5}$ and therefore the confidence level is $>$99.997\% ($>$4$\sigma$). However, considering that only $N=34$ spectra have enough S/N, the global random probability would further decrease to $P <  4\times 10^{-8}$ ($>$5.5$\sigma$).

From Table~B1 in Appendix~B we see that of the 23 sources with a Fanaroff \& Riley classification, 20 are identified as FR~IIs and only 3 as FR~Is. As discussed in \S2, the paucity of FR~Is can be explained as a selection effect due to the fact that we consider an X-ray selected flux-limited sample and they have a systematically lower luminosity than FR~IIs (e.g., Fabbiano et al.~1984; Hardcastle et al.~2009). Comparing with Table~1, we see that 6/20 ($\sim$30\%) of FR~IIs and 1/3 ($\sim$30\%) of FR~Is have outflows with observed velocities higher than 10,000~km~$^{-1}$, identified as UFOs, and only 1/3 of FR~Is (Centaurus~A) has an Fe K absorber with an observed low velocity of $\la$1,000~km~s$^{-1}$. However, if we limit only to the number of sources having at least one observation with enough S/N to detect UFOs if present, we find that 5/10 ($\sim$50\%) of FR~IIs and 1/2 ($\sim$50\%) of FR~Is show UFOs and 1/2 of FR~Is (Centaurus~A) show an Fe K absorber with an observed low velocity of $\la$1,000~km~s$^{-1}$. The very limited number of FR~Is compared to FR~IIs in this sample (3/20) does not allow to perform a significant comparison between these two populations. 

Finally, we note that the majority of the sources are classified as FR II (20) and only a few as FR I (3). Moreover, in the definition of the sample we specifically excluded those sources classified as blazars. Therefore, the estimate of the incidence of UFOs in radio-loud AGNs should be considered within the limit of this sample.

   \begin{figure}
   \centering
    \includegraphics[width=7cm,height=6cm,angle=0]{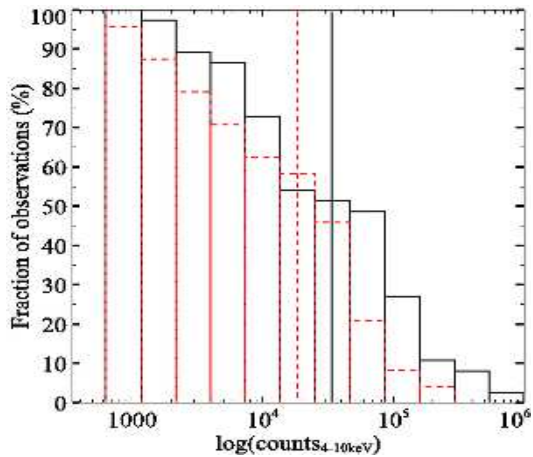}
   \caption{Cumulative distributions of the net 4--10~keV counts of the \emph{XMM-Newton} (red dashed lines) and \emph{Suzaku} (black solid lines) observations. The vertical lines refer to the values of $10^{4}$ counts (red dashed) and $2\times 10^{4}$ counts (black solid). These values correspond to the counts levels required for a significant detection of a typical UFO absorption line in the \emph{XMM-Newton} and \emph{Suzaku} spectra, respectively.}
    \end{figure}

\subsection{Parameters of the Fe K absorbers in radio-loud AGNs}

The distributions of the XSTAR parameters of the Fe K absorbers detected in the radio-loud AGN sample are reported in Figure~4. These values include also those sources with equivalent studies already reported in the literature and listed in Table~1. 

The upper panel of Figure~4 shows the distribution of the average outflow velocity of the Fe K absorbers detected for each source. The values are broadly distributed between $v_\mathrm{out}$$\la$1,000~km~$^{-1}$ (for Centaurus~A) and  $v_\mathrm{out}$$\simeq$0.4c. All the outflows, excluding only Centaurus~A, have velocities higher than 10,000~km~s$^{-1}$ and can be directly identified with UFOs. In general, the Fe K absorbers have mildly-relativistic mean and median velocities of $v_{out}$$\simeq$0.133c ($v_{out}$$\simeq$0.152c without Centaurus~A) and $v_{out}$$\simeq$0.117c ($v_{out}$$\simeq$0.119c without Centaurus~A), respectively. 

As already discussed in Tombesi et al.~(2011b), this velocity distribution has a bias that hampers the detection of outflows with high velocities. This is due to the fact that the sensitivity of the \emph{XMM-Newton} and \emph{Suzaku} CCDs quickly drops at energies higher than E$\simeq$8--9~keV and therefore outflows with velocities higher than $v_\mathrm{out}$$\sim$0.3--0.4c would be impossible to detect for these local sources. For instance, UFOs with even higher  velocities have been reported for a few bright quasars at much higher cosmological redshifts (e.g., Chartas et al.~2002, 2003, 2009; Lanzuisi et al.~2012). Moreover, these outflow velocities are only conservative estimates because they represent the projected wind velocity along the line of sight and therefore they depend on the inclination of the source and the wind with respect to the observer. 


The distribution of the ionization parameter shown in the middle panel of Figure~4 has mean and median values of log$\xi$$\simeq$4.18~erg~s$^{-1}$~cm (log$\xi$$\simeq$4.57~erg~s$^{-1}$~cm without 3C~445) and log$\xi$$\simeq$4.35~erg~s$^{-1}$~cm (log$\xi$$\simeq$4.36~erg~s$^{-1}$~cm without 3C~445), respectively.
This is consistent with the fact that the material is usually observed through highly ionized Fe~XXV--XXVI absorption lines. However, we note the presence of an outlier at log$\xi$$\simeq$1.5~erg~s$^{-1}$~cm due to 3C~445. As already discussed in Reeves et al.~(2010) and Braito et al.~(2011) this could be due to the fact that we are observing the wind in this source with a high inclination or the wind geometry is mainly equatorial. Indeed, this picture is consistent with the high inclination of its radio jet (Eracleous \& Halpern 1998; Sambruna et al.~2007) and the expectation of increasing/decreasing column density/ionization of accretion disk winds moving from low to high inclination angles (e.g., Proga \& Kallman 2004; Fukumura et al.~2010).

The distribution of the average observed column density of the Fe K absorbers for each source listed in Table~1 is shown in the lower panel of Figure~4. This is distributed in the range between log$N_\mathrm{H}$$\simeq$22~cm$^{-2}$ and log$N_\mathrm{H}$$\simeq$23.5~cm$^{-2}$. Considering only the values constrained within the errors (reported in gray), the mean and median values are log$N_\mathrm{H}$$=$22.7~cm$^{-2}$ and log$N_\mathrm{H}$$=$22.75~cm$^{-2}$, respectively. Instead, considering all the values including the lower limits (showed as diagonal crosses) the mean and median values of the column densities are $N_\mathrm{H}$$>$22.66~cm$^{-2}$ and $N_\mathrm{H}$$>$22.55~cm$^{-2}$, respectively.

We note that the parameters of the Fe K absorbers of the radio-loud AGN sample shown in Figure~4 are broadly consistent with those reported by similar other studies focused on radio-quiet sources (Tombesi et al.~2010; Gofford et al.~2013). A detailed comparison with the Fe K absorbers detected in radio-quiet AGNs will be discussed in a subsequent paper (Tombesi et al. in prep.).

   \begin{figure}
   \centering
    \includegraphics[width=6.5cm,height=5.5cm,angle=0]{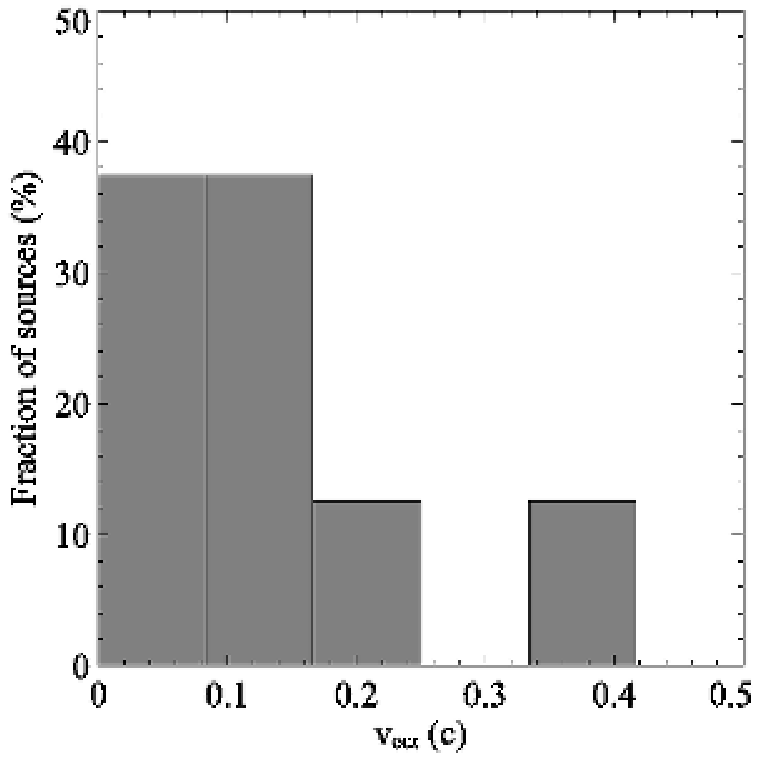}
    \includegraphics[width=6.5cm,height=5.5cm,angle=0]{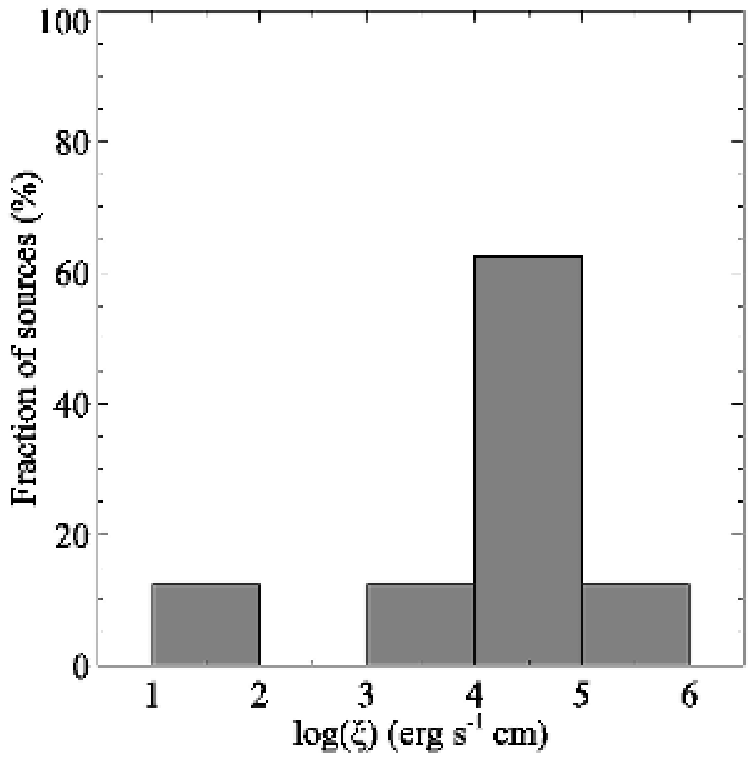}
    \includegraphics[width=6.5cm,height=5.5cm,angle=0]{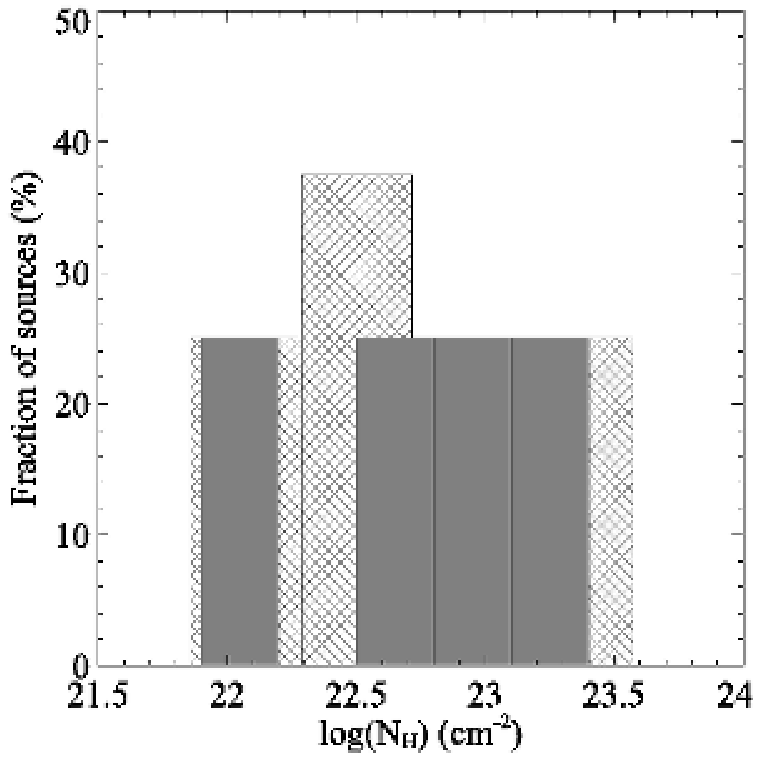}
   \caption{Distributions of mean Fe K absorber parameters for each source in the radio-loud AGN sample: logarithm of the mean outflow velocity (upper panel), logarithm of the mean ionization parameter (middle panel) and logarithm of the mean column density (lower panel). In the lower panel, the histogram derived using all the values including lower limits is indicated with diagonal crosses, instead the one derived using the measured values is in gray.}
    \end{figure}

\subsection{Variability of the Fe K absorbers}

As we can see from Table~1, there are three sources with multiple detections of UFOs in different observations. In 4C$+$74.26 the UFO is possibly detected in all the three observations taken respectively in June 2004, October 2007 and November 2011. However, we conservatively do not consider the last observation due to some modeling ambiguities. Comparing the first two observations, the ionization levels and column densities are consistent within the errors but the velocity changed on a time scale of about 3 years. 

The broad-line radio galaxy 3C~111 was observed five times with \emph{XMM-Newton} and \emph{Suzaku}, with the UFO being significantly detected twice. The intervals between these observations are $\sim$6~months, $\sim$2~years and one week. Including an XSTAR table with the typical velocity of $v_\mathrm{out}$$\simeq$0.1c and ionization parameter of log$\xi$$\simeq$4.5~erg~s$^{-1}$~cm for the UFOs from Figure~4 we can estimate the 90\% upper limits on the column density for the observations without a significant UFO detections. We obtain $N_\mathrm{H}$$<$$4\times 10^{22}$~cm$^{-2}$ for 7a,   $N_\mathrm{H}$$<$$8\times 10^{22}$~cm$^{-2}$ for 7c and  $N_\mathrm{H}$$<$$6\times 10^{22}$~cm$^{-2}$ for 7e, respectively. Given the limited statistics, the column density can only be constrained to be variable on timescales of less than $\sim$6~months and $\sim$2~years between observations 7a-7b and 7b-7c, respectively.  However, considering the EW of the relative absorption line, the UFO is inferred to be variable on timescales down to about one week (Tombesi et al.~2011a). Considering only the two observations with clear detections, the UFO is found to vary in both column density and velocity at the 90\% level on timescales of less than $\sim$2~years.     

The radio galaxy 3C~120 was observed for a total of 7 times, however three \emph{Suzaku} observations taken within two weeks were merged to increase their statistics (Tombesi et al.~2010a) and therefore only 5 spectra were analyzed. The intervals between these observations are $\sim$3.5~yrs, $\sim$2~weeks, $\sim$6~yrs and $\sim$1~week. Including an XSTAR table with typical UFO parameters, as already done previously for 3C~111, we can estimate the 90\% upper limits on the column density for the observations without a significant UFO detections. We obtain $N_\mathrm{H}$$<$$3\times 10^{22}$~cm$^{-2}$ for 9a,   $N_\mathrm{H}$$<$$2\times 10^{22}$~cm$^{-2}$ for 9b and  $N_\mathrm{H}$$<$$5\times 10^{22}$~cm$^{-2}$ for 9g, respectively. Given the limited statistics and the possible intrinsic weakness of the UFO in this source, the variability of the column density can only be roughly estimated to be on timescales shorter than 6 years comparing the two detections. Also the outflow velocity is variable on these timescales. However, we note that the detection of the first outflow was not unambiguously confirmed in a broad-band \emph{Suzaku} spectral analysis using different models by Gofford et al.~(2013). 

The galaxy 3C~390.3 was observed three times, at intervals of $\sim$9~days and $\sim$two years. The UFO was significantly detected in the last long \emph{Suzaku} observations. Also in this case, including an XSTAR table with typical UFO parameters we can estimate the 90\% upper limits on the column density for the observations without a significant UFO detections. We obtain $N_\mathrm{H}$$<$$7\times 10^{22}$~cm$^{-2}$ for 6a and $N_\mathrm{H}$$<$$5\times 10^{22}$~cm$^{-2}$ for 6b, respectively. Given the limited statistics, the variability of the column density can not be constrained, as the upper and lower limits are consistent between all three observations.

PKS~1549$-$79 was also observed twice, with \emph{XMM-Newton} and \emph{Suzaku}, with an interval between the two of about one month. Two Fe K absorption lines indicating a complex UFO with at least two velocity components of $\simeq$0.276c and $\simeq$0.427c was significantly observed in the first \emph{XMM-Newton} observation. However, their presence can not be excluded in the \emph{Suzaku} observation at the 90\% level. In particular, an absorption line at the same energy of E$\simeq$11~keV and consistent EW as in the \emph{XMM-Newton} spectrum is also independently detected in \emph{Suzaku}, although at a lower significance of $\sim$92\% which falls below our detection threshold. Finally, 3C~105 was observed only once with \emph{Suzaku} and the UFO was observed in this observation. The lack of additional observations of this source does not allow its variability to be investigated.

Regarding the detection of the Fe K absorber with low observed velocity of $\la$1,000~km~s$^{-1}$ in Centaurus~A, this has been clearly detected in three long \emph{Suzaku} observations taken within a period of one month in 2009. The values of the ionization and column density are consistent within the errors, indicating that the absorber did not vary on this relatively short time scale. However, there was no significant detection of Fe K absorption lines in the \emph{XMM-Newton} observations taken in 2001 and 2002. Therefore, the absorber can indeed be variable at least on time scales of less than $\sim$7~years.

\subsection{Comparison with the radio jet inclination}

In Table~1 we report the estimates of the inclination angle of the radio jet with respect to the line of sight for the sources with detected Fe K absorbers. Assuming that the jet is perpendicular to the accretion disk, this would represent also the angle with which we are looking at the accretion disk. Considering only the UFOs (i.e., excluding Centaurus~A), we note that the inclination angle spans a wide range of values from $\sim$10$^{\circ}$ to $\sim$70$^{\circ}$, indicating that these winds have a large opening angle and that they are not preferentially equatorial. This is consistent with the high covering fraction derived from the incidence of UFOs in the sample of $\sim$0.3--0.7. 

In Figure~5 we show a comparison between the jet inclination angle and the average column density, ionization parameter and outflow velocity of the UFOs for each source with a detection. Given the large uncertainties (mostly lower limits) on the column densities, we do not see any clear trend between log$N_H$ and jet inclination. Instead, the data are suggestive of a possible trend of decreasing ionization parameter and increasing outflow velocity going from low to intermediate or high inclination angles. However, we note that given the large uncertainties on the jet angles and the presently unknown angle between the jet and the wind, these plots should be considered only as illustrative.

   \begin{figure}
   \centering
    \includegraphics[width=5.5cm,height=7.5cm,angle=270]{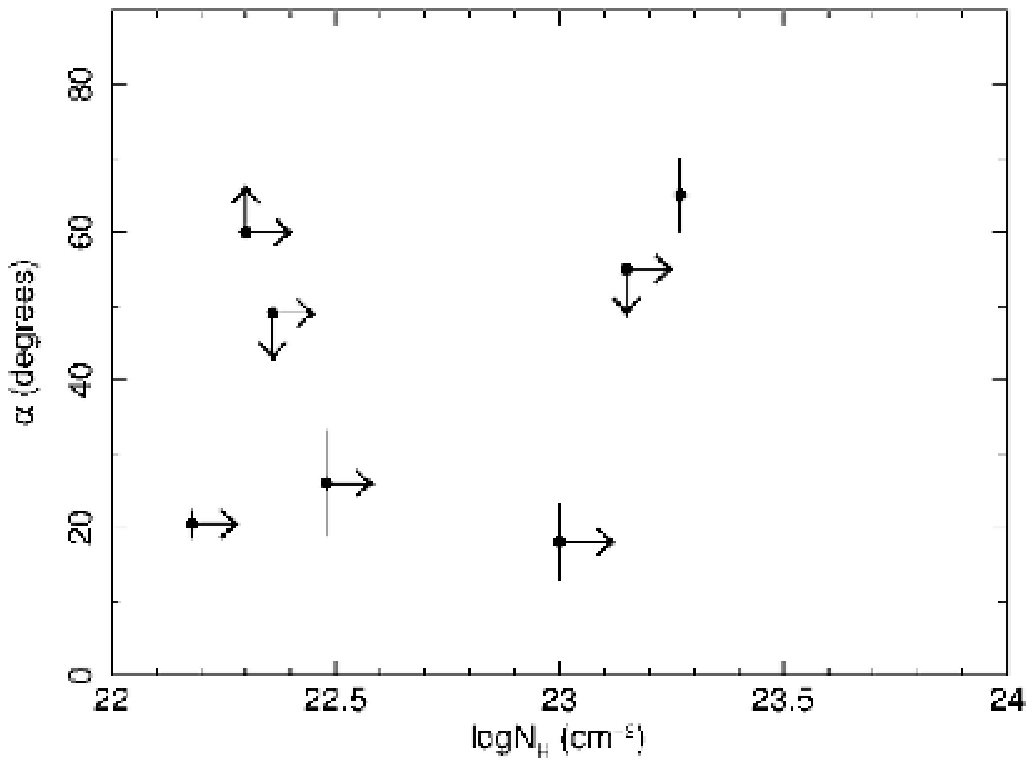}
    \includegraphics[width=5.5cm,height=7.5cm,angle=270]{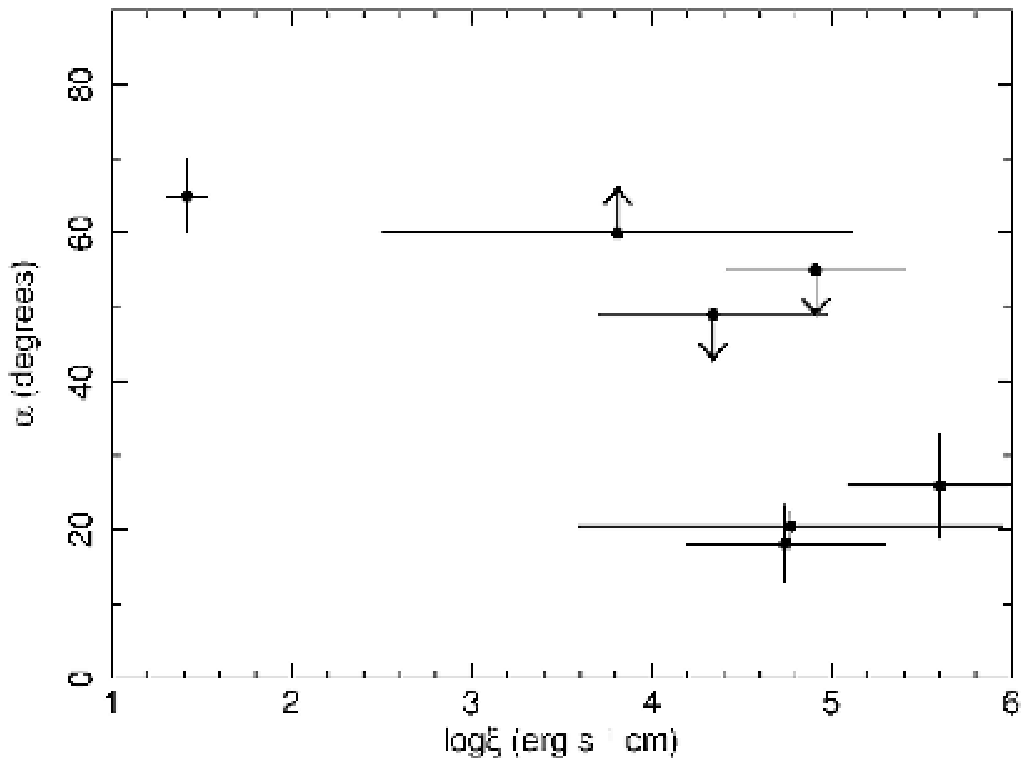}
    \includegraphics[width=5.5cm,height=7.5cm,angle=270]{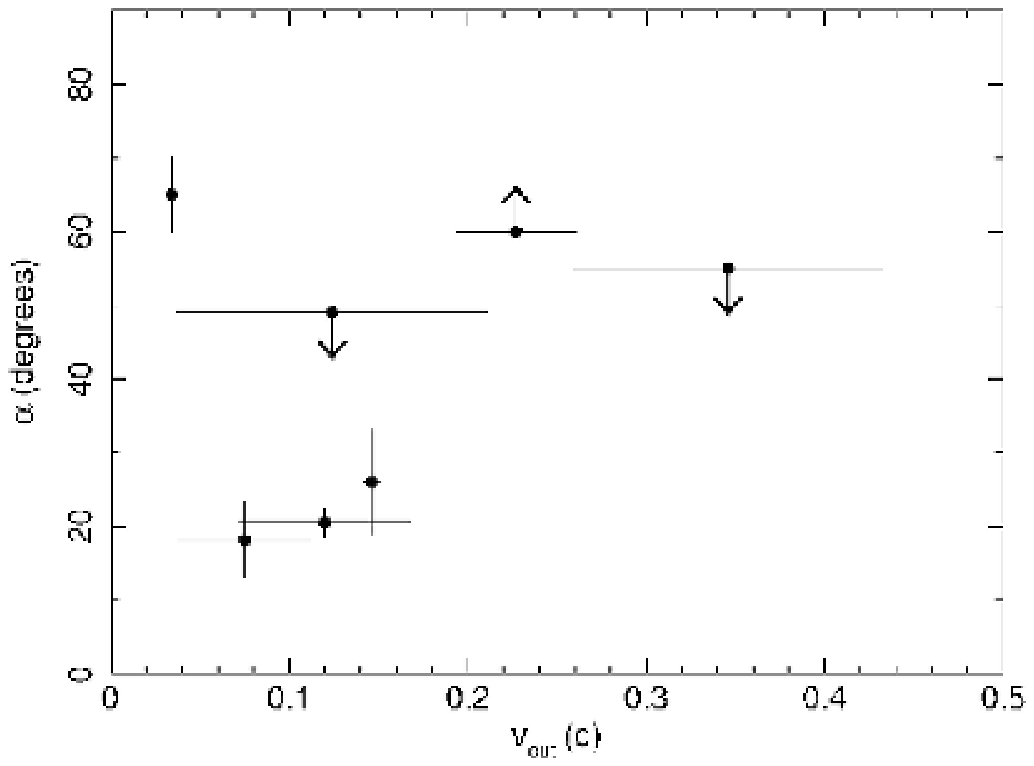}
   \caption{Comparison between the inclination of the radio jet and the average column density (upper panel), ionization parameter (middle panel) and outflow velocity (lower panel) of the UFOs for each source.}
    \end{figure}

\section{Discussion}

Combining the results of this analysis with those in the literature we find that UFOs are detected in $>$27\% of the sources. However, correcting for the number of spectra with insufficient signal-to-noise, we can estimate that the incidence of UFOs is likely in the range $f$$\simeq$$(50\pm20)$\%. From the fraction of radio-loud AGNs with detected UFOs and assuming a steady wind, we can roughly infer an estimate of the minimum covering fraction of $C$$=$$\Omega/4\pi$$\simeq$0.3--0.7, where $\Omega$ is the solid angle subtended by the wind with respect to the X-ray source. Thus, as an ensemble, the UFOs in radio-loud AGNs might cover a significant fraction of the sky as seen by the central X-ray source. This also provides an important geometric constraint indicating that the distribution of the absorbing material is not very collimated as in jets, thereby implying large opening angles. The UFOs are detected for sources with jet inclination angles in the wide range from $\sim$10$^{\circ}$ to $\sim$70$^{\circ}$. This further supports the idea that these winds have indeed a large opening angle and are not preferentially equatorial.

Interestingly, the detection fraction of UFOs is similar to that reported for complementary studies of large samples of radio-quiet AGNs (Tombesi et al.~2010b; Patrik et al.~2012; Gofford et al.~2013). This is also comparable to that of the slower and less ionized warm absorbers detected in the X-ray spectra of Seyfert 1 galaxies (Reynolds 1997; George et al.~1998; Blustin et al.~2005; McKernan et al.~2007; Tombesi et al.~2013b). In particular, warm absorbers have been reported in some X-ray bright broad-line radio galaxies as well (Reynolds 1997; Ballantyne 2005; Reeves et al.~2009, 2010; Torresi et al.~2010, 2012; Braito et al.~2011).

The fact that UFOs are observed in a similar fraction of both radio-loud and radio-quiet AGNs suggests that this dichotomy, which differentiates sources with and without strong radio jets, might actually not hold for UFOs or AGN winds in general. This would actually be expected from the fact that AGN winds should depend mainly, if not only, on the state of the accretion disk and not, for instance, on the black hole spin, as for relativistic jets (e.g., Blandford \& Znajek 1977; Tchekhovskoy, Narayan, \& McKinney 2010, 2011; Tchekhovskoy \& McKinney 2012). For instance, from a theoretical point of view, winds/outflows are ubiquitous in both cold, thin accretion disks and hot, think accretion flows and they can be driven by several mechanisms, such as thermal, radiative or MHD (e.g., Blandford \& Payne 1982; Blandford \& Begelman 1999; King \& Pounds 2003; Proga \& Kallman 2004; McKinney~2006; Ohsuga et al.~2009; Fukumura et al.~2010, 2014; Tchekhovskoy et al.~2011; Yuan, Bu \& Wu 2012; Sadowski et al.~2013, 2014). 

A statistically significant study of the variability of the UFOs is hampered by the limited number of sources with multiple observations and by their inhomogeneous monitorings. In fact, the variability study of the parameters of the UFOs was possible only for the 5 sources with more than one observation. In particular, 3/5 sources have multiple detections of UFOs. Combining the values derived in this study and those collected from the literature in Table~7, we find that the velocity is significantly variable in 3/3 cases, $N_\mathrm{H}$ in 2/3 cases and log$\xi$ in 1/3 cases. Comparing also the observations with and without detections, we find that UFOs can be variable even on timescales as short as a few days. This is consistent with the idea that these absorbers are preferentially observed at $\sim$sub-pc scales from the central SMBH in radio-loud AGNs and that they may also be intermittent, possibly being stronger during jet ejection or outburst events on longer timescales of the order of half to one year (Tombesi et al.~2011a, 2012a, 2013a).

Overall, the characteristics of these UFOs - they high ionization levels, column densities, variability and often mildly-relativistic velocities - are qualitatively in agreement with the picture of complex AGN accretion disk winds inferred from detailed simulations (e.g., King 2010; Ohsuga et al.~2009; Fukumura et al.~2010; Takeuchi et al.~2014). Indeed, the reported limits on the location of the UFOs in a few radio-loud AGNs indicate that they are at distances of $\sim$100--10,000~$r_s$ from the SMBH, therefore at accretion disk scales (e.g., Tombesi et al.~2010a, 2011a, 2012a; Reeves et al.~2010). However, a homogeneous and systematic estimate of the location, mass outflow rate and energetics of the UFOs in this sample, along with a comparison with those detected in radio-quiet AGNs, will be discussed in a subsequent paper (Tombesi et al. in prep.). This will allow to perform a more detailed discussion of the validity of the jet related radio-quiet/radio-loud dichotomy for winds and possibly to better constrain their acceleration mechanisms and energetics (e.g., King \& Pounds 2003; Proga \& Kallman 2004; Everett \& Ballantyne 2004; Everett 2005; Ohsuga et al.~2009; Fukumura et al.~2010, 2014; Kazanas et al.~2012; Ramirez \& Tombesi 2012; Reynolds 2012; Tombesi et al.~2013b).

\section{Conclusions}

We performed a systematic and homogeneous search for Fe K absorption lines in the \emph{XMM-Newton} and \emph{Suzaku} spectra of a large sample of 26 radio-loud AGNs selected from the \emph{Swift} BAT catalog. After an initial investigation of the E$=$3.5--10.5~keV spectra we performed a more detailed broad-band analysis of 8 observations of 5 sources showing possible Fe K absorption lines that were not already reported in the literature. The detection significance of these lines was further investigated through extensive Monte Carlo simulations and the parameters of the Fe K absorbers were derived using the XSTAR photo-ionization code. 

Combining the results of this analysis with those in the literature we find that UFOs are detected in $>$27\% of the sources. However, correcting for the number of spectra with insufficient signal-to-noise, we can estimate that the incidence of UFOs is likely in the range $f$$\simeq$$(50\pm20)$\%. This value is comparable to that of UFOs and warm absorbers in radio-quiet AGNs, suggesting that the jet related radio-quiet/radio-loud AGN dichotomy might not hold for AGN winds. However, we note that the majority of the sources in the present sample are classified as FR II (20) and only a few as FR I (3). Moreover, in the definition of the sample we specifically excluded those sources classified as blazars. Therefore, the estimate of the incidence of UFOs in radio-loud AGNs should be considered within the limit of this sample. 

From the XSTAR fits we derive a distribution of the main Fe K absorber parameters. The mean ($v_{out}$$\simeq$0.133c) and median ($v_\mathrm{out}$$\simeq$0.117c) outflow velocities are mildly-relativistic, but the observed values are broadly distributed between $v_\mathrm{out}$$\la$1,000~km~s$^{-1}$ (for the highly inclined radio galaxy Centaurus~A) and $v_\mathrm{out}$$\simeq$0.4c. The ionization of the material is high, with an average value of log$\xi$$\simeq$4.5~erg~s$^{-1}$~cm, and the lines are consistent with resonance absorption from Fe~XXV--XXVI. Finally, the column densities are larger than log$N_\mathrm{H}$$>$22~cm$^{-2}$.

The UFOs are detected for sources with jet inclination angles in the wide range from $\sim$10$^{\circ}$ to $\sim$70$^{\circ}$. This further supports the idea that these winds have a large opening angle and are not preferentially equatorial. Overall, these characteristics are consistent with the presence of complex accretion disk winds in a significant fraction of radio-loud AGNs. A detailed comparison with the Fe K absorbers detected in radio-quiet AGNs is discussed in a subsequent paper (Tombesi et al. in prep.). Finally, important improvements in these studies are expected from the upcoming ASTRO-H (Takahashi et al.~2012) and the future Athena observatories (Nandra et al.~2013; Cappi et al.~2013).

\section*{Acknowledgments}
FT would like to thank R. M. Sambruna for her contribution to the initial definition of the project and for the comments to the manuscript. FT thanks C.~S. Reynolds and G.~M. Madejski for the useful discussions. FT acknowledges support for this work by the National Aeronautics and Space Administration (NASA) under Grant No. NNX12AH40G issued through the Astrophysics Data Analysis Program, part of the ROSES 2010.

\clearpage

\newpage

\appendix

\section{Broad-band spectral analysis of the sources with absorption lines}

\subsection{4C$+$74.26}

\subsubsection{Observation 1a}

We analized the \emph{XMM-Newton} EPIC-pn spectrum of 4C$+$74.26 (observation number 1a) considering the observed energy range between E$=$0.5--10.5~keV. The best-fit model comprises a neutral absorbed power-law continuum with $\Gamma = 1.74\pm0.01$ and column density $N_H=(5\pm1)\times 10^{20}$~cm$^{-2}$. The neutral absorption component is required at $\gg$99.99\% using the F-test.

The energy of the Fe K$\alpha$ emission line of E$=$$6.50\pm0.04$~keV reported in Table~B2 in Appendix B suggests that the reflection should come from mildly ionized material instead of neutral material expected at E$=$6.4~keV. This evidence was already discussed in Ballantyne~(2005) and Gofford et al.~(2013). Indeed, the inclusion of an ionized reflection component modeled with \emph{xillver} (Garc{\'{\i}}a et al.~2010, 2013, 2014) provides a very significant fit improvement of $\Delta\chi^2/\Delta\nu = 35/1$ with respect to that using a neutral \emph{pexmon} component. The resultant ionization parameter is $\xi = 300^{+53}_{-46}$~erg~s$^{-1}$~cm.

As already reported in Ballantyne (2005) and Gofford et al.~(2013), the soft X-ray spectrum of 4C$+$74.26 shows intense O~VII and O~VIII absorption edges indicative of ionized material along the line of sight. These features are well parameterized with the inclusion of two warm absorber components. Therefore, we use two XSTAR tables with a typical $\Gamma = 2$ ionizing continuum and turbulent velocity of 100~km~s$^{-1}$. Given the relatively low energy resolution of the EPIC-pn spectrum in the soft X-rays, we can not reliably constrain their possible velocity shifts, which were fixed to zero. The fit improvement after including these two ionized warm absorbers is very high, $\Delta\chi^2/\Delta\nu = 476.2/4$, indicating a confidence level of $\gg$99.99\%.

The fit improvement after the inclusion of a Gaussian absorption line at the rest-frame energy of E$=$$7.28\pm0.05$~keV is $\Delta\chi^2/\Delta\nu = 13.0/2$, corresponding to an F-test probability of 99.8\%. Given that the energy of the line is not directly consistent with that expected from ionized iron at rest, we also performed Monte Carlo simulations in order to better estimate its detection level as described in \S4. The resultant Monte Carlo probability is 99.2\%, which is higher than the selection threshold of 95\%. The final model provides a very good representation of the data, as shown in Figure~A1. Instead, the contour plot in the E$=$5--10~keV energy band  with respect to the baseline model without the inclusion of the absorption line is reported in Figure~A2. The best-fit parameters are listed in Table~A1.

If identified as Fe~XXV He$\alpha$ or Fe~XXVI Ly$\alpha$ the absorption line would indicate a significant blue-shifted velocity of $\simeq$0.04--0.08c ($\simeq$12,000--24,000~km~s$^{-1}$). We checked that this narrow feature can not be modeled with a photo-electric edge. The upper limit of the width of the absorption line is $\sigma_v$$<$12,000~km~s$^{-1}$.

We performed a more physically self-consistent fit of this absorption line using the XSTAR table and the method described in \S4. Given that the velocity width of the line is not well constrained, we performed fits using three XSTAR tables with turbulent velocities of 1,000~km~s$^{-1}$, 3,000~km~s$^{-1}$ and 5,000~km~s$^{-1}$, respectively. For each case, we find a single solution indicating a preferred identification of the absorption line as Fe~XXVI Ly$\alpha$. However, the best-fits using the three different turbulent velocities are equivalent at the 90\% level. We took into account this degeneracy in turbulent velocity calculating the average of the best-fit parameters. The best-fit values of the column density, ionization parameter and outflow velocity are $N_\mathrm{H} > 5\times 10^{22}$~cm$^{-2}$, log$\xi$$=$$4.62\pm0.25$~erg~s$^{-1}$~cm and $v_\mathrm{out} = 0.045\pm0.008$c, respectively. The outflow velocity of $13,500\pm2,400$~km~s$^{-1}$ is consistent with a UFO. This modeling provides a good representation of the data, with a fit improvement with respect to the baseline model of $\Delta\chi^2/\Delta\nu = 13.7/3$, which corresponds to a confidence level of 99.6\%. The best-fit XSTAR parameters are reported in the main text in Table~1.   

We checked that the data do not significantly require the inclusion of a rest-frame neutral or ionized partial covering component using \emph{zpcfabs} or \emph{zxipcf} in XSPEC (Reeves et al.~2008). Moreover, as we can see from the lack of emission residuals in the contour plot in Figure~A2, the data do not require any broad emission line indicative of disk reflection. Therefore, these possible model complexities do not affect the results on the Fe K absorption lines for this observation.

   \begin{figure}
   \centering
    \includegraphics[width=8cm,height=6cm,angle=0]{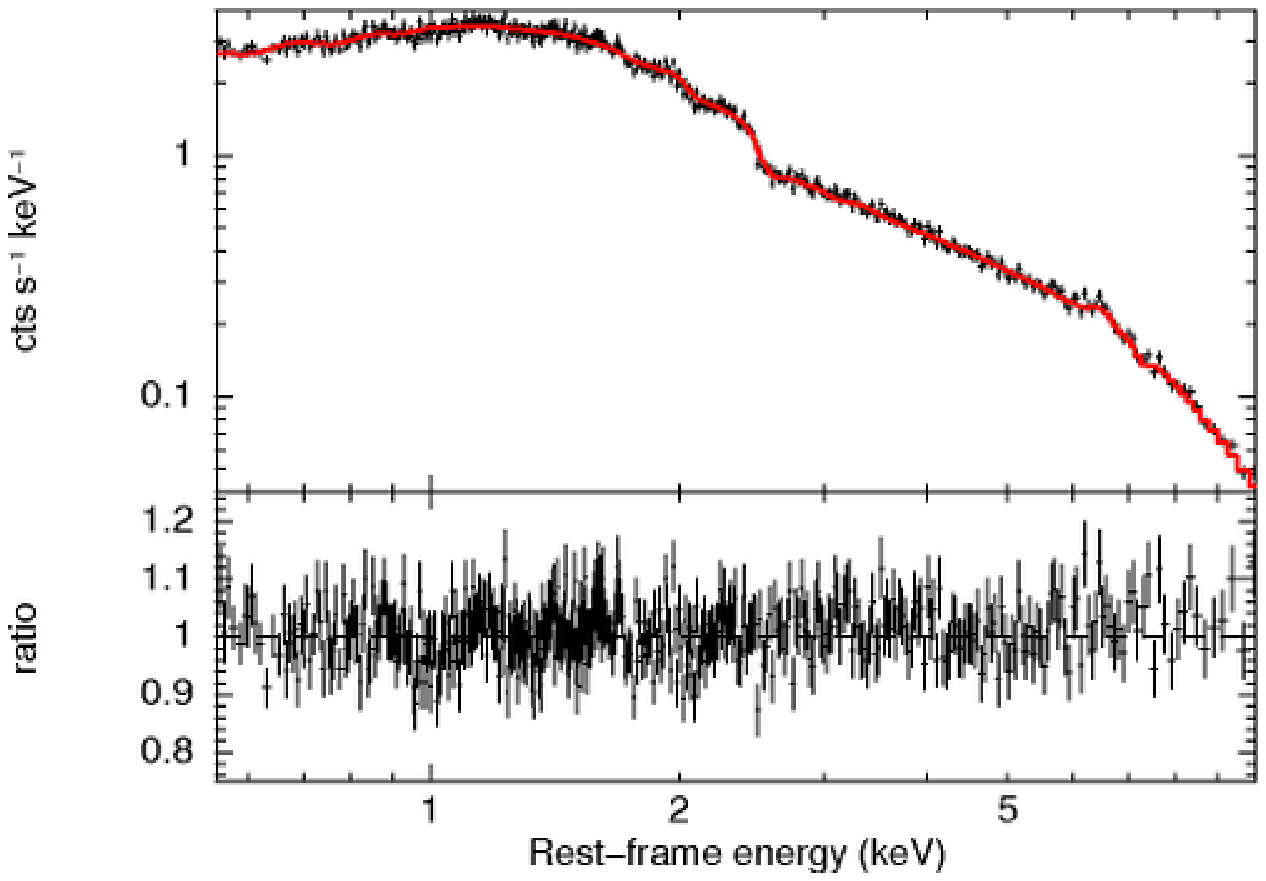}
   \caption{Broad-band \emph{XMM-Newton} EPIC-pn spectrum and ratios against the best-fit model (red) of 4C$+74.26$ (number 1a) in the E$=$0.5--10~keV energy band.}
    \end{figure}

   \begin{figure}
   \centering
    \includegraphics[width=7.5cm,height=5.5cm,angle=0]{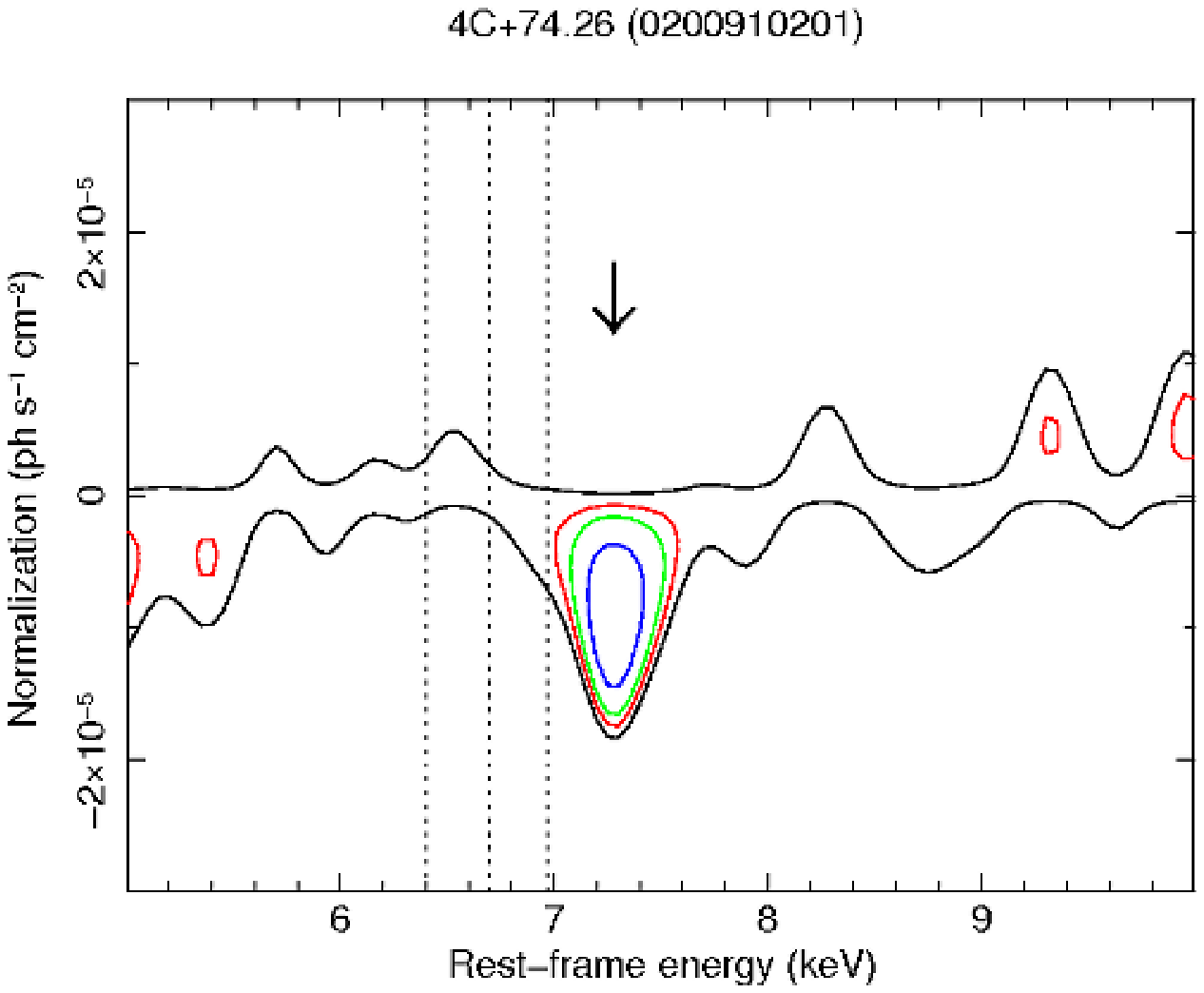}
   \caption{Energy--intensity contour plot in the interval E$=$5--10~keV calculated using the broad-band spectrum of 4C$+$74.26 (number 1a), see text for more details. The arrow points the absorption line. The vertical lines refer to the energies of the neutral Fe K$\alpha$, Fe~XXV and Fe~XXVI lines at E$=$6.4~keV, 6.7~keV and 6.97~keV, respectively.}
    \end{figure}

\begin{table}
\centering
\begin{minipage}{55mm}
\caption{Best-fit broad-band model of 4C$+$74.26 (number 1a).}
\begin{tabular}{l c}
\hline\hline
\multicolumn{2}{l}{Neutral absorption} \\
\hline
$N_H$ ($10^{22}$~cm$^{-2}$) & $0.05\pm0.01$\\
\hline\hline
\multicolumn{2}{l}{Warm absorber 1} \\
\hline
$N_H$ ($10^{22}$~cm$^{-2}$) & $0.40^{+0.10}_{-0.06}$\\[+4pt]
$log\xi$ (erg~s$^{-1}$~cm) & $2.53^{+0.04}_{-0.02}$\\[+4pt]
$z$ & 0.104\\
\hline\hline
\multicolumn{2}{l}{Warm absorber 2} \\
\hline
$N_H$ ($10^{22}$~cm$^{-2}$) & $0.17^{+0.02}_{-0.03}$\\[+4pt]
$log\xi$ (erg~s$^{-1}$~cm) & $1.06^{+0.04}_{-0.09}$\\[+4pt]
$z$ & 0.104\\
\hline\hline
\multicolumn{2}{l}{Power-law continuum} \\
\hline
$\Gamma_\mathrm{Pow}$ & $1.74\pm0.01$ \\
 \hline\hline
\multicolumn{2}{l}{Ionized reflection (\emph{xillver})} \\
\hline
$\Gamma_\mathrm{Xil}$ & $\Gamma_\mathrm{Pow}$\\[+4pt]
$\xi$ (erg~s$^{-1}$~cm) & $300^{+53}_{-46}$\\[+4pt]
$N$ ($10^{-7}$) & $2.2^{+0.6}_{-0.4}$\\[+4pt]
$A_\mathrm{Fe}$ (Solar) & 1\\
\hline\hline
\multicolumn{2}{l}{Absorption line} \\
\hline
$E$ (keV) & $7.28\pm0.05$\\
$\sigma$ (eV) & 100\\ 
$I$ ($10^{-5}$ph~s$^{-1}$cm$^{-2}$) & $-0.90\pm0.20$\\
$EW$ (eV) & $-35\pm8$ \\
\hline\hline
\multicolumn{2}{l}{Absorption line statistics}\\
\hline
$\Delta\chi^2/\Delta\nu$ & 13.0/2\\
F-test (\%) & 99.8\\
MC (\%) & 99.2\\
\hline\hline
\multicolumn{2}{l}{Broad-band model statistics}\\
\hline
$\chi^2/\nu$ & 1451.9/1428\\
\hline   
\end{tabular}
\end{minipage}
\end{table}

\subsubsection{Observation 1c}

We analyzed the combined \emph{Suzaku} XIS-FI and PIN broad-band spectrum 4C$+$74.26 (observation number 1c) in the E$=$0.6--70~keV energy band (excluding the E$=$1.5--2~keV interval). The best-fit model comprises a Galactic absorbed power-law continuum with $\Gamma = 1.76\pm0.01$. From the E$=$3.5--10.5~keV analysis reported in Table~B2 in Appendix B we see that the energy of the Fe K$\alpha$ emission line in this observation of E$=$$6.38\pm0.02$~keV is consistent with neutral material. Therefore, an ionized reflection component is not required in this observation and we include a \emph{pexmon} component with an inclination of $i = 30^{\circ}$, high-energy cut-off of $E_C = 300$~keV and standard Solar abundances. We obtain a reflection fraction of $R = 0.36\pm0.03$.

As already reported in the literature, the soft X-ray spectrum of 4C$+$74.26 shows intense O~VII and O~VIII absorption edges indicative of ionized material along the line of sight (Ballantyne 2005; Gofford et al.~2013). As already discussed in the previous section, these features are well parameterized with the inclusion of two warm absorber components. Therefore, we use again two XSTAR tables with a typical $\Gamma = 2$ ionizing continuum and turbulent velocity of 100~km~s$^{-1}$. Given the relatively low energy resolution of the XIS spectrum in the soft X-rays, we can not reliably constrain their possible velocity shifts, which were fixed to zero. The fit improvement after including these two ionized warm absorbers is very high, $\Delta\chi^2/\Delta\nu = 251.2/4$, indicating a confidence level of $\gg$99.99\%. It should be noted that a neutral absorption component is not formally required in this fit, but the value of the ionization parameter of the second warm absorber is much lower than in the previous observation.

As we can see in the contour plot in Figure~C1, there is evidence for a broad absorption trough at energies of E$\simeq$7--10~keV. Including a Gaussian absorption line at the rest-frame energy of E$\simeq$8~keV with free width we find a high fit improvement ($\Delta\chi^2/\Delta\nu = 38.7/3$), which corresponds to an F-test confidence level of 5.5$\sigma$. The line has a rest-frame energy of E$=$$8.32\pm0.18$~keV, a width of $\sigma = 900^{+200}_{-160}$~eV, intensity $I = (-4.0^{+0.8}_{-0.9})\times 10^{-5}$ ph~s$^{-1}$~cm$^{-2}$ and EW$= -174\pm37$~eV. Given that the energy of the line is not directly consistent with that expected from ionized iron at rest, we also performed Monte Carlo simulations in order to better estimate its detection level as described in \S4. The resultant Monte Carlo probability is $>$99.9\%, which is much higher than the selection threshold of 95\%. The final model provides a very good representation of the data, as shown in Figure~A3. The contour plot in the E$=$5--10~keV energy band with respect to the baseline model without the inclusion of the absorption line is shown in Figure~A4. The best-fit parameters are listed in Table~A2.

We checked the possible alternative phenomenological modeling of the broad absorption feature with a photo-electric edge. The inclusion of an edge provides a slightly higher fit improvement of $\Delta\chi^2/\Delta\nu = 42.2/2$. However, the resultant rest-frame energy of E$= 7.12\pm0.06$~keV is consistent with neutral iron with a maximum optical depth of $\tau = 0.097^{+0.013}_{-0.016}$, which would indicate a large column density of $N_\mathrm{H}\sim 10^{23}$~cm$^{-2}$. This is at odds with the fact that such neutral absorption component is not required when fitting the broad-band spectrum ($\Delta\chi^2/\Delta\nu = 1.4/1$), with a stringent upper limit on the column density of $N_\mathrm{H} < 5\times 10^{20}$~cm$^{-2}$. We also checked the possibile identification of an ionized iron edge including another XSTAR component used for the modeling of the warm absorbers. The resultant fit improvement is negligible ($\Delta\chi^2/\Delta\nu = 2.9/2$). Therefore, even if the phenomenological modeling with a single edge provides a fit comparable to that of a broad Gaussian absorption line, this possibility is rejected because the data do not require any further neutral or lowly ionized absorption component. 

If identified with Fe~XXV He$\alpha$ or Fe~XXVI Ly$\alpha$ the broad absorption line would indicate a significant blue-shifted velocity of $\simeq$0.16--0.19c. If only due to a velocity broadening, the width of the line would suggest a very large velocity of $\sigma_v$$\simeq$32,400~km~s$^{-1}$, which corresponds to FWHM$\simeq$76,400~km~s$^{-1}$.

   \begin{figure}
   \centering
    \includegraphics[width=8cm,height=6cm,angle=0]{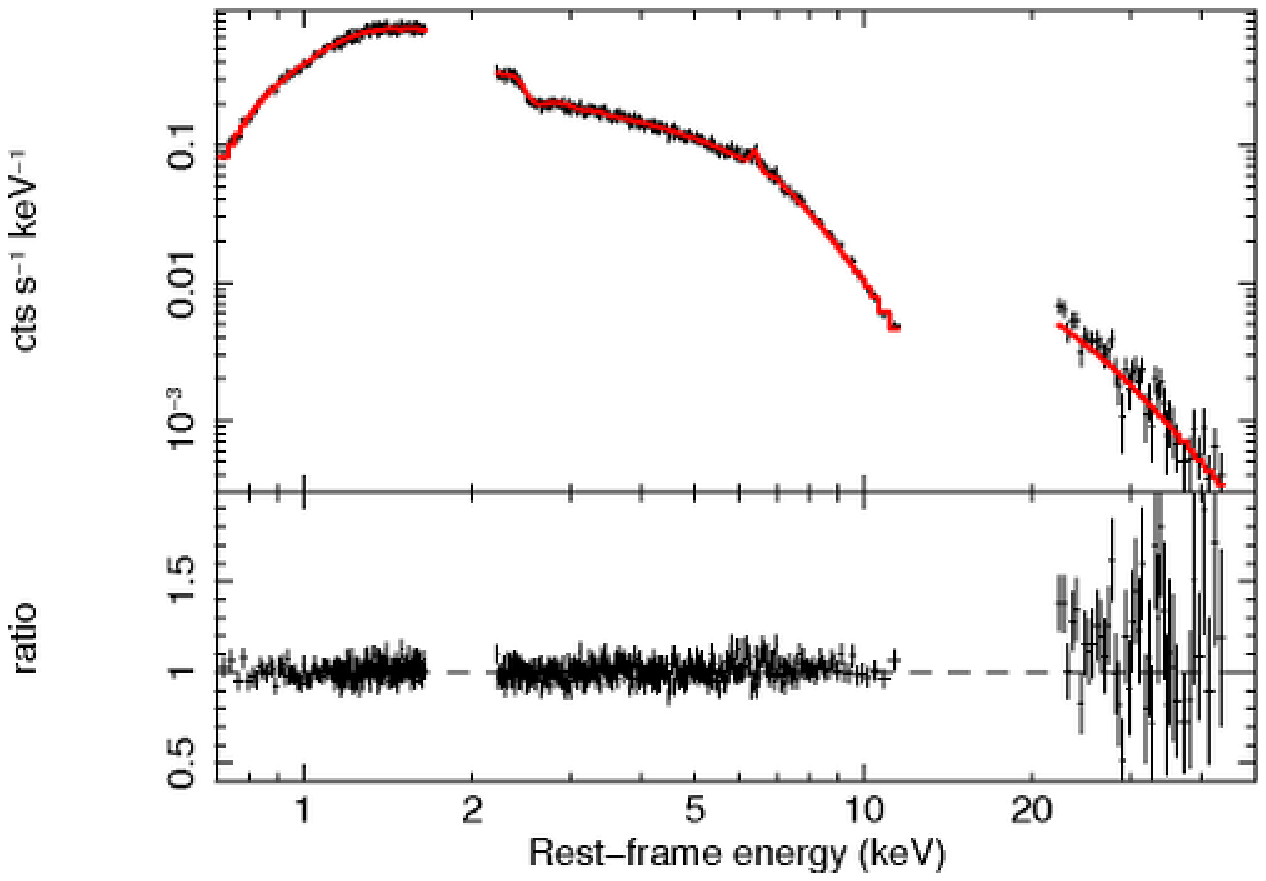}
   \caption{Combined broad-band \emph{Suzaku} XIS-FI and PIN spectrum and ratios against the best-fit model (red) of 4C$+74.26$ (number 1c) in the E$=$0.6--70~keV energy band.}
    \end{figure}

   \begin{figure}
   \centering
    \includegraphics[width=7.5cm,height=5.5cm,angle=0]{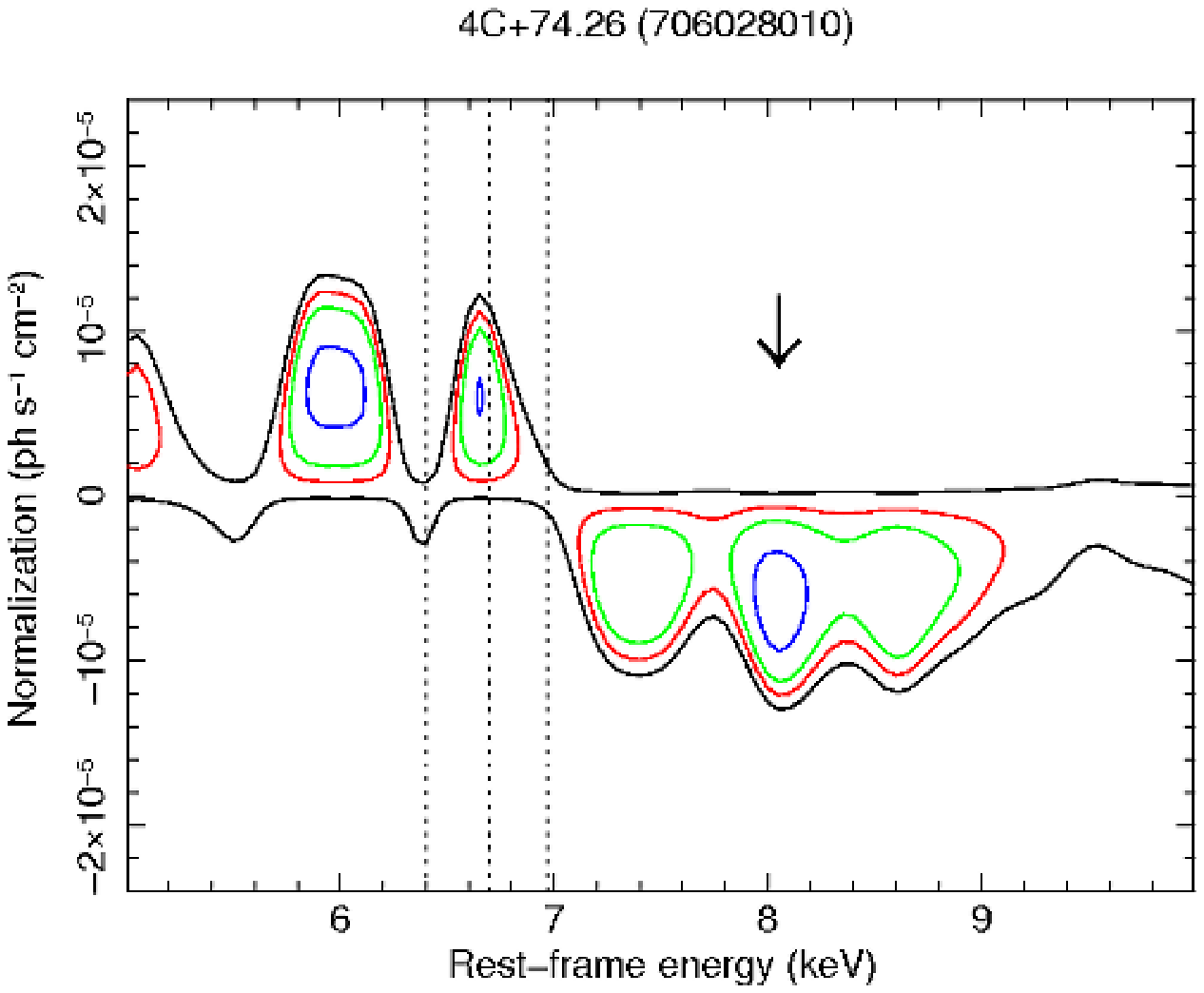}
   \caption{Energy--intensity contour plot in the interval E$=$5--10~keV calculated using the broad-band spectrum of 4C$+$74.26 (number 1c), see text for more details. The arrows points the absorption line. The vertical lines refer to the energies of the neutral Fe K$\alpha$, Fe~XXV and Fe~XXVI lines at E$=$6.4~keV, 6.7~keV and 6.97~keV, respectively.}
    \end{figure}

We checked the fit using a single XSTAR table with a high turbulent velocity of 5,000~km~s$^{-1}$. 
The fit improvement is $\Delta\chi^2/\Delta\nu = 15.8/3$, corresponding to a confidence level of 99.86\%. The absorber has a high column of $N_\mathrm{H} = (1.92^{+7.12}_{-1.28})\times 10^{23}$~cm$^{-2}$ and it is highly ionized, with a ionization parameter of log$\xi$$=$$4.97^{+0.40}_{-1.04}$~erg~s$^{-1}$~cm, indicating that most of the absorption comes from a combination of Fe~XXV--XXVI. The derived outflow velocity is mildly-relativistic, $v_\mathrm{out} = 0.145\pm0.008$c.

However, this XSTAR component alone is not sufficient to explain the width of the line and does not provide a fit comparable to that using the broad Gaussian. Therefore, we checked the possibility that the broad absorption is due to a highly ionized absorber with a wide range of velocities including a second and then a third component. We tie together the column density and ionization parameters of the absorbers but leave their velocity free. The inclusion of a second highly ionized absorber provides a fit improvement of $\Delta\chi^2/\Delta\nu = 16.4/1$, corresponding to a probability of 99.99\%. The best-fit parameters now are log$\xi$$=$$5.20^{+0.15}_{-0.78}$~erg~s$^{-1}$~cm and $N_\mathrm{H} = (4.29^{+2.46}_{-3.56})\times 10^{23}$~cm$^{-2}$. The two outflow velocities are $v_\mathrm{out} = 0.146\pm0.007$c and $v_\mathrm{out} = 0.061\pm0.014$c, respectively.

The inclusion of a third component provides a further fit improvement of $\Delta\chi^2/\Delta\nu = 11.4/1$, corresponding to a confidence level of 99.9\%. The new ionization parameter and column density are log$\xi$$=$$4.98^{+0.30}_{-0.61}$~erg~s$^{-1}$~cm and $N_\mathrm{H} = (1.95^{+3.94}_{-1.09})\times 10^{23}$~cm$^{-2}$. The outflow velocities range from $v_\mathrm{out} = 0.069^{+0.017}_{-0.011}$c to $v_\mathrm{out} = 0.144\pm0.007$c and $v_\mathrm{out} = 0.209\pm0.008$c.

\begin{table}
\centering
\begin{minipage}{55mm}
\caption{Best-fit broad-band model of 4C$+$74.26 (number 1c).}
\begin{tabular}{l c}
\hline\hline
\multicolumn{2}{l}{Warm absorber 1} \\
\hline
$N_H$ ($10^{22}$~cm$^{-2}$) & $0.30^{+0.06}_{-0.09}$\\[+4pt]
$log\xi$ (erg~s$^{-1}$~cm) & $2.71^{+0.04}_{-0.05}$\\[+4pt]
$z$ & 0.104\\
\hline\hline
\multicolumn{2}{l}{Warm absorber 2} \\
\hline
$N_H$ ($10^{22}$~cm$^{-2}$) & $0.07\pm{0.01}$\\[+4pt]
$log\xi$ (erg~s$^{-1}$~cm) & $<0.17^{a}$\\[+4pt]
$z$ & 0.104\\
\hline\hline
\multicolumn{2}{l}{Neutral reflection (\emph{pexmon})} \\
\hline
$\Gamma$ & 1.76$\pm$0.01 \\
$R$ & $0.36\pm0.03$\\
$E_C$ (keV) & 300\\
$A_\mathrm{Fe}$ (Solar) & 1\\
$i$ (deg) & 30\\ 
\hline\hline
\multicolumn{2}{l}{Absorption line} \\
\hline
$E$ (keV) & $8.32\pm0.18$\\
$\sigma$ (eV) & $900^{+200}_{-160}$\\ 
$I$ ($10^{-5}$ph~s$^{-1}$cm$^{-2}$) & $-4.0^{+0.8}_{-0.9}$\\
$EW$ (eV) & $-174\pm37$ \\
\hline\hline
\multicolumn{2}{l}{Absorption line statistics}\\
\hline
$\Delta\chi^2/\Delta\nu$ & 38.7/3\\
F-test (\%) & $\gg$99.99\\
MC (\%) & $>$99.9\\
\hline\hline
\multicolumn{2}{l}{Broad-band model statistics}\\
\hline
$\chi^2/\nu$ & 2132.3/2117\\
\hline   
\end{tabular}
{\em Note.} $^a$ 90\% upper limit. 
\end{minipage}
\end{table}

Therefore, the best-fit of the broad absorption line at E$\simeq$8~keV is provided by a highly ionized absorber with a complex velocity structure, which we modeled as three XSTAR components with mildly-relativistic velocities in the range 0.069--0.209c. These velocities are consistent with those of UFOs. This XSTAR fit provides a global improvement of $\Delta\chi^2/\Delta\nu = 43.6/5$, which corresponds to a high detection level of 5.5$\sigma$. This fit improvement is equivalent to the one derived using a broad Gaussian absorption line. The best-fit XSTAR parameters are reported in the main text in Table~1.

In the scenario of a wind ejected from the accretion disk surrounding the central black hole in 4C$+$74.26, this range of velocities could suggest that the wind is ejected from a relatively wide range of disk radii. 
In fact, assuming that the outflow velocity is comparable to the Keplerian velocity at the wind launching radius (e.g., Fukumura et al.~2010), we obtain the range 25--200~$r_g$ ($r_g = G M_\mathrm{BH}/c^2$). Indeed, this is consistent with the launching region of accretion disk winds (e.g., Proga \& Kallman 2004; Oshuga et al.~2009; Fukumura et al.~2010).

\subsubsection{Spectral model complexities for observation 1c}

From the contour plot with respect to the baseline model for observation 1c of 4C$+$74.26 in Figure~A4 we note that besides the broad absorption residuals at E$\simeq$7--9~keV, there are also emission residuals in the energy range  E$\simeq$6--7~keV. This contour plot shows higher Fe K band emission/absorption complexities with respect to the other cases discussed in Appendix~A.
In panel 1 and 2 of Figure~A5 we show the contour plots calculated after the inclusion in the baseline model of the broad absorption line or the three XSTAR absorption components, respectively. We note that the residuals in emission at E$\simeq$6--7~keV are now much weaker than in Figure~A4, indicating that their presence is possibly dependent on the modeling of the broad absorption at higher energies. 

However, it is important to check the alternative interpretation of these emission residuals as due to reflection from the accretion disk and the check the effect on the modeling of the broad absorption. Initially, we model these emission residuals with a broad Gaussian emission line. The line parameters are E$=$$6.21^{+0.12}_{-0.25}$~keV, $\sigma=438^{+81}_{-0.18}$~eV, $I = (2.1^{+0.9}_{-0.3})$ ph~s$^{-1}$~cm$^{-2}$ and EW$=$$52^{+17}_{-13}$~eV. As we can see from panel 3 of Figure~A5 the inclusion of this broad line is able to model the residuals in emission at E$\simeq$6--7~keV, but it still leaves absorption at higher energies. In fact, the best-fit statistics in this case of $\chi^2/\nu = 2153.7/2117$ is  worse than the one derived including only the broad absorption line of $\chi^2/\nu = 2132.2/2117$.

The energy of the emission line indicates a possible identification with redshifted neutral Fe K$\alpha$ emission originating from reflection off the accretion disk surrounding the central black hole. Therefore, we test a more physical model replacing the Gaussian emission line with a relativistically blurred (\emph{kdblur} in XSPEC)  \emph{pexmon} component. We assume a maximally spinning black hole ($a = 1$), a standard emissivity index $\beta = 3$, an outer disk radius of $r_\mathrm{out}$$=$$400$~$r_\mathrm{g}$, an inclination angle of 30$^{\circ}$ and Solar abundances. Given the limited quality of the dataset, equivalent results are obtained using more complex models (Dauser et al.~2010, 2013; Garc{\'{\i}}a et al.~2014). We estimate the inner radius of the reflecting surface of the disk of $r_\mathrm{in} = 33^{+12}_{-8}$~$r_\mathrm{g}$ and a moderate reflection fraction of $R = 0.43\pm0.06$. This is consistent with the picture of a truncated disk in broad-line radio galaxies, as previously reported for this (Larsson et al.~2008) and other sources (Ballantyne et al.~2004; Kataoka et al.~2007; Sambruna et al.~2009, 2011; Tombesi et al.~2011a, 2013a; Lohfink et al.~2013; Tazaki et al.~2013). Due to the contribution from disk reflection in both intensity of the Fe K$\alpha$ and high-energy hump, the reflection fraction from distant material is slightly lower than the value reported in Table~A2 of $R = 0.18\pm0.05$. We obtain a good fit with $\chi^2/\nu = 2119.5/2118$. 

From the contour plots in panel 4 of Figure~A5 we note that broad residuals in absorption at E$>$7~keV are still present. In fact, the inclusion of a broad absorption line is still required at $\Delta\chi^2/\Delta\nu = 32.2/2$ with parameters consistent at the 1$\sigma$ level with those listed in Table~A2. We also checked that the inclusion of the three XSTAR absorption components is still required at a statistical level of $\Delta\chi^2/\Delta\nu = 34/5$ (corresponding to $\simeq$4.5~$\sigma$) and the parameters are consistent at the 90\% level with those discussed in \S~A1.2. The best-fit statistics in this case is $\chi^2/\nu = 2085.5/2113$. Therefore, even including the possible presence of disk reflection, we find that the presence of the highly ionized absorber is still required with high significance. 

Evidence for hard excesses at E$>$10~keV possibly due to Compton-thick partial covering absorbers has been recently reported in a number of Type 1 AGNs (Tatum et al.~2013). From the ratios in Figure~A3 we do not see a significant hard excess at E$>$10~keV in the \emph{Suzaku} PIN data of this source, limiting the possible presence of such Compton-thick clouds along the line of sight of this source to have very low covering fractions. In fact, in \S~A1.2 we already discussed the fact that the broad absorption at E$>$7~keV can not be adequately modeled with a fully covering neutral or ionized absorber at the source rest frame. However, it is important to check at what level the data can be explained by a partial covering absorber.

At first we include a neutral partial covering absorber modeled with \emph{zpcfabs} in XSPEC. The resultant parameters are a column density is $N_\mathrm{H} = (1.92^{+0.80}_{-0.43})\times 10^{24}$~cm$^{-2}$ and a low covering fraction of $C_\mathrm{f} = 0.20\pm0.04$. The fit improvement of $\Delta\chi^2/\Delta\nu = 24.7/2$ is lower than in the case of the broad gaussian absorption. Moreover, as we can see from the contours in panel 5 of Figure~A5, there are still absorption residuals at E$>$7~keV. In fact, the inclusion of a broad absorption line with parameters consistent at the 1$\sigma$ level with those reported in Table~A2 is still required and provides a high fit improvement of $\Delta\chi^2/\Delta\nu = 39.3/2$. 

Finally, we checked the possible modeling with an ionized partial covering absorber using \emph{zxipcf} in XSPEC (Reeves et al.~2008). We obtain a high column density of $N_\mathrm{H} = (1.34^{+0.07}_{-0.11})\times 10^{24}$~cm$^{-2}$, a relatively high ionization parameter of log$\xi$$=$$2.69^{+0.06}_{-0.17}$~erg~s$^{-1}$~cm and a low covering fraction of $C_\mathrm{f} = 0.33\pm0.03$. As we can see from the panel 6 in Figure~A5, this model provides a relatively good representation of the data, leaving some emission/absorption residuals in the Fe K band. The best-fit statistics is $\chi^2/\nu = 2102.9/2117$. The goodness of fit parameter of the model including the disk reflection and broad Gaussian absorption or three XSTAR components is still slightly better. However, these are not nested models. 

In order to get a more statistically reliable comparison, we start from the model including the disk reflection and include the ionized partial covering component. We obtain a fit improvement of $\Delta\chi^2/\Delta\nu = 26.9/3$. Instead, considering the alternative inclusion of a broad absorption line, we obtain a slightly higher fit improvement of $\Delta\chi^2/\Delta\nu = 33.6/3$. The statistical difference between the two is not very high. 

Therefore, the \emph{Suzaku} spectrum of the observation number 1c of 4C$+$74.26 shows complexities in the Fe K band both in emission and absorption which could be modeled as disk reflection and a fast ionized outflow or a single rest-frame partial covering ionized absorber. Given that we can not strongly distinguish between these two models, we conservatively do not include the possible Fe K outflows detected in this observation in Table~1.

   \begin{figure}
   \centering
    \includegraphics[width=6cm,height=19cm,angle=0]{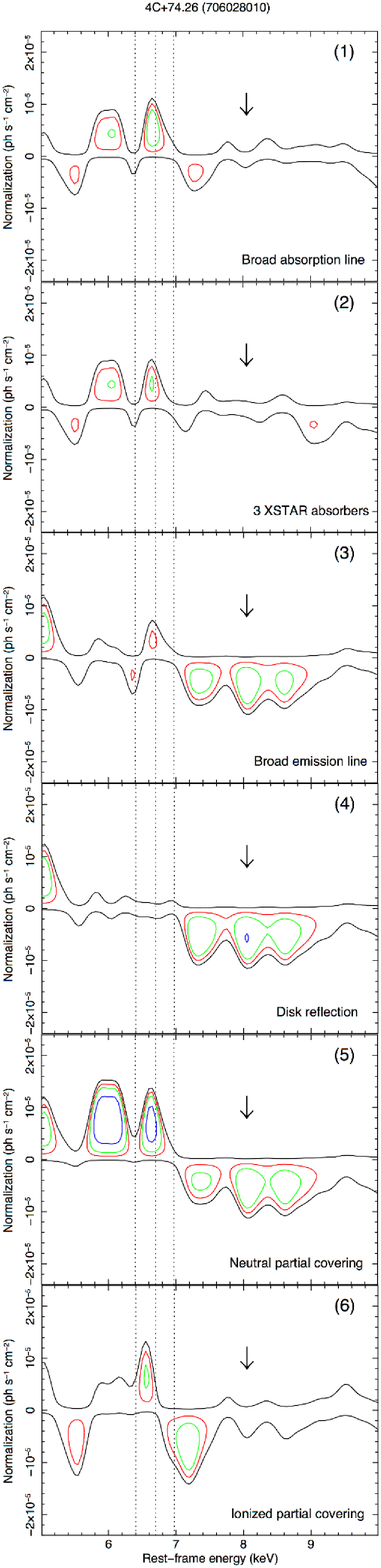}
   \caption{Energy--intensity contour plots of the spectrum of 4C$+$74.26 (number 1c) in the interval E$=$5--10~keV. The different panels indicate the contours calculated using the broad-band model including a broad Gaussian absorption line (1), three XSTAR absorbers (2), a broad Gaussian emission line (3), disk reflection (4), a neutral (5) or ionized (6) partial covering absorber, respectively. The arrows points the absorption line. The vertical lines refer to the energies of the neutral Fe K$\alpha$, Fe~XXV and Fe~XXVI lines at E$=$6.4~keV, 6.7~keV and 6.97~keV, respectively.}
    \end{figure}

\subsection{3C~120}

We analyzed the combined \emph{Suzaku} XIS-FI and PIN broad-band spectrum 3C~120 (observation number 9f) in the E$=$0.6--70~keV energy band (excluding the interval E$=$1.5--2~keV). The best-fit model comprises a Galactic absorbed power-law continuum with $\Gamma = 1.78\pm0.01$. There is no requirement for neutral absorption intrinsic to the source. 

The neutral Fe K$\alpha$ emission line is self-consistently modeled with a \emph{pexmon} neutral reflection component with an inclination of $i = 30^{\circ}$, high-energy cut-off of $E_C = 300$~keV and standard Solar abundances. We derive a relatively low reflection fraction of $R = 0.39\pm0.02$, consistent with previous estimates (Ballantyne et al.~2004; Kataoka et al.~2007; Gofford et al.~2013; Lohfink et al.~2013). The width of the emission line is resolved and therefore we applied a multiplicative Gaussian smoothing of $\sigma = 110\pm10$~eV. 

We find an additional narrow emission line at the rest-frame energy of E$=$$6.85\pm0.05$~keV, possibly due to Fe~XXV--XXVI. There is also a broad excess of emission in the soft X-rays which can be adequately parameterized with a black body component with $kT = 0.38\pm0.03$~keV. 

The fit improvement after the inclusion of a Gaussian absorption line at the rest-frame energy of E$=$$8.23\pm0.04$~keV is $\Delta\chi^2/\Delta\nu = 12.5/2$, corresponding to an F-test probability of 99.6\%. Given that the energy of the line is not directly consistent with that expected from ionized iron at rest, we also performed Monte Carlo simulations in order to better estimate its detection level as described in \S4. The resultant Monte Carlo probability is 96.5\%. The final model provides a very good representation of the data, as shown in Figure~A6. The contour plot  in the E$=$5--10~keV energy band with respect to the baseline model without the inclusion of the absorption line is shown in Figure~A7. The best-fit parameters are reported in Table~A3.

If identified with Fe~XXV He$\alpha$ or Fe~XXVI Ly$\alpha$ the absorption line would indicate a significant blue-shifted velocity of $\simeq$0.15--0.18c. We checked that the possible alternative modeling with a photoelectric edge provides a much worse fit ($\Delta\chi^2/\Delta\nu = 3.8/2$).  The width of the absorption line can be loosely constrained at $\sigma_E = 110^{+70}_{-62}$~eV, which corresponds to a velocity width of $\sigma_v = 4,000^{+2,600}_{-2,300}$~km~s$^{-1}$.

We performed a more physically self-consistent fit of this absorption line using the XSTAR table and the method described in \S4. The velocity width of the line is consistent with the values of 3,000~km~s$^{-1}$ and 5,000~km~s$^{-1}$ within the errors. Therefore, we used both XSTAR tables with these turbulent velocities. We find two equivalent fits at the 90\% and calculate the average. The best-fit column density is $N_\mathrm{H} > 2\times 10^{22}$~cm$^{-2}$, the ionization parameter is log$\xi$$=$$4.91\pm1.03$~erg~s$^{-1}$~cm and the outflow velocity is $v_\mathrm{out} = 0.161\pm0.006$c. The absorption is dominated by Fe~XXV--XXVI ions and the velocity is consistent with a UFO. This modeling provides a good representation of the data, with a fit improvement with respect to the baseline model of $\Delta\chi^2/\Delta\nu = 15.3/3$, which corresponds to a confidence level of 99.5\%. The best-fit XSTAR parameters are reported in the main text in Table~1.

We checked that the data do not significantly require the inclusion of a rest-frame neutral or ionized partial covering component using \emph{zpcfabs} or \emph{zxipcf} in XSPEC (Reeves et al.~2008). Therefore, these possible model complexities do not affect the results on the Fe K absorption lines for this observation. 

The inclusion of a disk reflection component modeled with \emph{pexmon} and relativistically blurred with \emph{kdblur} in XSPEC provides only a marginal improvement to the fit of $\Delta\chi^2/\Delta\nu = 8.6/2$, corresponding to a confidence level of 98\%. We assume a maximally spinning black hole ($a = 1$), a standard emissivity index $\beta = 3$, an outer disk radius of $r_\mathrm{out}$$=$$400$~$r_\mathrm{g}$, an inclination angle of 30$^{\circ}$ and Solar abundances. The inner radius of the disk reflecting surface is suggested to be relatively far away from the black hole at $r_{in} > 25$~$r_g$ and subsequently the reflection fraction is low $R\simeq 0.2$. This is consistent with previous evidences of a truncated disk in this source (e.g., Ballantyne et al.~2004; Kataoka et al.~2007), possibly linked with the disk-jet coupling (Lohfink et al.~2013). 

Even including the possible weak disk reflection component, we find that the parameters of the Fe K absorber are consistent within the 1$\sigma$ uncertainties. Therefore, this possible additional model component does not affect the results on the Fe K absorption lines for this observation.

   \begin{figure}
   \centering
    \includegraphics[width=8cm,height=6cm,angle=0]{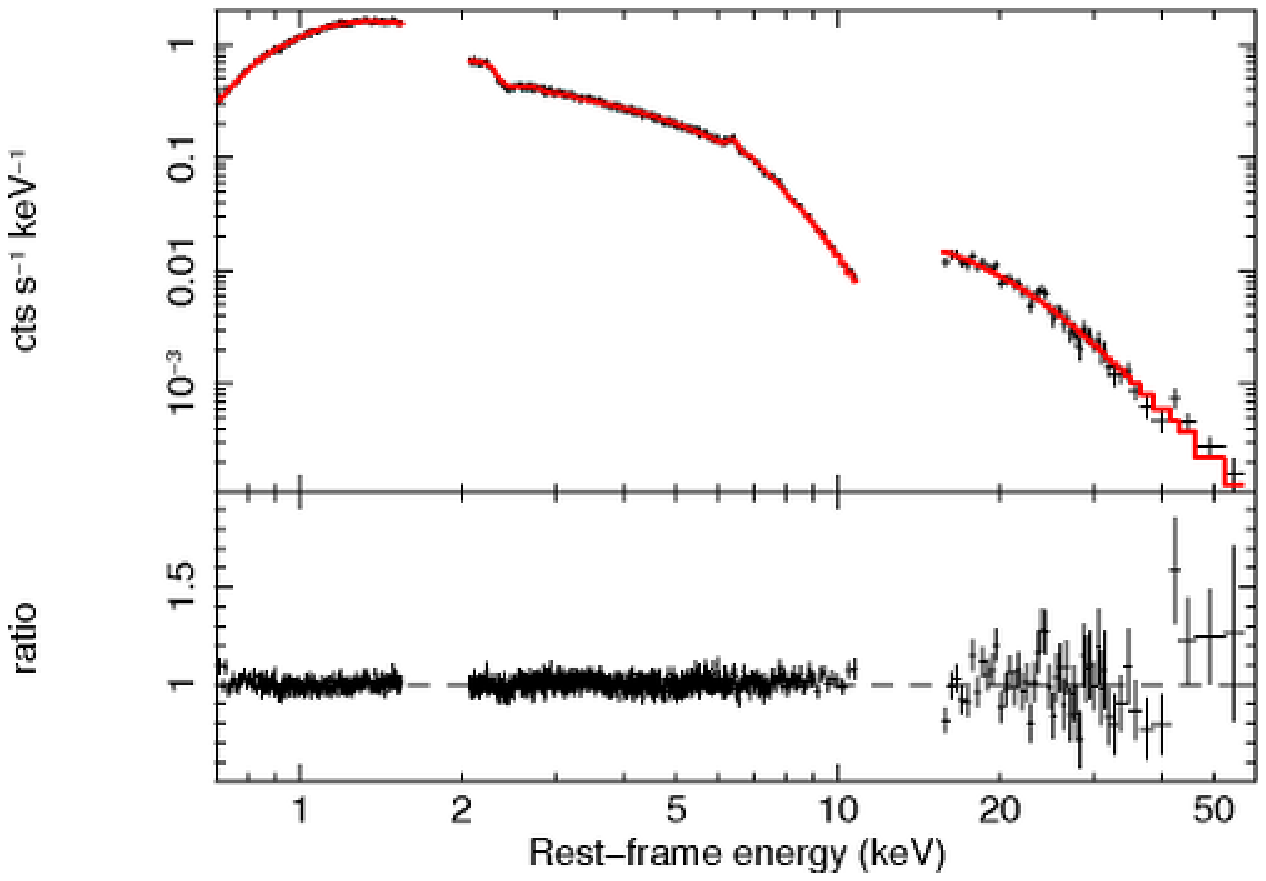}
   \caption{Combined broad-band \emph{Suzaku} XIS-FI and PIN spectrum and ratios against the best-fit model (red) of 3C~120 (number 9f) in the E$=$0.6--70~keV energy band.}
    \end{figure}

   \begin{figure}
   \centering
    \includegraphics[width=7.5cm,height=5.5cm,angle=0]{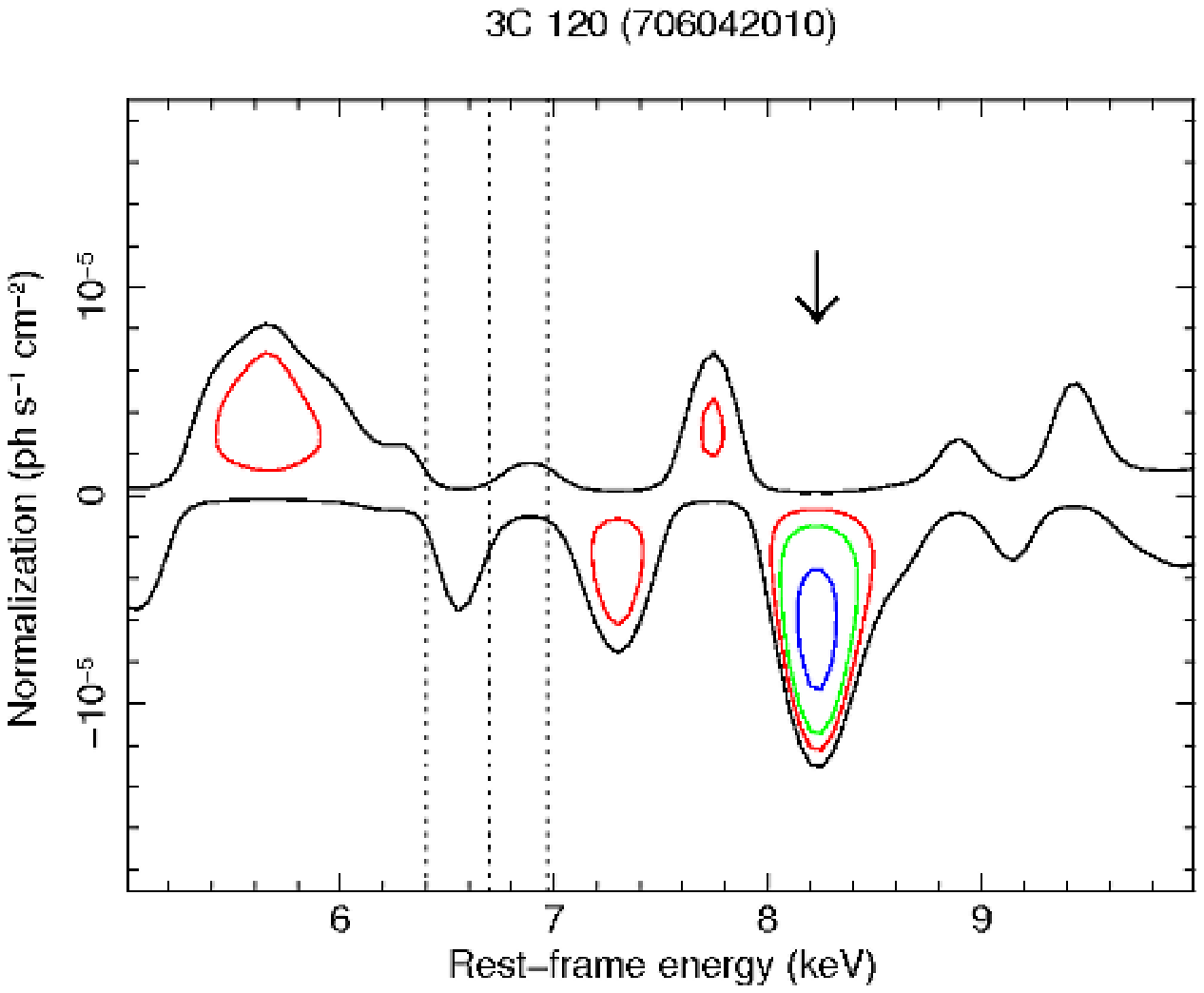}
   \caption{Energy--intensity contour plot in the interval E$=$5--10~keV calculated using the broad-band spectrum of 3C~120 (number 9f), see text for more details. The arrows points the absorption line. The vertical lines refer to the energies of the neutral Fe K$\alpha$, Fe~XXV and Fe~XXVI lines at E$=$6.4~keV, 6.7~keV and 6.97~keV, respectively.}
    \end{figure}

\begin{table}
\centering
\begin{minipage}{55mm}
\caption{Best-fit broad-band model of 3C~120 (number 9f).}
\begin{tabular}{l c}
\hline\hline
\multicolumn{2}{l}{Black body} \\
\hline
$kT$ (keV) & $0.38\pm0.03$\\
$N$ ($10^{-5}$) & $1.3\pm0.3$ \\
\hline\hline
\multicolumn{2}{l}{Neutral reflection (\emph{pexmon})} \\
\hline
$\Gamma$ & 1.78$\pm$0.01 \\
$R$ & $0.39\pm0.02$\\
$E_C$ (keV) & 300\\
$A_\mathrm{Fe}$ (Solar) & 1\\
$i$ (deg) & 30\\ 
$\sigma$ (eV) & $110\pm10$\\
 \hline\hline
\multicolumn{2}{l}{Emission line} \\
\hline
$E$ (keV) & 6.85$\pm$0.04\\
$\sigma$ (eV) & 10\\ 
$I$ ($10^{-5}$ph~s$^{-1}$cm$^{-2}$) & $-0.60\pm0.15$\\
$EW$ (eV) & $12\pm3$ \\
\hline\hline
\multicolumn{2}{l}{Absorption line} \\
\hline
$E$ (keV) & 8.23$\pm$0.06\\
$\sigma$ (eV) & $110^{+70}_{-62}$\\ 
$I$ ($10^{-5}$ph~s$^{-1}$cm$^{-2}$) & $-0.70\pm0.19$\\
$EW$ (eV) & $-20\pm5$ \\
\hline\hline
\multicolumn{2}{l}{Absorption line statistics}\\
\hline
$\Delta\chi^2/\Delta\nu$ & 12.5/2\\
F-test (\%)  & 99.6\\
MC (\%) & 96.5\\
\hline\hline
\multicolumn{2}{l}{Broad-band model statistics}\\
\hline
$\chi^2/\nu$ & 2601.2/2510\\
\hline   
\end{tabular}
\end{minipage}
\end{table}

\subsection{PKS~1549$-$79}

We analyzed the \emph{XMM-Newton} EPIC-pn spectrum of PKS~1549$-$79 (observation number 12a) considering the observed energy range between E$=$0.5--10.5~keV. No emission lines are detected in the Fe K band. The upper limits of the intensity and equivalent width of the 6.4~keV emission line are $I<2.2\times 10^{-6}$~ph~s$^{-1}$~cm$^{-2}$ and EW$<$18~eV. Therefore, a neutral reflection component was not included in the spectrum. The best-fit model comprises a Galactic absorbed power-law continuum with $\Gamma=1.61\pm0.03$, intrinsic neutral absorption with column density $N_\mathrm{H}=(5.2\pm0.1)\times 10^{22}$~cm$^{-2}$ and a scattered continuum with a fraction of $\simeq$2\% with respect to the direct one to model the soft excess. 

As already reported in the E$=$3.5--10.5~keV analysis, we find the presence of two absorption lines at the rest-frame energies of E$=$$9.15\pm0.04$~keV and E$=$$11.0\pm0.04$~keV, respectively. The F-test significance level is 99.9\% for each of them. This is slightly reduced to 99.1\% and 98.8\% when compared to 1000 Monte Carlo simulations, respectively. The final model including these lines provides a very good representation of the data, as shown in Figure~A8. Given the relatively high redshift of PKS~1549$-$79 ($z = 0.1522$) and the high energies of the lines, in Figure~A9 we show the contour plot with respect to this model without the inclusion of absorption lines in the wider interval E$=$5--12~keV. The best-fit parameters are reported in Table~A4. 

We checked that these narrow features can not be modeled with two photo-electric edges. If identified with Fe~XXV He$\alpha$ or Fe~XXVI Ly$\alpha$ these two absorption lines would indicate a high blue-shifted velocity of $\simeq$0.25c and $\simeq$0.4c, respectively. Given the presence of two absorption lines, we checked their possible mutual identification as Fe~XXV He$\alpha$ (E$=$6.7~keV) and Fe~XXVI Ly$\alpha$ (E$=$6.97~keV) or Fe~XXV He$\alpha$ and He$\beta$ (E$=$7.88~keV) or Fe~XXVI Ly$\alpha$ and Ly$\beta$ (E$=$8.25~keV) with a common blue-shift. We included two absorption lines in the model with energies fixed to those expected for these transitions and let their common energy shift to vary. However, we find that the fit with two separated Gaussian absorption lines still provides a much better representation of the data, with a fit improvement higher than $\Delta\chi^2/\Delta\nu = 10.5/1$ with respect to the case with two lines with fixed energies and common shift.

   \begin{figure}
   \centering
    \includegraphics[width=8cm,height=6cm,angle=0]{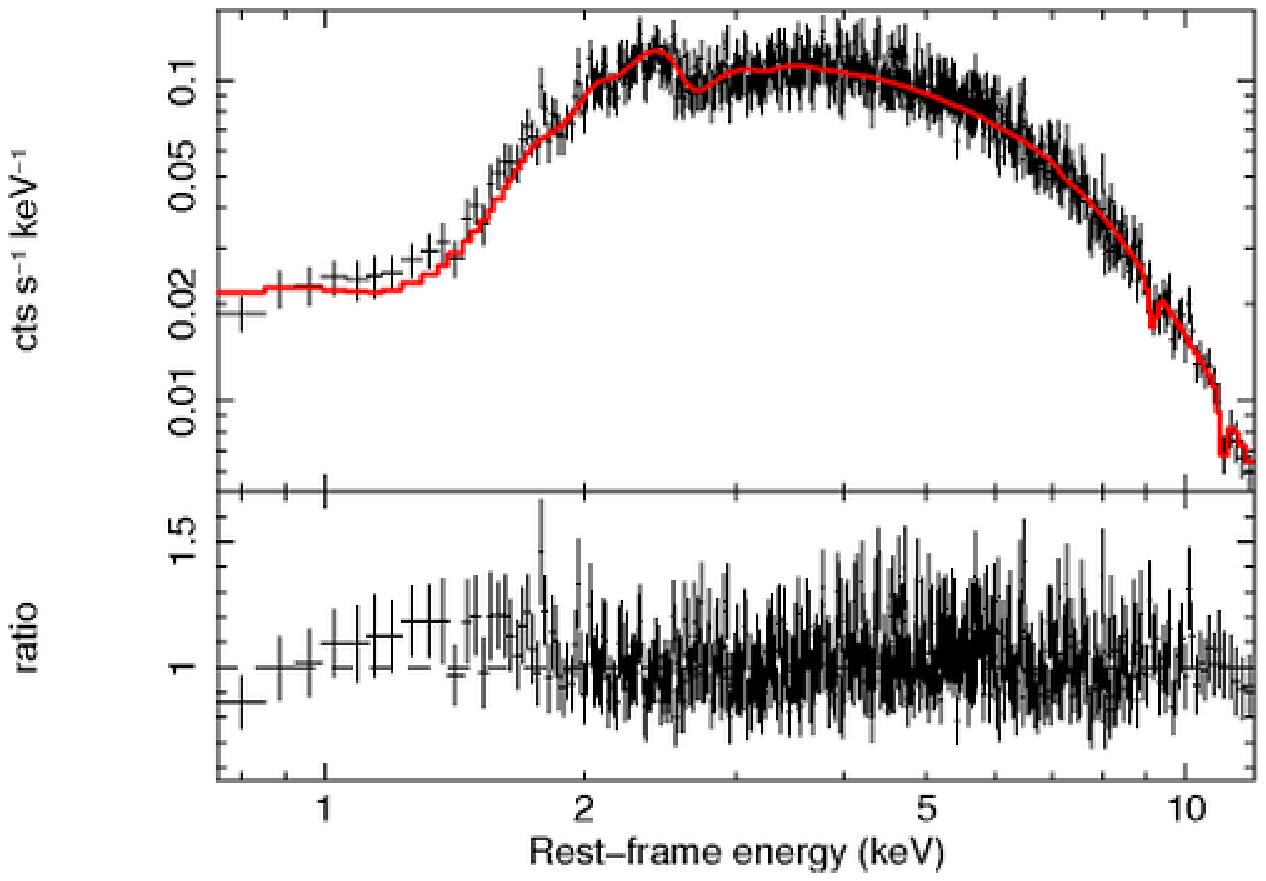}
   \caption{Broad-band \emph{XMM-Newton} EPIC-pn spectrum and ratios against the best-fit model (red) of PKS~1549$-$79 (number 12a) in the E$=$0.7--12~keV energy band.}
    \end{figure}

   \begin{figure}
   \centering
    \includegraphics[width=7.5cm,height=5.5cm,angle=0]{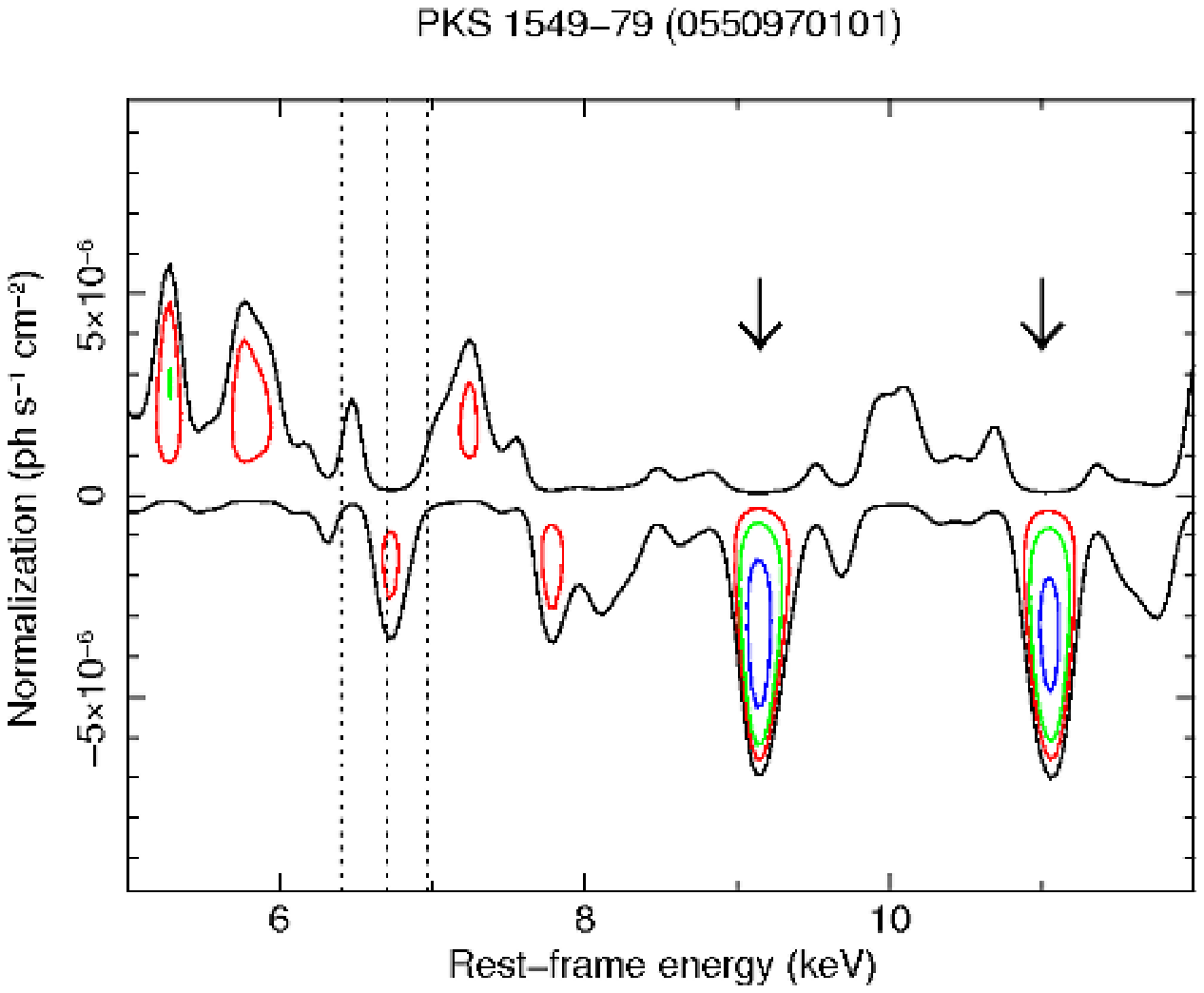}
   \caption{Energy--intensity contour plot in the interval E$=$5--12~keV calculated using the broad-band spectrum of PKS~1549$-$79 (number 12a), see text for more details. The arrows points the absorption lines. The vertical lines refer to the energies of the neutral Fe K$\alpha$, Fe~XXV and Fe~XXVI lines at E$=$6.4~keV, 6.7~keV and 6.97~keV, respectively.}
    \end{figure}

We performed a more physically self-consistent fit of this absorption line using the XSTAR table and the method described in \S4. Given that we can place only an upper limit of $\sigma_v$$<$6,000~km~s$^{-1}$ on the velocity width of these lines, we test three XSTAR tables with turbulent velocities of 1,000~km~s$^{-1}$, 3,000~km~s$^{-1}$ and 5,000~km~s$^{-1}$, respectively.  The fits using these three different turbulent velocities are equivalent at the 90\% level. Therefore, we took into account this degeneracy considering the average of their best-fit parameters. We started with the inclusion of one XSTAR component to model the first line at E$=$9.15~keV. This provides a fit improvement of $\Delta\chi^2/\Delta\nu = 16.1/3$ with respect to the baseline model, corresponding to a probability of 99.9\%. The best-fit values of the column density, ionization parameter and outflow velocity are $N_\mathrm{H} > 1\times 10^{23}$~cm$^{-2}$, log$\xi$$=$$4.91\pm0.53$~erg~s$^{-1}$~cm and $v_\mathrm{out} = 0.278\pm0.006$c, respectively.

\begin{table}
\centering
\begin{minipage}{55mm}
\caption{Best-fit broad-band model of PKS~1549$-$79 (number 12a).}
\begin{tabular}{l c}
\hline\hline
\multicolumn{2}{l}{Neutral absorption} \\
\hline
$N_H$ ($10^{22}$~cm$^{-2}$) & $5.2\pm0.1$\\
\hline\hline
\multicolumn{2}{l}{Power-law continuum} \\
\hline
$\Gamma_\mathrm{Pow}$ & 1.61$\pm$0.03 \\
 \hline\hline
\multicolumn{2}{l}{Scattered continuum} \\
\hline
$\Gamma_\mathrm{Scatt}$ & $\Gamma_\mathrm{Pow}$\\
$f$ (\%) & $\simeq$$2$\\
\hline\hline
\multicolumn{2}{l}{Absorption line 1} \\
\hline
$E$ (keV) & 9.15$\pm$0.04\\
$\sigma$ (eV) & 10\\ 
$I$ ($10^{-5}$ph~s$^{-1}$cm$^{-2}$) & $-0.4\pm0.1$\\
$EW$ (eV) & $-59\pm15$ \\
\hline\hline
\multicolumn{2}{l}{Absorption line 1 statistics}\\
\hline
$\Delta\chi^2/\Delta\nu$ & 13.9/2\\
F-test (\%) & 99.9\\
MC (\%) & 99.1\\
\hline\hline
\multicolumn{2}{l}{Absorption line 2} \\
\hline
$E$ (keV) & 11.0$\pm$0.04\\
$\sigma$ (eV) & 10\\ 
$I$ ($10^{-5}$ph~s$^{-1}$cm$^{-2}$) & $-0.4\pm0.1$\\
$EW$ (eV) & $-80^{+21}_{-25}$ \\
\hline\hline
\multicolumn{2}{l}{Absorption line 2 statistics}\\
\hline
$\Delta\chi^2/\Delta\nu$ & 13.3/2\\
F-test (\%) & 99.9\\
MC (\%) & 98.8\\
\hline\hline

\multicolumn{2}{l}{Broad-band model statistics}\\
\hline
$\chi^2/\nu$ & 847.8/930\\
\hline   
\end{tabular}
\end{minipage}
\end{table}

However, including another XSTAR table provides a better representation of both absorption lines. Similarly to the case of observation number 1c of 4C$+$74.26 discussed in \S4.1.2, the inclusion of another table with tied ionization parameter and column density provides a fit improvement higher than $\Delta\chi^2/\Delta\nu = 8.9/1$ with respect to the single XSTAR model, corresponding to a probability of 99.9\%. The new best-fit ionization and column density are log$\xi$$=$$4.91\pm0.49$~erg~s$^{-1}$~cm and $N_\mathrm{H} > 1.4\times 10^{23}$~cm$^{-2}$, respectively. The two outflow velocities are $v_\mathrm{out} = 0.276\pm0.006$c and $v_\mathrm{out} = 0.427\pm0.005$c, respectively. This provides a good modeling of both absorption lines. The fit improvement including both XSTAR tables with respect to the baseline model is high, $\Delta\chi^2/\Delta\nu = 25/4$, corresponding to a confidence level of 99.99\% ($\ge 4 \sigma$). The best-fit XSTAR parameters are reported in the main text in Table~1. 

Therefore, the best-fit of the two absorption lines at E$\simeq$9.15~keV and E$\simeq$11~keV is provided by a highly ionized absorber with a somewhat complex velocity structure, which we modeled with two XSTAR components with high outflow velocities of 0.276c and 0.427c. These mildly-relativistic velocities indicate the presence of a complex UFO in this source. 

PKS~1549$-$79 was also observed with \emph{Suzaku} (observation number 12b) about one month after \emph{XMM-Newton}. The presence of the two blue-shifted absorption lines indicative of UFOs can not be excluded in the \emph{Suzaku} observation at the 90\% level. In particular, an absorption line consistent with the one detected in the \emph{XMM-Newton} spectrum at with E$=$$11.1\pm0.1$~keV and EW$=$$-110\pm51$~eV is also independently detected in \emph{Suzaku}, although at a lower significance of $\sim$92\% which falls below our detection threshold. This feature can be seen in the contour plots in Figure~C1 in Appendix C.

The inclusion of a possible rest-frame neutral partial covering component using \emph{zpcfabs} in XSPEC with column density $N_H = (3.8^{+0.5}_{-0.4})\times 10^{22}$~cm$^{-2}$ and $C_f = 0.84^{+0.13}_{-0.20}$ does not affect the results on the Fe K absorption lines. Moreover, as we can see from the lack of emission residuals in the contour plot in Figure~A9, the data do not require any broad emission line indicative of disk reflection.

   \begin{figure}
   \centering
    \includegraphics[width=8cm,height=6cm,angle=0]{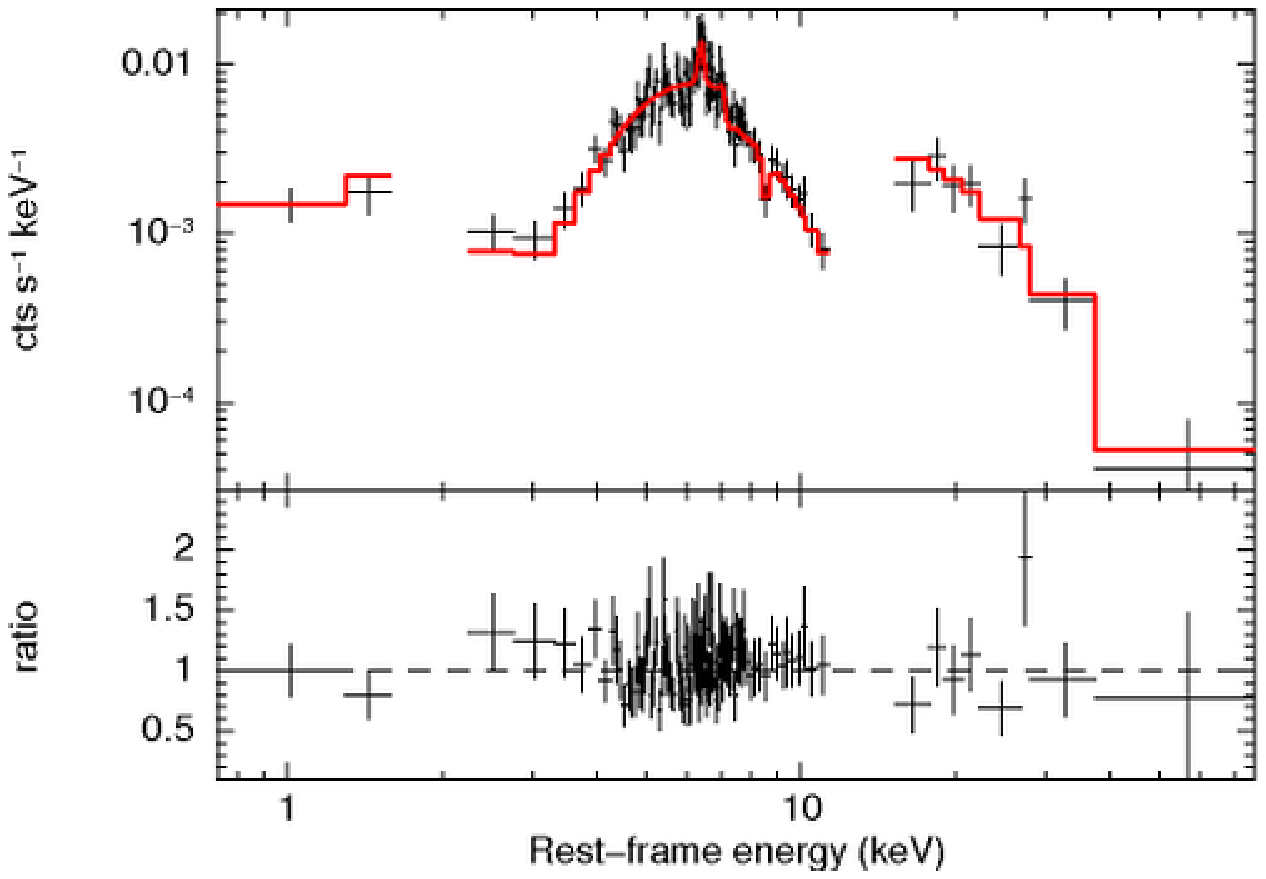}
   \caption{Combined broad-band \emph{Suzaku} XIS-FI and PIN spectrum and ratios against the best-fit model (red) of 3C~105 (number 16) in the E$=$0.6--70~keV energy band.}
    \end{figure}

   \begin{figure}
   \centering
    \includegraphics[width=7.5cm,height=5.5cm,angle=0]{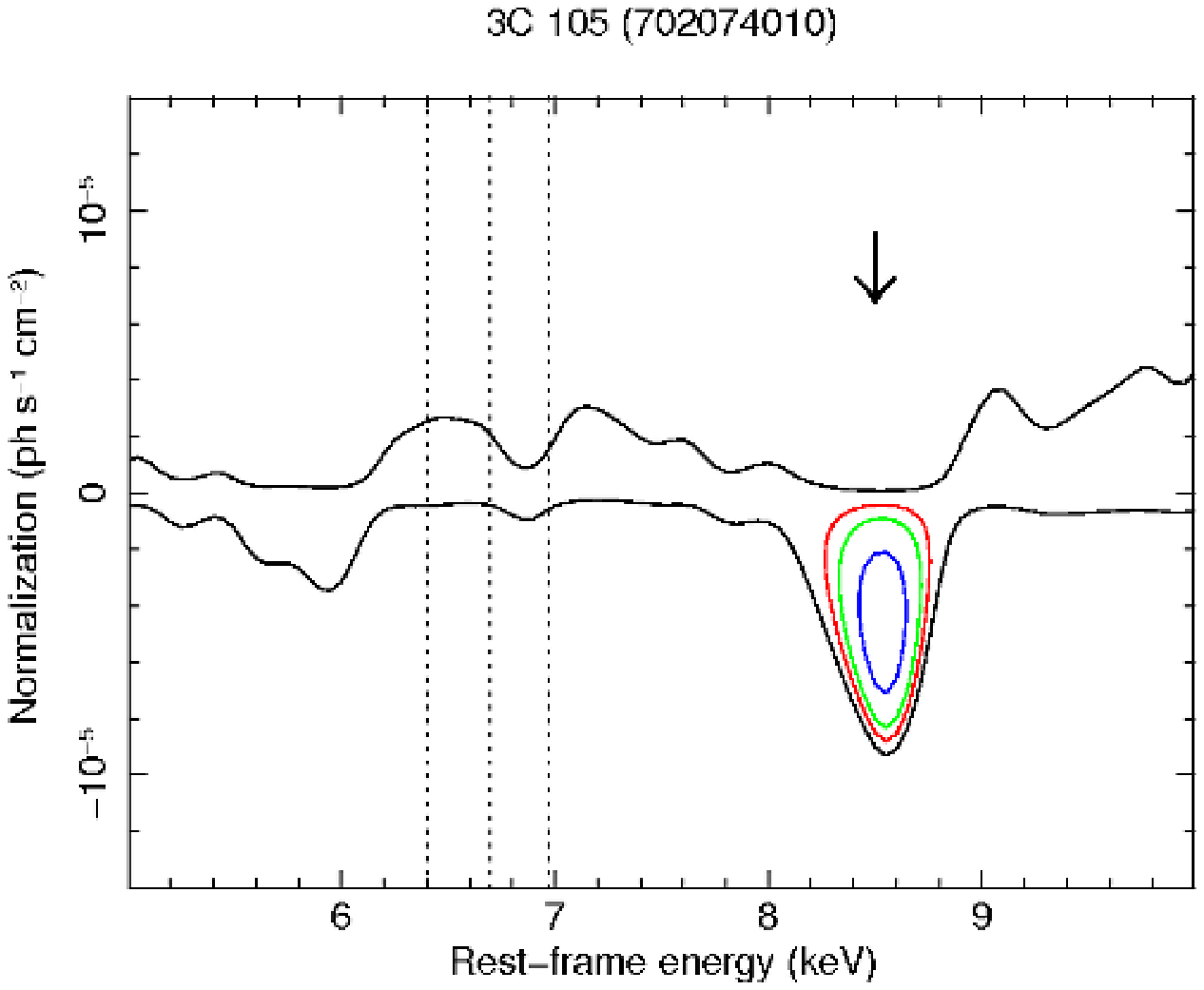}
   \caption{Energy--intensity contour plot in the interval E$=$5--10~keV calculated using the broad-band spectrum of 3C~105 (number 16), see text for more details. The arrows points the absorption line. The vertical lines refer to the energies of the neutral Fe K$\alpha$, Fe~XXV and Fe~XXVI lines at E$=$6.4~keV, 6.7~keV and 6.97~keV, respectively.}
    \end{figure}

\subsection{3C~105}

We analyzed the combined \emph{Suzaku} XIS-FI and PIN broad-band spectrum 3C~105 (observation number 16) in the E$=$0.6--70~keV energy band (excluding the interval E$=$1.5--2~keV). The best-fit model comprises a power-law continuum with $\Gamma = 1.80\pm0.10$ absorbed by high column of neutral material with $N_H = (4.13\pm0.27)\times 10^{23}$~cm$^{-2}$. A scattered continuum with a fraction of $\simeq$1\% with respect to the direct one is present in the soft X-rays. The neutral Fe K$\alpha$ emission line is self-consistently modeled with a \emph{pexmon} neutral reflection component with an inclination of $i = 60^{\circ}$, high-energy cut-off of $E_C = 300$~keV and standard Solar abundances. We derive a reflection fraction of $R = 1.26^{+0.34}_{-0.30}$.

\begin{table}
\centering
\begin{minipage}{55mm}
\caption{Best-fit broad-band model of 3C~105 (number 16).}
\begin{tabular}{l c}
\hline\hline
\multicolumn{2}{l}{Neutral absorption} \\
\hline
$N_H$ ($10^{22}$~cm$^{-2}$) & $41.3\pm2.7$\\
\hline\hline
\multicolumn{2}{l}{Neutral reflection (\emph{pexmon})} \\
\hline
$\Gamma_\mathrm{Pow}$ & 1.80$\pm$0.10 \\
$R$ & $1.26^{+0.34}_{-0.30}$\\
$E_C$ (keV) & 300\\
$A_\mathrm{Fe}$ (Solar) & 1\\
$i$ (deg) & 60\\ 
 \hline\hline
\multicolumn{2}{l}{Scattered continuum} \\
\hline
$\Gamma_\mathrm{Scatt}$ & $\Gamma_\mathrm{Pow}$\\
$f$ (\%) & $\simeq$1\\
\hline\hline
\multicolumn{2}{l}{Absorption line} \\
\hline
$E$ (keV) & 8.53$\pm$0.05\\
$\sigma$ (eV) & 100\\ 
$I$ ($10^{-5}$ph~s$^{-1}$cm$^{-2}$) & $-0.85\pm0.24$\\
$EW$ (eV) & $-153\pm41$ \\
\hline\hline
\multicolumn{2}{l}{Absorption line statistics}\\
\hline
$\Delta\chi^2/\Delta\nu$ & 11.6/2\\
F-test (\%) & 99.7\\
MC (\%) & 99.2\\
\hline\hline
\multicolumn{2}{l}{Broad-band model statistics}\\
\hline
$\chi^2/\nu$ & 224.7/254\\
\hline   
\end{tabular}
\end{minipage}
\end{table}

The fit improvement after the inclusion of a Gaussian absorption line at the rest-frame energy of E$=$$8.53\pm0.05$~keV is $\Delta\chi^2/\Delta\nu = 11.6/2$, corresponding to an F-test probability of 99.7\%. Given that the energy of the line is not directly consistent with that expected from ionized iron at rest, we also performed Monte Carlo simulations in order to better estimate its detection level as described in \S4. The resultant Monte Carlo probability is 99.2\%, higher than the selection threshold of 95\%. The final model provides a very good representation of the data, as shown in Figure~A10. The contour plot in the E$=$5--10~keV energy band with respect to the baseline model without the inclusion of the absorption line is shown in Figure~A11. The best-fit parameters are reported in Table~A5.

If identified with Fe~XXV He$\alpha$ or Fe~XXVI Ly$\alpha$ the absorption line would indicate a significant blue-shifted velocity of $\simeq$0.2c. We checked that the possible alternative modeling with a photo-electric edge provides a much worse fit ($\Delta\chi^2/\Delta\nu = 1.5/2$). 

The width of the absorption line in the \emph{Suzaku} spectrum of 3C~105 is not well constrained and we can place an upper limit of $\sigma_v$$<$6,000~km~s$^{-1}$. Therefore, we performed fits using three XSTAR tables with turbulent velocities of 1,000~km~s$^{-1}$, 3,000~km~s$^{-1}$ and 5,000~km~s$^{-1}$, respectively. We obtain equivalent fits at the 90\% level in the three cases. For each of them we also find a degeneracy in the identification of the line with Fe~XXV or Fe~XXVI. Therefore, we consider an average among these six solutions with colum density $N_\mathrm{H} > 2\times 10^{22}$~cm$^{-2}$, ionization parameter log$\xi$$=$$3.81\pm1.30$~erg~s$^{-1}$~cm and outflow velocity of $v_\mathrm{out} = 0.227\pm0.033$c. This highly ionized absorber has a velocity that is consistent with a UFO. The model provides a good representation of the data, with a fit improvement of $\Delta\chi^2/\Delta/\nu = 10.5/3$, corresponding to a confidence level of 99\%. The best-fit XSTAR parameters are reported in the main text in Table~1. 

We checked that the data do not significantly require the inclusion of a rest-frame neutral or ionized partial covering component using \emph{zpcfabs} or \emph{zxipcf} in XSPEC (Reeves et al.~2008). Moreover, as we can see from the lack of emission residuals in the contour plot in Figure~A11, the data do not require any broad emission line indicative of disk reflection. Therefore, these possible model complexities do not affect the results on the Fe K absorption lines for this observation.

  \begin{figure}
   \centering
    \includegraphics[width=7cm,height=15cm,angle=0]{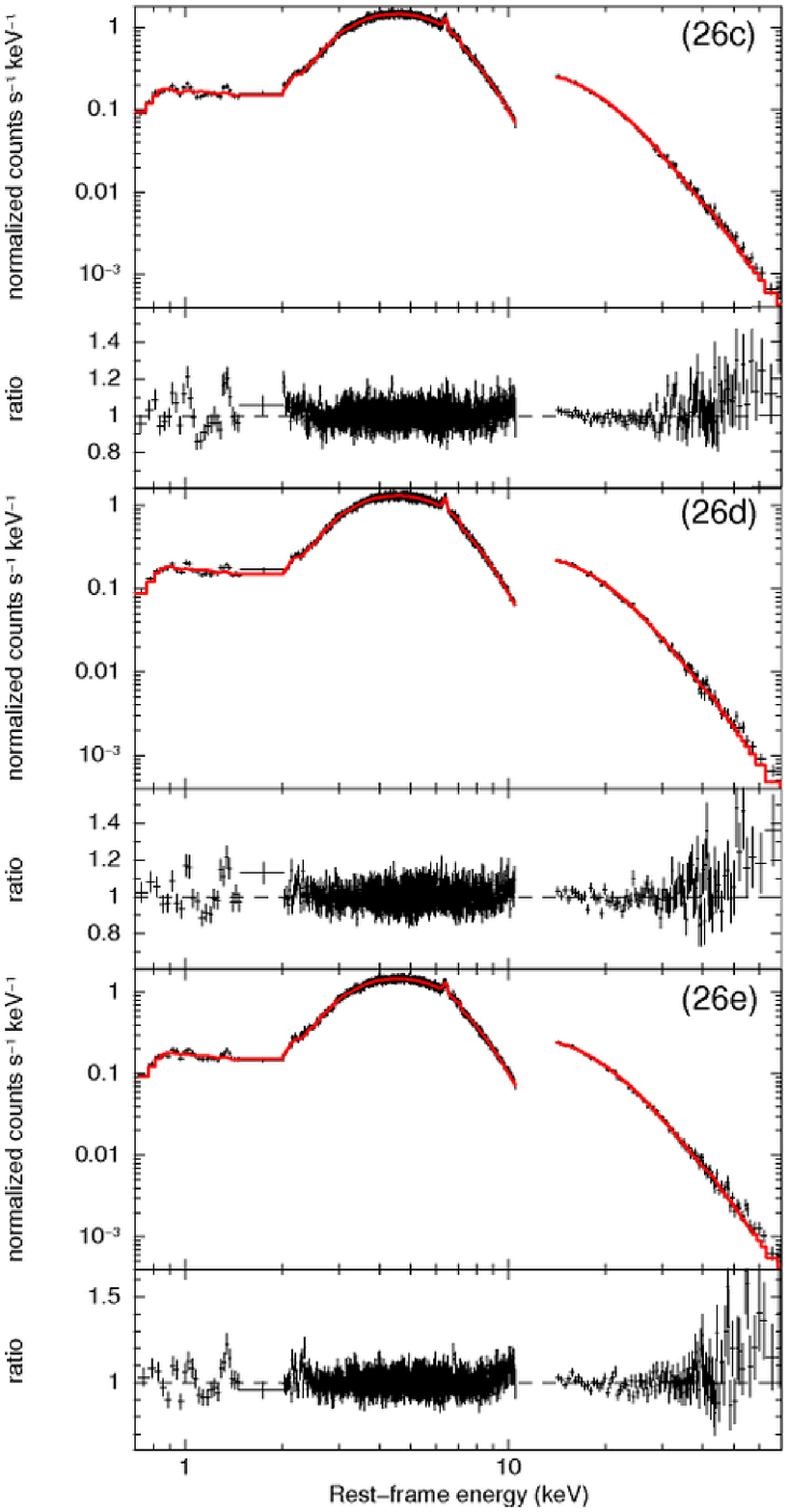}
   \caption{Combined broad-band \emph{Suzaku} XIS-FI and PIN spectrum and ratios against the best-fit model (red) of Centaurus~A (numbers 26c, 26d, 26e) in the E$=$0.6--70~keV energy band.}
    \end{figure}

\subsection{Centaurus~A}

We analyze the combined \emph{Suzaku} XIS-FI and PIN broad-band spectrum of Centaurus~A (observation number 26c) in the energy band E$=$0.6--70~keV (excluding the interval E$=$1.5--2~keV). As already reported for the E$=$3.5--10.5~keV analysis in Table~A2 in Appendix A, the spectrum is well described by a neutral absorbed power-law continuum with $N\mathrm{H}=(10.66\pm0.04)\times 10^{22}$~cm$^{-2}$ and $\Gamma = 1.850\pm0.075$. 

The neutral Fe K$\alpha$ emission line is self-consistently modeled with a \emph{pexmon} neutral reflection component with an inclination fixed to $i = 60^{\circ}$ and high-energy cut-off of $E_C = 300$~keV (e.g., Dadina 2007). We obtain a reflection fraction of $R = 0.92\pm 0.04$ and an iron abundance lower than Solar, of $A_\mathrm{Fe} = 0.36\pm0.02$. The soft X-ray part of the spectrum requires a scattered continuum component with a fraction of $\simeq$1\% and a hot plasma emission modeled with \emph{apec} in XSPEC with a temperature of $kT = 0.65\pm 0.02$~keV. 

We find the presence of two narrow absorption lines at the rest-frame energies of E$=$$6.66\pm0.02$~keV and E$=$$6.95\pm0.02$~keV, respectively. Their significance level is very high, being $\gg$99.99\% for both of them. The energies of these features clearly favor an interpretation as resonance absorption lines from Fe~XXV He$\alpha$ at E$=$6.7~keV and Fe~XXVI Ly$\alpha$ at E$=$6.97~keV, respectively. Given that the lines are consistent with the expected values, the standard F-test can be applied and there is no need to perform additional Monte Carlo simulations. The final model provides a very good representation of the data, as shown in Figure~A12. The contour plot in the E$=$5--10~keV energy band with respect to the baseline model without the inclusion of the absorption lines is shown in Figure~A13. The best-fit parameters are reported in Table~A6. 

The best-fit model derived for observation number 26c is also applicable to the subsequent \emph{Suzaku} observations 23d and 23e. These observations were performed in a period of about three weeks. The fit parameters are equivalent. In particular, the two narrow absorption lines are still observed at the same energies with a statistical confidence level always higher than 99.5\%. There is only a slight change in the equivalent width. The final models for observations 26d and 26e provide a very good representation of the data, as shown in Figure~A12. The contour plots in the E$=$5--10~keV band are shown in Figure~A13 and their best-fit parameters are reported in Table~A6.

The energies of the detected absorption lines are directly consistent with Fe~XXV He$\alpha$ and Fe~XXVI Ly$\alpha$. The absorption lines are narrow and we can place a 90\% upper limit of $\sigma_v$$<$5,000~km~s$^{-1}$. We performed a more physically self-consistent fit using the XSTAR table and the method described in \S4. We test XSTAR tables with turbulent velocities of 1,000~km~s$^{-1}$, 3,000~km~s$^{-1}$ and 5,000~km~s$^{-1}$, respectively. For each observation we average the values of the parameters for XSTAR solutions with different turbulent velocities but equivalent best-fits at the 90\% level. Given that the energy of the lines is consistent with the expected values, we fixed the shift of the absorption components to zero. The best-fit values of the column density and ionization parameter for observation number 26c are $N_\mathrm{H}= (4.2\pm0.4)\times 10^{22}$~cm$^{-2}$ and log$\xi$$=$$4.33\pm0.03$~erg~s$^{-1}$~cm, respectively. The fit improvement is $\Delta\chi^2/\Delta\nu = 126.5/2$, corresponding to a very high confidence level of $\gg$99.99\%. 

The best-fit parameters for the observation number 26d are $N_\mathrm{H}= (4.1^{+0.9}_{-1.0})\times 10^{22}$~cm$^{-2}$ and log$\xi$$=$$4.37^{+0.03}_{-0.05}$~erg~s$^{-1}$~cm. The fit improvement is $\Delta\chi^2/\Delta\nu = 54.7/2$, corresponding to a high confidence level of $\gg$99.99\%. Finally, the best-fit values of the column density and ionization parameter for observation number 26e are $N_\mathrm{H}= (4.0\pm1.1)\times 10^{22}$~cm$^{-2}$ and log$\xi$$=$$4.31\pm0.10$~erg~s$^{-1}$~cm, respectively. The fit improvement is $\Delta\chi^2/\Delta\nu = 105.3/2$, corresponding to a very high confidence level of $\gg$99.99\%. Leaving the outflow velocity free, we derive upper limits of $v_\mathrm{out}$$<$1,100~km~s$^{-1}$, $v_\mathrm{out}$$<$1,500~km~s$^{-1}$ and $v_\mathrm{out}$$<$800~km~s$^{-1}$ for the observation number 26c, 26d and 26e, respectively. The best-fit XSTAR parameters of these three observations are consistent with each other and they are reported in the main text in Table~1.  

We note that this spectral analysis of this dataset was already reported by Fukazawa et al.~(2011). The authors did not perform a detailed search for Fe K absorption lines, but indeed suggested that the spectral variability of this source may be influenced by a complex nuclear absorber.

   \begin{figure}
   \centering
    \includegraphics[width=6.5cm,height=12cm,angle=0]{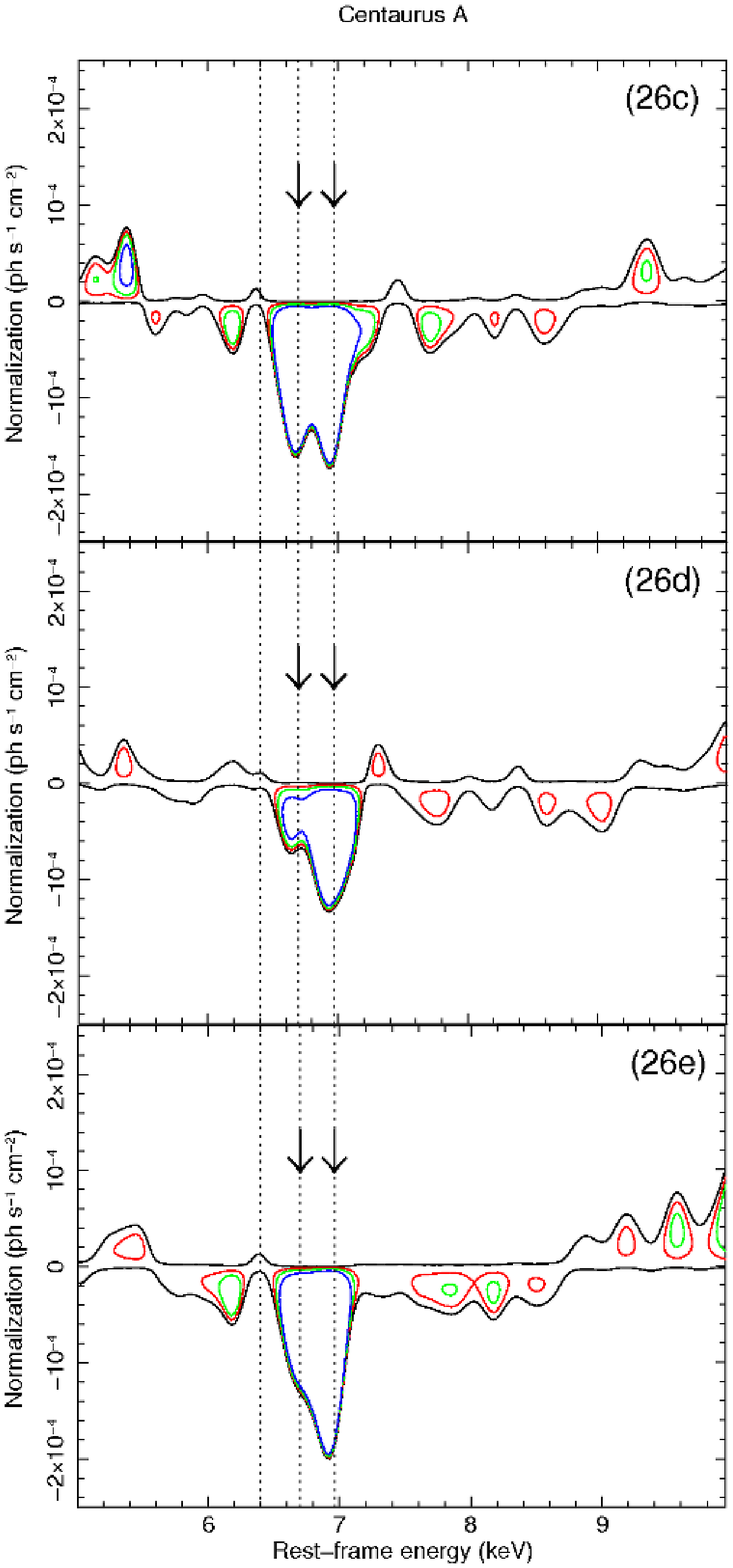}
   \caption{Energy--intensity contour plot in the interval E$=$5--10~keV calculated using the broad-band spectrum of Centaurus~A (numbers 26c, 26d and 26e), see text for more details. The arrows points the absorption lines. The vertical lines refer to the energies of the neutral Fe K$\alpha$, Fe~XXV and Fe~XXVI lines at E$=$6.4~keV, 6.7~keV and 6.97~keV, respectively.}
    \end{figure}

\begin{table*}
\centering
\begin{minipage}{95mm}
\caption{Best-fit broad-band model of Centaurus~A (number 26c, 26d, 26e).}
\begin{tabular}{l c c c}
\hline\hline
Number & 26c & 26d & 26e\\
\hline\hline
\multicolumn{4}{l}{Neutral absorption} \\
\hline
$N_H$ ($10^{22}$~cm$^{-2}$) & $10.66\pm0.04$ & $10.80\pm0.05$ & $10.61\pm0.04$\\
\hline\hline
\multicolumn{4}{l}{Neutral reflection (\emph{pexmon})} \\
\hline
$\Gamma_\mathrm{Pow}$ & 1.850$\pm$0.075 & $1.849\pm0.009$ & $1.831\pm0.008$\\
$R$ & $0.92\pm0.04$ & $1.04\pm0.005$ & $0.99\pm0.04$\\
$E_C$ (keV) & 300 & 300 & 300\\
$A_\mathrm{Fe}$ (Solar) & $0.36\pm0.02$ & $0.41\pm0.02$ & $0.34\pm0.02$\\
$i$ (deg) & 60 & 60 & 60\\ 
\hline\hline
\multicolumn{4}{l}{Scattered continuum} \\
\hline
$\Gamma_\mathrm{Scatt}$ & $\Gamma_\mathrm{Pow}$ & $\Gamma_\mathrm{Pow}$ & $\Gamma_\mathrm{Pow}$\\
$f$ (\%) & $\simeq$1 & $\simeq$1 & $\simeq$1\\
\hline\hline
\multicolumn{4}{l}{Hot plasma (\emph{apec})} \\
\hline
$kT$ (keV) & $0.65\pm0.02$ & $0.65\pm0.02$ & $0.73\pm0.02$\\
$N$ ($10^{-4}$) & $2.90\pm0.15$ & $2.90\pm0.16$ & $2.77\pm0.14$\\
\hline\hline
\multicolumn{4}{l}{Absorption line 1} \\
\hline
$E$ (keV) & 6.66$\pm$0.02 & $6.64\pm0.03$ & $6.67\pm0.02$\\
$\sigma$ (eV) & 10 & 10 & 10\\ 
$I$ ($10^{-5}$ph~s$^{-1}$cm$^{-2}$) & $-7.7\pm0.9$ & $-3.4\pm1.0$ & $-5.6\pm1.0$\\
$EW$ (eV) & $-13\pm2$ & $-7\pm2$ & $-9\pm2$\\
\hline\hline
\multicolumn{4}{l}{Absorption line 1 statistics}\\
\hline
$\Delta\chi^2/\Delta\nu$ & 63.7/2 & 11.3/2 & 30.0/2\\
F-test (\%) & $\gg$99.99 & 99.5 & $>$99.99\\
\hline\hline
\multicolumn{4}{l}{Absorption line 2} \\
\hline
$E$ (keV) & 6.95$\pm$0.02 & $6.95\pm0.02$ & $6.93\pm0.02$\\
$\sigma$ (eV) & 10 & 10 & 10\\ 
$I$ ($10^{-5}$ph~s$^{-1}$cm$^{-2}$) & $-8.4\pm0.9$ & $-6.6\pm1.0$ & $-9.6\pm1.0$\\
$EW$ (eV) & $-16\pm2$ & $-14\pm2$ & $-18\pm2$\\
\hline\hline
\multicolumn{4}{l}{Absorption line 2 statistics}\\
\hline
$\Delta\chi^2/\Delta\nu$ & 64.0/2 & 37.6/2 & 82.3/2\\
F-test (\%) & $\gg$99.99 & $>$99.99 & $\gg$99.99\\
\hline\hline
\multicolumn{4}{l}{Broad-band model statistics}\\
\hline
$\chi^2/\nu$ & 2830.7/2680 & 2858.0/2659 & 2848.8/2696\\
\hline   
\end{tabular}
\end{minipage}
\end{table*}

\clearpage

\newpage

\onecolumn{

\section{Long tables}

\begin{table}
\centering
\begin{minipage}{165mm}
\caption{List of radio-loud AGNs with relative \emph{XMM-Newton} and \emph{Suzaku} observations.}

\begin{tabular}{@{\hspace{0.1cm}}l l c@{\hspace{0.2cm}} c@{\hspace{0.2cm}} c@{\hspace{0.2cm}} c c c c c c c@{\hspace{0.2cm}} l@{\hspace{0.0cm}}}

\hline\hline       
Num & Source & $z$ & Sy & FR & Obs & OBSID & Date & Counts & Exp & Flux & Lum & Abs\\ 
\scriptsize{(1)} & \scriptsize{(2)} & \scriptsize{(3)} & \scriptsize{(4)} & \scriptsize{(5)} & \scriptsize{(6)} & \scriptsize{(7)} & \scriptsize{(8)} & \scriptsize{(9)}& \scriptsize{(10)} & \scriptsize{(11)} & \scriptsize{(12)}  & \scriptsize{(13)} \\
\hline

1a & 4C+74.26 & 0.1040 & 1 & II & X & 0200910201 & 2004-02-06 & 27800/112 & 26.6 & 15.0 & 40.0 & \\

1b &          &        & & & S & 702057010  & 2007-10-28 & 58631/1751 & 91.6 & 19.0 & 51.0 & Y$^d$\\

1c &          &        & & & S & 706028010  & 2011-11-23 & 67814/706 & 101.4 & 16.0 & 44.0 & \\

2a & PKS~0558-504 & 0.1372 & 1 & - &  X & 0129360201 & 2000-10-10 & 3399/38 & 7.5 & 6.2 & 32.0 & \\

2b &              &        & & & X & 0137550201 & 2001-06-26 & 1779/51 & 10.5 & 6.5 & 33.0 &\\

2c &              &        & & & X & 0137550601 & 2001-10-19 & 3288/60 & 10.2 & 12.0 & 63.0 & \\

2d &              &        & & & X & 0555170201 & 2008-09-07 & 30651/1112 & 71.2 & 5.8 & 30.0 & \\

2e &              &        & & & X & 0555170301 & 2008-09-09 & 31709/913 & 78.7 & 5.5 & 28.0 & \\

2f &              &        & & & X & 0555170401 & 2008-09-11 & 48304/2223 & 86.4 & 7.6 & 39.0 & \\

2g &              &        & & & S & 701011010  & 2007-01-17 & 5493/277 & 20.6 & 6.0 & 31.0 & \\

2h &              &        & & & S & 701011020  & 2007-01-18 & 6927/274 & 18.9 & 8.1 & 42.0 & \\

2i &              &        & & & S & 701011030  & 2007-01-19 & 5368/288 & 21.3 & 5.8 & 29.0 & \\

2l &              &        & & & S & 701011040  & 2007-01-20 & 8137/286 & 19.8 & 9.4 & 48.0 & \\
				
2m &              &        & & & S & 701011050  & 2007-01-21 & 7211/269 & 19.5 & 8.3 & 43.0 & \\

3 & 3C~332 & 0.1510 & 1 & II & S & 704038010  & 2009-08-08 & 9362/870 & 59.9 & 3.8 & 22.0 & \\

4a & 3C~382 & 0.0578 & 1 & II & X & 0506120101 & 2008-04-28 & 36824/675 & 22.0 & 23.0 & 18.0 & \\

4b &        &        & & & S & 702125010  & 2007-04-27 & 145739/2370 & 130.6 & 27.0 & 21.0 & N$^{a,d,e}$\\

5 & 3C~059  & 0.1097 & 1.8 & - & X & 0205390201 & 2004-07-30 & 17939/255 & 47.1 & 5.5 & 16.0 & \\

6a & 3C~390.3 & 0.0561 & 1.5 & II & X & 0203720201 & 2004-10-08 & 60103/607 & 36.1 & 24.0 & 18.0 & \\

6b &          &        & & & X & 0203720301 & 2004-10-17 & 32896/265 & 22.5 & 21.0 & 15.0 & \\

6c &          &        & & & S & 701060010  & 2006-12-14 & 81187/1236 & 99.4 & 20.0 & 15.0 & Y$^{a,d}$\\

7a & 3C~111 & 0.0485 & 1 & II & X & 0552180101 & 2009-02-15 & 116320/1532 & 55.0 & 31.0 & 17.0 &  N$^{b}$\\

7b &        &        & & & S & 703034010 & 2008-08-22 & 55621/2499 & 122.4 & 13.0 & 7.4 & Y$^{a,b,d}$\\ 

7c &        &        & & & S & 705040010 & 2010-09-02 & 100435/1117 & 80.6 & 30.0 & 17.0 &  N$^{c,d}$\\

7d &        &        & & & S & 705040020 & 2010-09-09 & 126335/1147 & 79.4 & 38.0 & 21.0 & Y$^{c,d}$\\

7e &        &        & & & S & 705040030 & 2010-09-14 & 123357/1246 & 80.4 & 36.0 & 20.0 & N$^{c,d}$\\

8 & 3C~206 & 0.0860 & 1.2 & II & S & 705007010  & 2010-05-08 & 17729/1012 & 80.6 & 6.4 & 12.0 & \\

9a & 3C~120 & 0.0330 & 1.5 & I & X & 0152840101 & 2003-08-26 & 166430/1681 & 85.9 & 28.0 & 6.9 & \\

9b &        &        & & & S & 700001010  & 2006-02-09 & 80076/1885 & 41.9 & 28.0 & 7.0 & N$^{a,d}$\\

9c$^*$ &   &         & & & S & 700001020  & 2006-02-16 & 69754/1789 & 41.6 & 25.0 & 6.2 & Y$^{a,d}$\\

9d$^*$ &   &         & & & S & 700001030  & 2006-02-23 & 69241/1775 & 40.9 & 25.0 & 6.2 & Y$^{a,d}$\\

9e$^*$ &   &         & & & S & 700001040  & 2006-03-02 & 66260/1629 & 40.9 & 25.0 & 6.0 & Y$^{a,d}$\\

9f     &   &         & & & S & 706042010  & 2012-02-09 & 227512/5834 & 183.0 & 30.0 & 7.5 & \\ 

9g     &   &         & & & S & 706042020  & 2012-02-14 & 144163/3697 & 118.0 & 29.0 & 7.3 & \\

10a & 4C+50.55 & 0.0200 & 1 & II & X & 0306320201 & 2005-11-06 & 50289/101 & 21.6 & 36.0 & 3.3 & \\

10b &          &        & & & S & 702027010	 & 2007-04-16 & 142711/2764 & 85.0 & 51.0 & 4.9 & \\

11  & Pictor~A  & 0.0351 & 3b & II & X & 0206390101 & 2005-01-14 & 12356/304 & 40.7 & 6.5 & 1.8 & \\

\hline

\end{tabular}
\end{minipage}
\end{table}

\begin{table}
\centering
\begin{minipage}{165mm}
\contcaption{-- List of radio-loud AGNs with relative \emph{XMM-Newton} and \emph{Suzaku} observations.}

\begin{tabular}{@{\hspace{0.1cm}}l l c@{\hspace{0.2cm}} c@{\hspace{0.2cm}} c@{\hspace{0.2cm}} c c c c c c c@{\hspace{0.2cm}} l@{\hspace{0.0cm}}}

\hline\hline       
Num & Source & $z$ & Sy & FR & Obs & OBSID & Date & Counts & Exp & Flux & Lum & Abs\\ 
\scriptsize{(1)} & \scriptsize{(2)} & \scriptsize{(3)} & \scriptsize{(4)} & \scriptsize{(5)} & \scriptsize{(6)} & \scriptsize{(7)} & \scriptsize{(8)} & \scriptsize{(9)} & \scriptsize{(10)} & \scriptsize{(11)} & \scriptsize{(12)}  & \scriptsize{(13)} \\
\hline   


12a & PKS~1549-79  & 0.1522 & 1i & II & X & 0550970101 & 2008-09-21 & 14525/296 & 48.2 & 4.3 & 29.0 & \\

12b &              &        & & & S & 703059010  & 2008-10-13 & 12205/751 & 82.2 & 3.6 & 25.0 & \\

13 & 3C~234 & 0.1850 & 1.9 & II & X & 0405340101 & 2006-04-24 & 1894/132 & 26.6 & 1.2 & 15.0 & \\

14  & PKS~0707-35  & 0.1108 & - & II & S & 706008010  & 2011-04-02 & 11373/474 & 81.2 & 3.9 & 13.0 & \\

15  & PKS~2356-61  & 0.0963 & 2 & II & S & 801016010  & 2006-12-06 & 10765/639 & 100.9 & 2.9 & 9.4 & \\

16  & 3C~105 & 0.0890 & 2 & II & S & 702074010  & 2008-02-05 & 2126/414  & 37.4 & 2.1 & 8.0 & \\	

17  & PKS~0326-288 & 0.1088 & 2 & - & S & 704039010  & 2010-01-30 & 1692/332 & 58.0 & 0.9 & 6.5 & \\

18a & 3C~445   & 0.0558	& 1.5 & II & X & 0090050601 & 2001-12-16 & 5676/63 & 15.4 & 5.9 & 5.8 & \\

18b &          &        & & & S & 702056010	 & 2007-05-25 & 30259/1459 & 139.7 & 6.1 & 7.0 & Y$^{a,d, f, g}$\\

19  & 3C~433   & 0.1016	& 2 & II & S & 705050010  & 2010-11-06 & 4686/326 & 62.6 & 2.2 & 6.3 & \\

20a & 3C~452   & 0.0811	& 2 & II & X & 0552580201 & 2008-11-30 & 6770/304 & 54.2 & 2.2 & 5.5 & \\

20b &          &        & & & S & 702073010	 & 2007-06-16 & 3274/368 & 66.7 & 2.0 & 4.3 & \\

21a & 3C~33 & 0.0597 & 2 & II & X & 0203280301 & 2004-01-21 & 870/20 & 6.3 & 2.5 & 3.9 &\\

21b &       &        & & & S & 702059010  & 2007-12-26 & 9165/1191 & 127.4 & 2.6 & 4.0 &\\

22a & NGC~612      & 0.0298 & 2 & II & X & 0312190201 & 2006-06-26 & 778/90 & 9.6 & 1.7 & 2.1 & \\

22b &              &         & & & S & 703016010  & 2008-05-20 & 1502/280 & 48.5 & 1.5 & 1.4 & \\

23 & 4C+29.30 & 0.0647 & 2 & II & X & 0504120101 & 2008-04-11 & 630/106 & 17.5 & 0.6 & 1.3 & \\

24  & 3C~403   & 0.0590  & 2 & II & S & 704011010  & 2009-04-08 & 760/288  & 47.9 & 0.6 & 0.9 & \\

25  & IC~5063      & 0.0114 & 1h & I & S & 704010010  & 2009-04-24 & 9236/569 & 45.2 & 7.3 & 0.4 & \\

26a & Centaurus A  & 0.0018 & 2 & I & X & 0093650201 & 2001-02-02 & 26596/160 & 16.7 & 170.0 & 0.2 & \\

26b &              &        & & & X	& 0093650301 & 2002-02-06 & 13687/83 & 7.9 & 180.0 & 0.2 & \\

26c &              &        & & & S & 704018010  & 2009-07-20 & 584087/9497 & 62.6 & 280.0 & 0.3 & \\

26d &              &        & & & S	& 704018020  & 2009-08-05 & 431570/7464 & 51.3 & 250.0 & 0.2 & \\

26e &              &        & & & S & 704018030  & 2009-08-14 & 527093/8571 & 55.9 & 280.0 & 0.3 & \\

\hline   
\end{tabular}
{\em Notes:} (1) Object number. Letters are used to enumerate multiple observations. (2) Object name. (3) Cosmological redshift. (4) Optical identification and reference, by referring V{\'e}ron-Cetty \& V{\'e}ron (2010). Sy1i and Sy1h show the sources with broad Paschen lines in the infrared and with broad polarized Balmer lines, respectively. The source identified as S3b is Seyfert 3 of liner with broad Balmer lines. (5) Fanaroff-Riley (FR) type. (6) Observatory, X if {\it XMM-Newton} and S if {\it Suzaku}, respectively. (7) Observation ID. (8) Start date of the observation in yyyy-mm-dd. (9) Source/background counts in the E$=$4--10~keV band. Data are relative to the EPIC-pn or XIS-FI if corresponding to {\it XMM-Newton} or {\it Suzaku}, respectively. (10) Net exposure in units of ks. (11) Observed flux in the E$=$4--10~keV band in units of $10^{-12}$~erg~s$^{-1}$~cm. (12) Absorption corrected luminosity in the E$=$4--10~keV band in units of $10^{43}$~erg~s$^{-1}$. (13) Y (or N) indicate if (or if not) a detailed Fe K absorption line search was already reported in the literature and if any line was (was not) detected. $^a$, $^b$, $^c$, $^d$, $^e$, $^f$ and $^g$ indicate that the observation was already analyzed in detail by Tombesi et al.~(2010a), Ballo et al.~(2011), Tombesi et al.~(2011a), Gofford et al.~(2013), Sambruna et al.~(2011), Braito et al.~(2011) or Reeves et al.~(2010) respectively. $^{*}$ indicates that the consecutive observations have been added together and were considered as a single one in the spectral analysis.
\end{minipage}
\end{table}



\begin{table}
\centering
\begin{minipage}{153mm}
\caption{Baseline model parameters of the spectra in the E$=$3.5--10.5~keV band. We consider only observations without Fe K absorption line searches reported in the literature.}

\begin{tabular}{l@{\hspace{0.0cm}}|l|@{\hspace{0.1cm}}c@{\hspace{0.0cm}}c@{\hspace{0.2cm}}c@{\hspace{0.2cm}}c@{\hspace{0.2cm}}c@{\hspace{0.2cm}}c@{\hspace{0.2cm}}c@{\hspace{0.0cm}}|c@{\hspace{0.0cm}}|c@{\hspace{0.0cm}}}

\hline\hline       
Num & Source & $N_\mathrm{H, Gal}$ & $\Gamma$ & $N_\mathrm{H}$ & $E$ & $\sigma$ & $I$ & EW & $\chi^2/\nu$ & Abs\\ 
\scriptsize{(1)} & \scriptsize{(2)} & \scriptsize{(3)} & \scriptsize{(4)} & \scriptsize{(5)} & \scriptsize{(6)} & \scriptsize{(7)} & \scriptsize{(8)} & \scriptsize{(9)} & \scriptsize{(10)} & \scriptsize{(11)}\\
\hline   
1a & 4C$+$74.26 & 12 & $1.67\pm0.02$ & & $6.50\pm0.04$ & 100 & $1.8\pm0.3$ & $56\pm10$ & 849/830 & Y\\[+4pt] 
1c &            &    & $1.79\pm0.01$ & & $6.38\pm0.02$ & 100 & $2.6\pm0.2$ & $74^{+6}_{-4}$ & 1304/1334 & Y\\[+4pt]
2a & PKS~0558$-$504 & 4 & $2.12\pm0.06$ & &            &     &             &               & 127/157 & N\\[+4pt]
2b & & & $2.02\pm0.08$ & & & & & & 72/88 & N\\[+4pt]
2c & & & $2.19\pm0.06$ & & & & & & 142/156 & N\\[+4pt]
2d & & & $2.04\pm0.02$ & & & & & & 834/852 & N\\[+4pt]
2e & & & $2.03\pm0.02$ & & & & & & 800/859 & N\\[+4pt]
2f & & & $2.03\pm0.02$ & & & & & & 1082/1000 & N\\[+4pt]
2g & & & $2.10\pm0.05$ & & & & & & 258/255 & N\\[+4pt]
2h & & & $2.21\pm0.04$ & & & & & & 310/314 & N\\[+4pt]
2i & & & $2.01\pm0.05$ & & & & & & 273/252 & N\\[+4pt]
2l & & & $2.09\pm0.04$ & & & & & & 344/362 & N\\[+4pt]
2m & & & $2.18\pm0.04$ & & & & & & 308/325 & N\\[+4pt]
3  & 3C~332 & 2  & $1.56\pm0.04$ & & $6.42\pm0.08$ & 100 & $0.5\pm0.1$ & $58^{+14}_{-16}$ & 416/413 & N\\[+4pt]
4a & 3C~382 & 7  & $1.76\pm0.01$ & & $6.42\pm0.01$ & $87\pm18$ & $2.6\pm0.3$ & $52\pm5$ & 1631/1599 & N\\[+4pt]
5  & 3C~059 & 6 & $1.55\pm0.03$ & & & & & & 580/642 & N\\[+4pt]
6a & 3C~390.3 & 4 & $1.62\pm0.02$ & & $6.49^{+0.09}_{-0.05}$ & $186^{+149}_{-84}$ & $2.2^{+1.0}_{-0.6}$ & $50\pm9$ & 1113/1084 & N\\[+4pt]
6b &          &   & $1.62\pm0.02$ & & $6.44\pm0.02$ & $70\pm33$ & $2.3\pm0.4$ & $58\pm9$ & 1012/896 & N\\[+4pt]
8  & 3C~206 & 6  & $1.75\pm0.03$ & &               &  &  & & 670/687 & N\\[+4pt]
9a & 3C~120 & 10 & $1.74\pm0.01$ & & $6.42\pm0.01$ & $86\pm14$ & $3.1\pm0.3$ & $62\pm5$ & 1323/1314 & N\\[+4pt]
   &        &    &               & & $6.96\pm0.02$ & 10        & $0.9\pm0.2$ & $21\pm4$ &           &  \\[+4pt]
9f &        &    & $1.76\pm0.01$ & & $6.37\pm0.01$ & $92\pm16$ & $3.9\pm0.3$ & $71\pm5$ & 1780.0/1737 & Y\\[+4pt]
   &        &    &               & & $6.85\pm0.03$ & $125^{+35}_{-32}$ & $1.5\pm0.3$ & $31\pm8$ & &\\[+4pt]
9g &        &    & $1.81\pm0.01$ & & $6.37\pm0.01$ & $108\pm19$ & $3.8\pm0.3$ & $72^{+5}_{-7}$ & 1573/1603 & N\\
   &        &    &               & & $6.91\pm0.03$ & 10 & $1.0\pm0.2$ & $21\pm4$ & &\\[+4pt]
10a & 4C$+$50.55 & 100 & $1.35\pm0.02$ & & $6.40\pm0.04$ & 100 & $1.7\pm0.5$ & $27\pm8$ & 1134/1066 & N\\[+4pt]
10b &            &     & $1.71\pm0.02$ & $2\pm1$ & $6.37\pm0.03$ & 100 & $2.7\pm0.4$ & $28\pm5$ & 1687/1601 & N\\[+4pt]
11 & Pictor~A & 3 & $1.73\pm0.03$ & & $6.44\pm0.05$ & 100 & $0.7\pm0.2$ & $61\pm17$ & 454/479 & N\\[+4pt]

\hline

\end{tabular}
\end{minipage}
\end{table}

\begin{table}
\centering
\begin{minipage}{153mm}
\contcaption{-- Baseline model parameters of the spectra in the E$=$3.5--10.5~keV band. We consider only observations without Fe K absorption line searches reported in the literature.}

\begin{tabular}{l@{\hspace{0.0cm}}|l|@{\hspace{0.1cm}}c@{\hspace{0.0cm}}c@{\hspace{0.2cm}}c@{\hspace{0.2cm}}c@{\hspace{0.2cm}}c@{\hspace{0.2cm}}c@{\hspace{0.2cm}}c@{\hspace{0.0cm}}|c@{\hspace{0.0cm}}|c@{\hspace{0.0cm}}}

\hline\hline       
Num & Source & $N_\mathrm{H, Gal}$ & $\Gamma$ & $N_\mathrm{H}$ & $E$ & $\sigma$ & $I$ & EW & $\chi^2/\nu$ & Abs\\ 
\scriptsize{(1)} & \scriptsize{(2)} & \scriptsize{(3)} & \scriptsize{(4)} & \scriptsize{(5)} & \scriptsize{(6)} & \scriptsize{(7)} & \scriptsize{(8)} & \scriptsize{(9)} & \scriptsize{(10)} & \scriptsize{(11)}\\
\hline   
12a & PKS~1549$-$79 & 10 & $1.78\pm0.06$ & $6\pm1$ & & & & & 511/540 & Y\\[+4pt]
12b &               &    & $1.93\pm0.07$ & $6\pm1$ & & & & & 466/493 & N\\[+4pt]
13  & 3C~234 & 2  & $1.40\pm0.23$ & $25\pm5$ & $6.43\pm0.04$ & $94\pm42$ & $0.9\pm0.2$ & $216\pm40$ & 89/83 & N\\[+4pt]
14 & PKS~0707$-$35 & 16 & $1.75\pm0.08$ & $5\pm1$ & $6.39\pm0.02$ & 10 & $0.6\pm0.1$ & $66^{+14}_{-12}$ & 477/463 & N\\[+4pt]
15 & PKS~2356$-$61 & 2 & $1.85\pm0.08$ & $17\pm2$ & & & & & 467/436 & N\\[+4pt]
16  & 3C~105 & 10.2 & $1.35\pm0.23$ & $35\pm4$ & $6.39\pm0.03$ & 100 & $1.7\pm0.3$ & $214^{+94}_{-33}$ & 111/101 & Y\\[+4pt] 
17 & PKS~0326$-$288 & 1  & $1.83\pm0.31$ & $47\pm7$ & $6.40\pm0.04$ & $138^{+66}_{-54}$ & $1.3\pm0.3$ & $275^{+77}_{-57}$ & 127/81 & N\\[+4pt]
18a & 3C~445 & 5 & $1.22\pm0.12$ & $15\pm2$ & $6.38\pm0.03$ & $107^{+43}_{-37}$ & $2.1\pm0.4$ & $145^{+31}_{-35}$ & 230/227 & N\\[+4pt]
19  & 3C~433 & 12 & $1.69\pm0.13$ & $6\pm2$ & $6.46\pm0.03$ & 10 & $0.4\pm0.1$ & $83^{+27}_{-22}$ & 195/214 & N\\[+4pt]
20a & 3C~452 & 10 & $0.80\pm0.20$ & $21\pm3$ & $6.40\pm0.01$ & $71\pm21$ & $1.4\pm0.2$ & $285^{+38}_{-35}$ & 124/145 & N\\[+4pt]
20b &        &    & $0.87\pm0.12$ & $29\pm3$ & $6.42\pm0.02$ & $92\pm23$ & $1.6\pm0.2$ & $245^{+52}_{-22}$ & 271/265 & N\\[+4pt]
21a & 3C~33 & 4 & $1.25\pm0.31$ & $34\pm6$ & $6.40\pm0.05$ & $95^{+59}_{-46}$ & $2.0\pm0.6$ & $243^{+53}_{-59}$ & 35/32 & N\\[+4pt]
21b &       &   & $1.44\pm0.11$ & $32\pm2$ & $6.38\pm0.01$ & 100 & $1.6\pm0.1$ & $180^{+20}_{-10}$ & 420/385 & N\\[+4pt]
22a & NGC~612 & 2 & $1.67\pm0.36$ & $98\pm12$ & & & & & 32/34 & N\\[+4pt]
22b &         &   & $1.38\pm0.29$ & $84\pm8$ & $6.40\pm0.02$ & 10 & $2.8\pm0.5$ & $246\pm45$ & 45/68 & N\\[+4pt]
23 & 4C$+$29.30 & 4 & $1.34\pm0.44$ & $38\pm9$ & $6.41\pm0.04$ & $118^{+45}_{-42}$ & $1.0\pm0.2$ & $437^{+196}_{-107}$ & 24/26 & N\\[+4pt]
24  & 3C~403   & 15 & $1.20\pm0.50$ & $31\pm11$ & $6.44\pm0.02$ & $85\pm31$ & $1.3\pm0.2$ & $69^{+19}_{-10}$ & 48/40 & N\\[+4pt]
25 & IC~5063 & 6 & $1.84\pm0.10$ & $25\pm2$ & $6.40\pm0.01$ & $84\pm21$ & $4.2\pm0.5$ & $194^{+26}_{-20}$ & 347/362 & N\\[+4pt]
26a & Centaurus~A$^{*}$ & 8 & $1.91\pm0.08$ & $11.7\pm0.8$ & $6.41\pm0.02$ & $10$ & $35.3\pm3.7$ & $88^{+12}_{-8}$ & 672/740 & N\\[+4pt]
26b &          &   & $1.83\pm0.10$ & $12.4\pm1.1$ & $6.40\pm0.02$ & $10$ & $50.4\pm5.7$ & $115^{+17}_{-9}$ & 426/409 & N\\[+4pt]
26c &             &   & $1.82\pm0.01$ & $10.7\pm0.2$ & $6.388\pm0.003$ & $10$ & $35.8\pm1.1$ & $58\pm2$ & 2042/1909 & Y\\[+4pt]
26d &             &   & $1.83\pm0.01$ & $11.3\pm0.2$ & $6.393\pm0.003$ & $10$ & $38.4\pm1.1$ & $69\pm2$ & 2004/1889 & Y\\[+4pt]
26e &           &   & $1.80\pm0.01$ & $10.9\pm0.2$ & $6.391\pm0.003$ & $10$ & $36.0\pm1.1$ & $58\pm2$ & 2009/1908 & Y\\[+4pt]

\hline

\end{tabular}
{\em Notes:} (1) Object number. Letters are used to enumerate multiple observations. (2) Object name. (3) Galactic column density in units of $10^{20}$~cm$^{-2}$. (4) Power-law photon index. (5) Neutral absorption column density in units of $10^{22}$~cm$^{-2}$. (6)--(7)--(8)--(9) Emission line energy, width, intensity and equivalent width in units of keV, eV, $10^{-5}$~ph~$s^{-1}$~cm$^{-2}$ and eV, respectively. (10) Best-fit $\chi^2$ over degrees of freedom $\nu$. (11) Presence (or not) of absorption lines at rest-frame energies E$>$6~keV in the energy-intensity contour plots. $^{*}$ Given the high flux and S/N of the observations of Centaurus A, we included neutral reflection \emph{pexrav} component in XSPEC with reflection fraction $R=1$, high-energy cut-off $E_\mathrm{C} = 300$~keV and inclination of $i$$=$60$^{\circ}$.
\end{minipage}
\end{table}

\clearpage



\section{Ratios and contour plots}


   \begin{figure}
   \centering

    \includegraphics[width=4.8cm,height=6.7cm,angle=0]{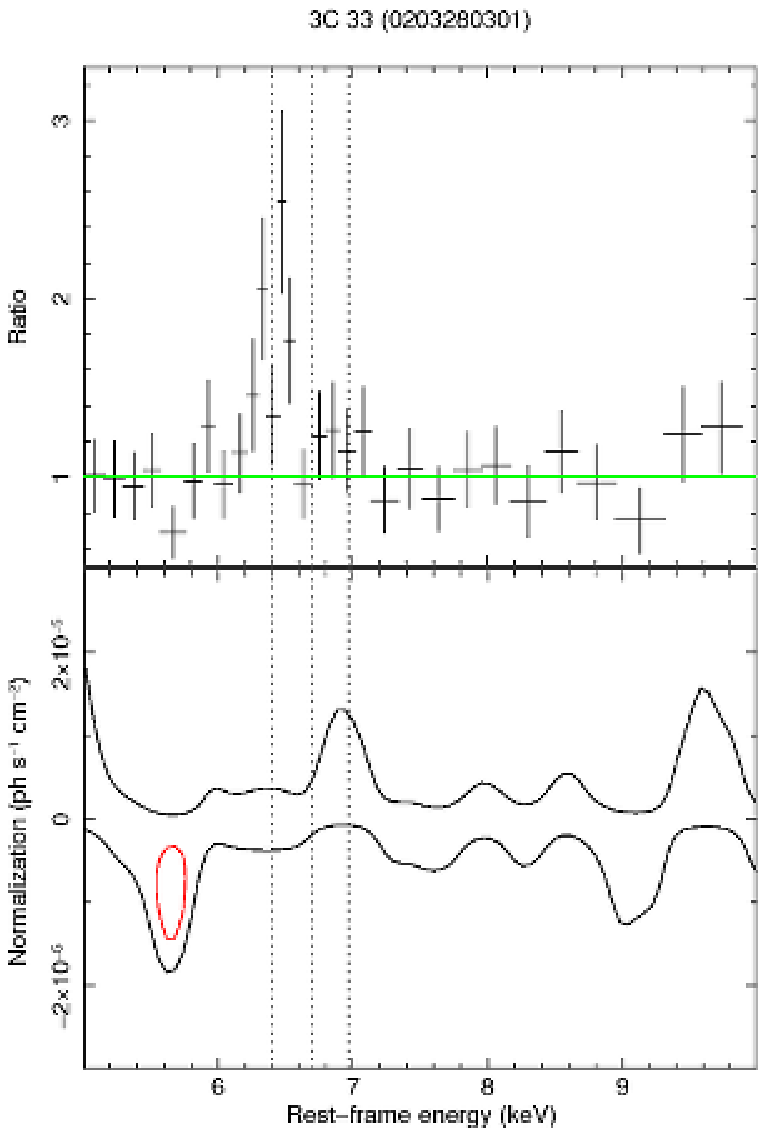}
\hspace{0.3cm}
    \includegraphics[width=4.8cm,height=6.7cm,angle=0]{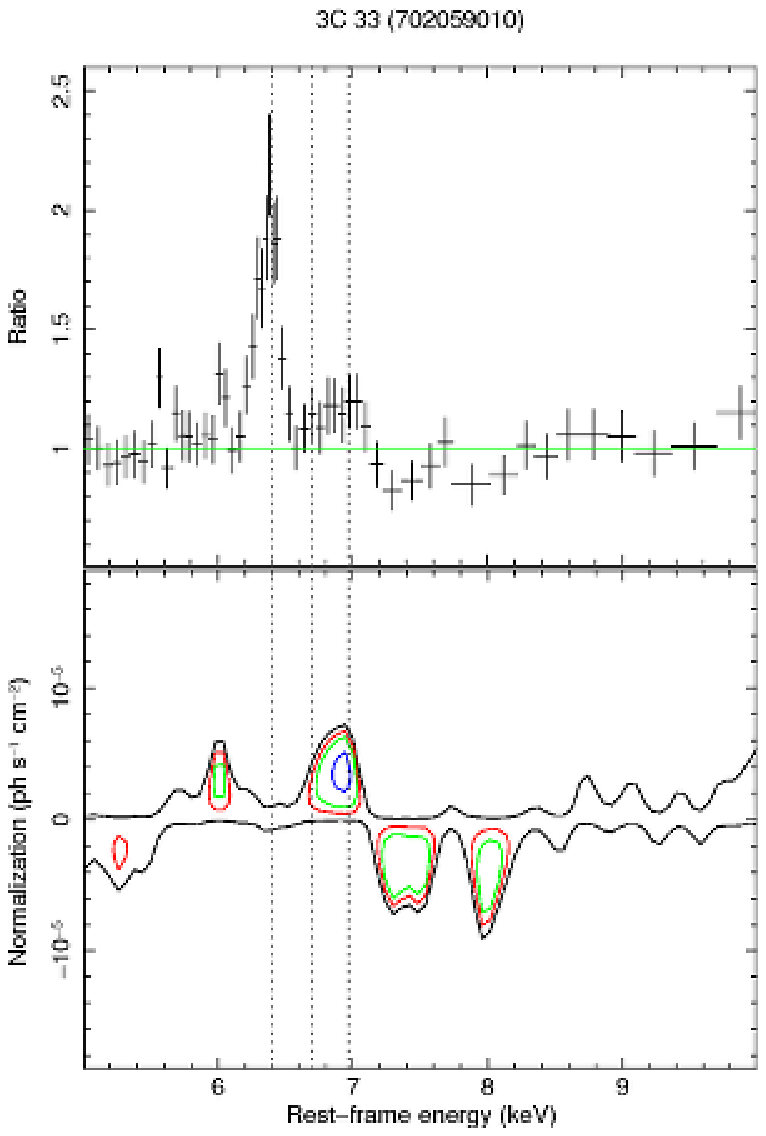}
\hspace{0.3cm}
    \includegraphics[width=4.8cm,height=6.7cm,angle=0]{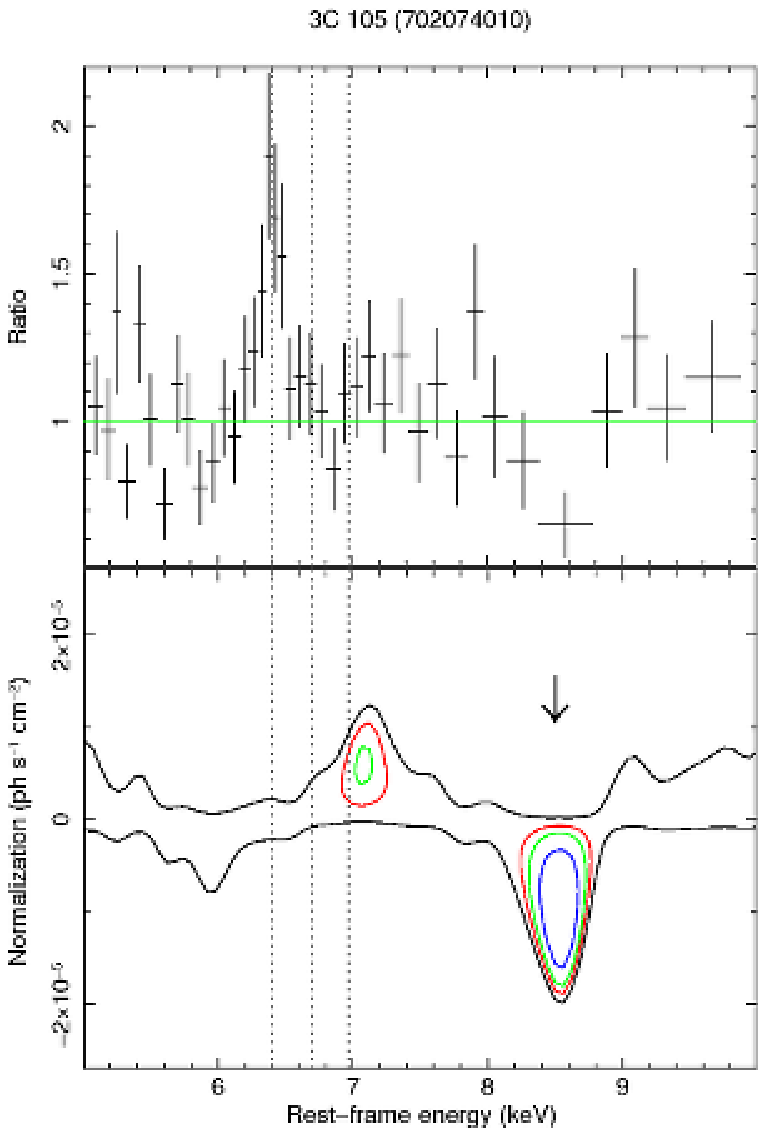}

\vspace{0.2cm}
    \includegraphics[width=4.8cm,height=6.7cm,angle=0]{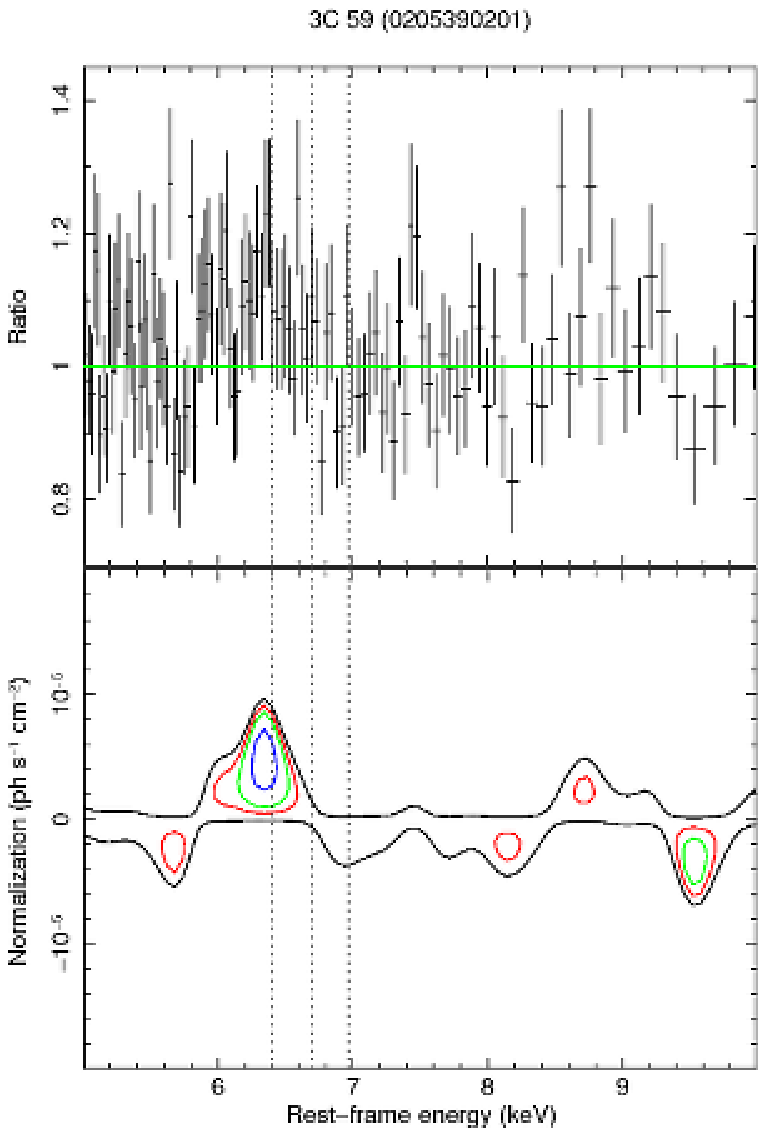}
\hspace{0.3cm} 
    \includegraphics[width=4.8cm,height=6.7cm,angle=0]{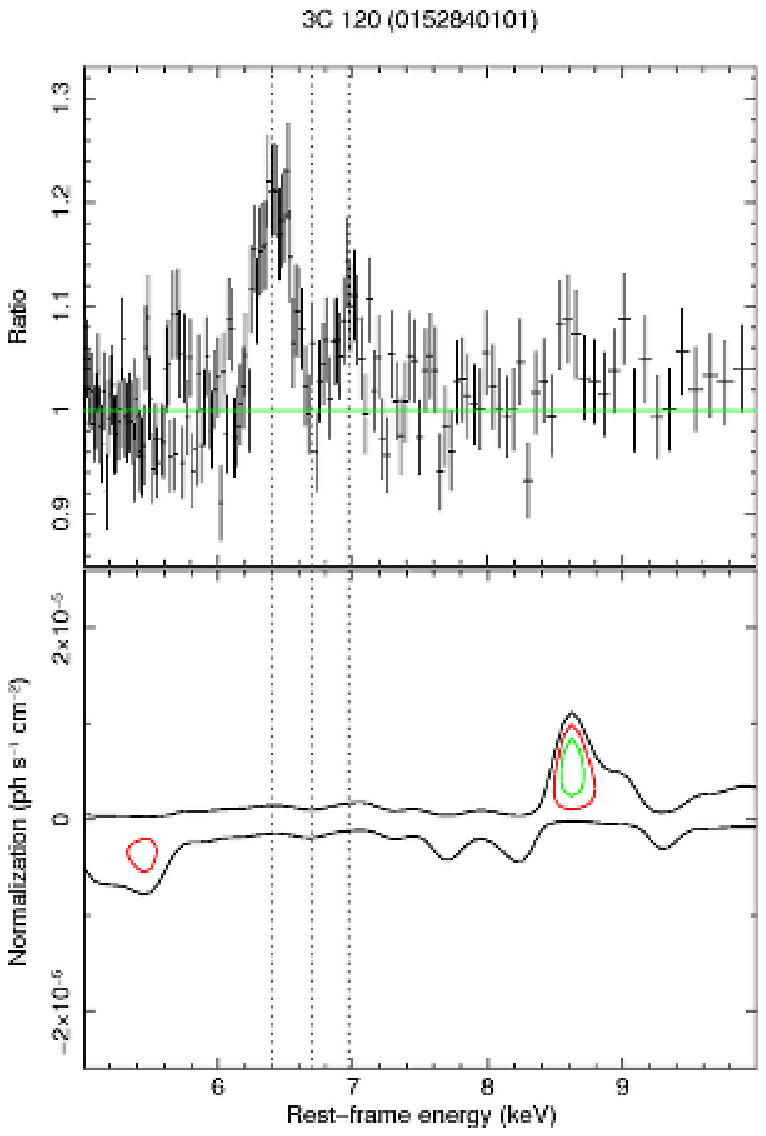}
\hspace{0.3cm}
    \includegraphics[width=4.8cm,height=6.7cm,angle=0]{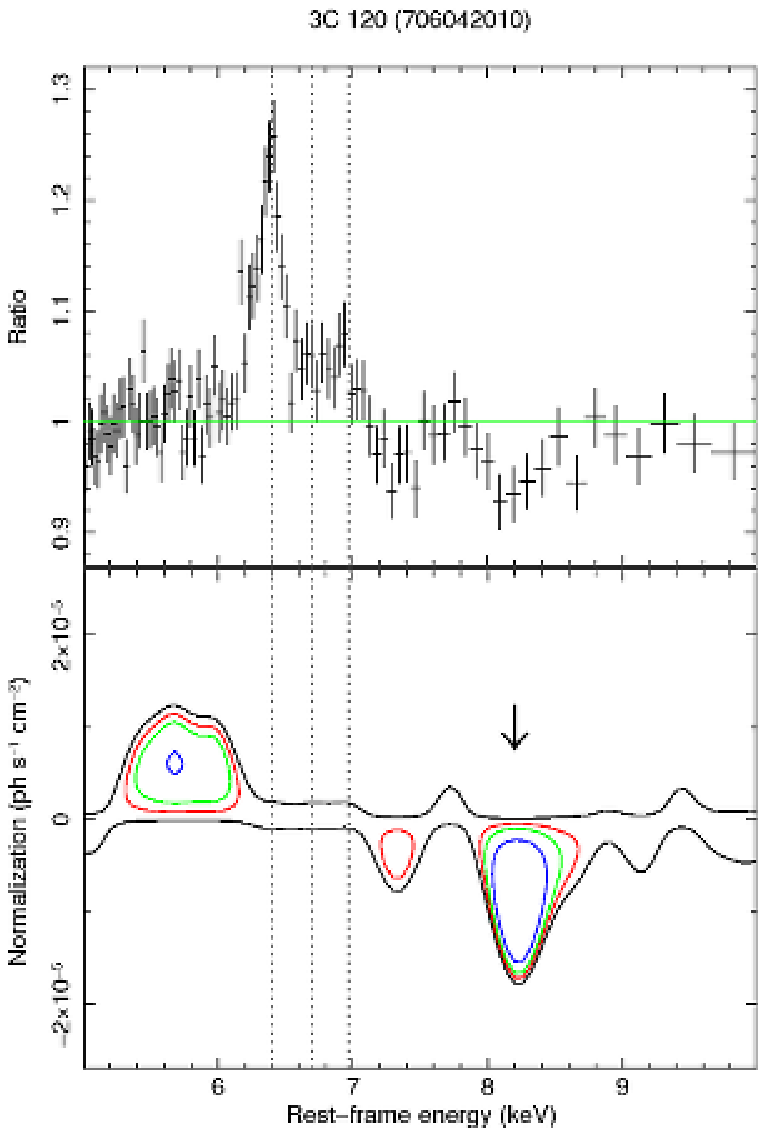}

    \caption{Ratio against the continuum (\emph{upper panel}) and contour plots with respect to the best-fit baseline model (\emph{lower panel}; 68\% (red), 90\% (green), 99\% (blue) levels) for the radio-loud sources. The The Fe K absorption lines are pointed by arrows. The vertical lines refer to the neutral Fe K$\alpha$, Fe~XXV and Fe~XXVI lines at E$=$6.4~keV, 6.7~keV and 6.97~keV, respectively.}
    \end{figure}


   \begin{figure}
   \centering

    \includegraphics[width=4.8cm,height=6.7cm,angle=0]{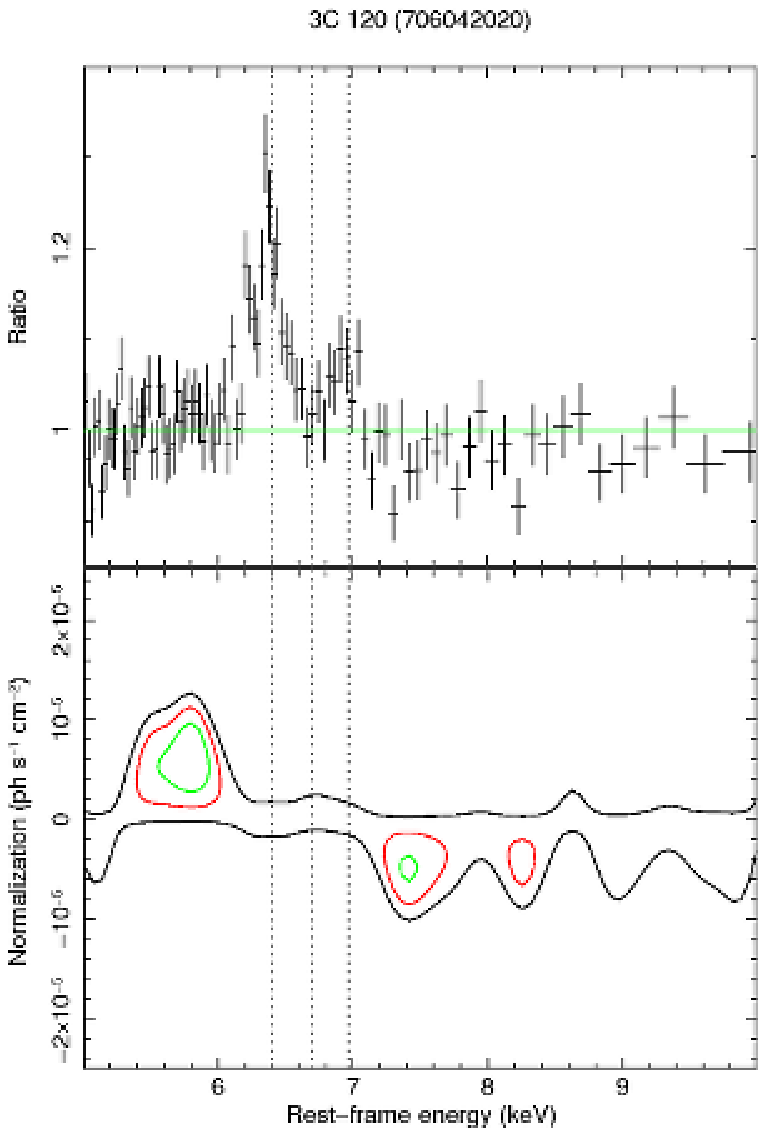}
\hspace{0.3cm}
    \includegraphics[width=4.8cm,height=6.7cm,angle=0]{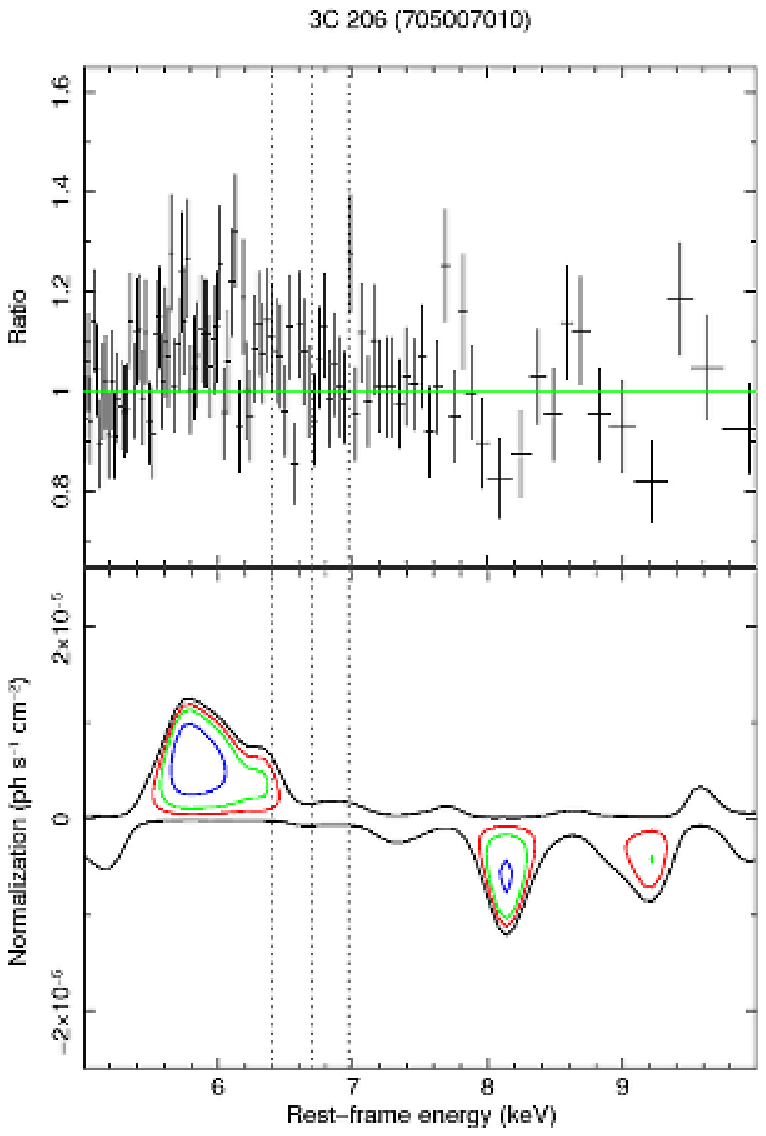}
\hspace{0.3cm}
    \includegraphics[width=4.8cm,height=6.7cm,angle=0]{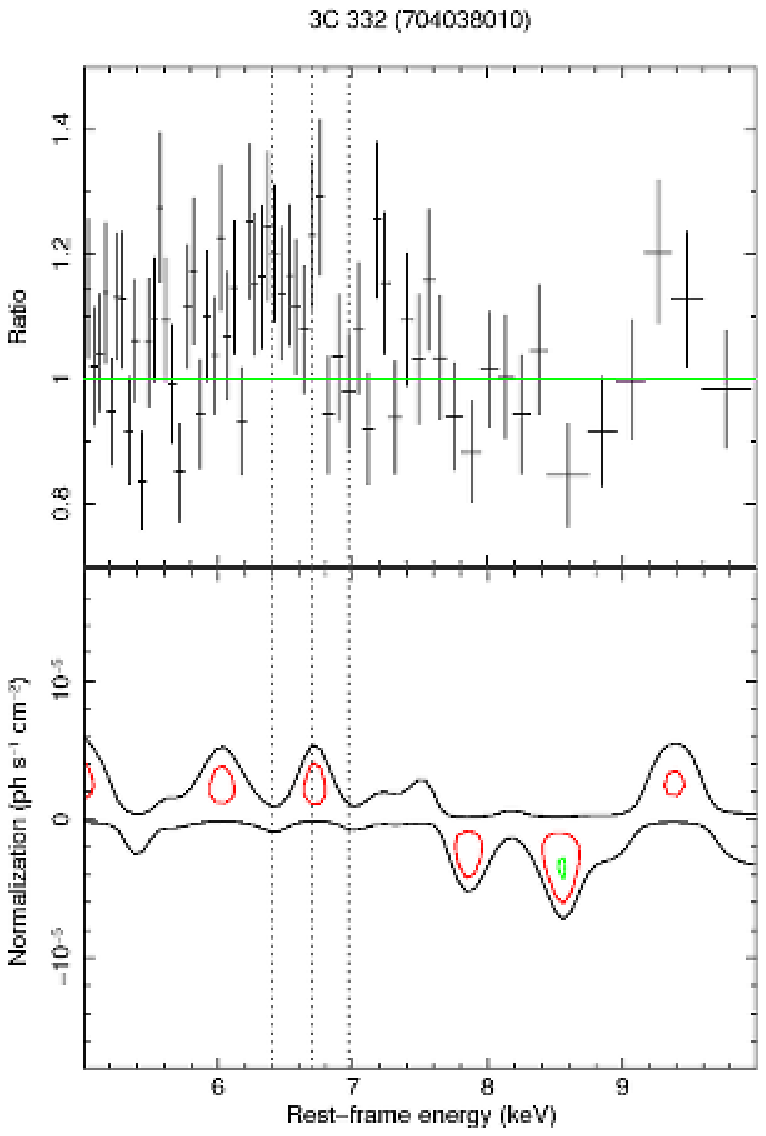}

\vspace{0.2cm}
   \includegraphics[width=4.8cm,height=6.7cm,angle=0]{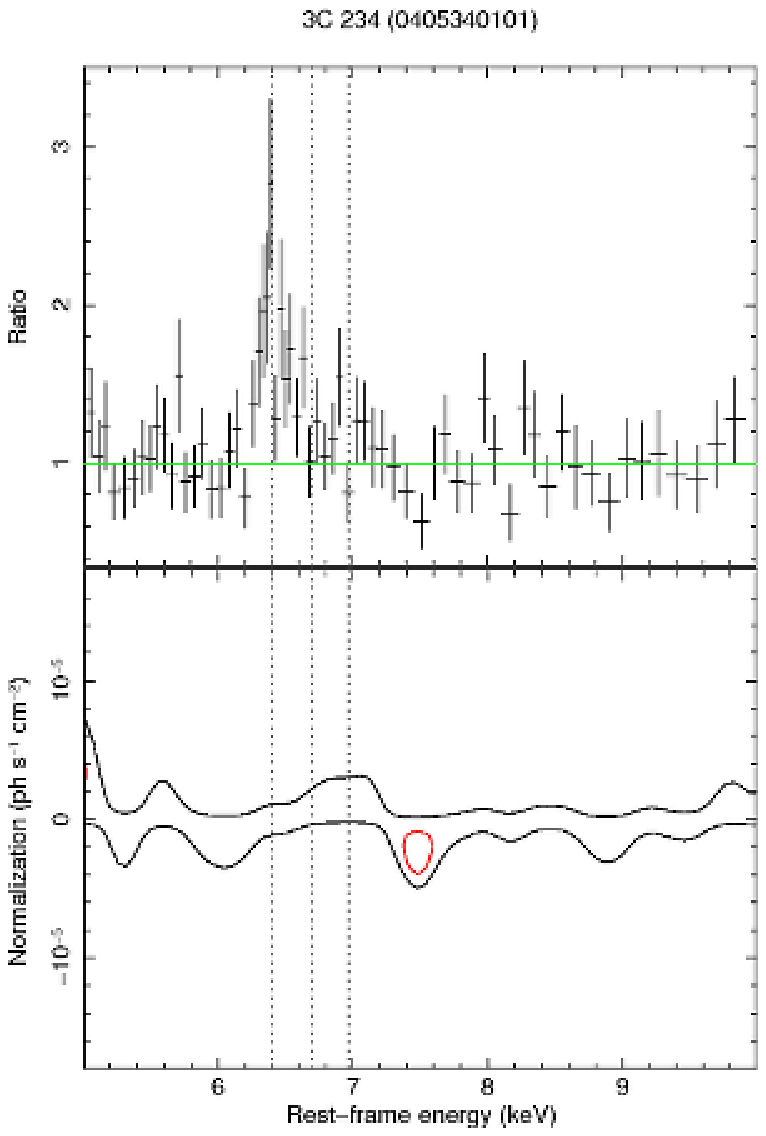}
\hspace{0.3cm} 
   \includegraphics[width=4.8cm,height=6.7cm,angle=0]{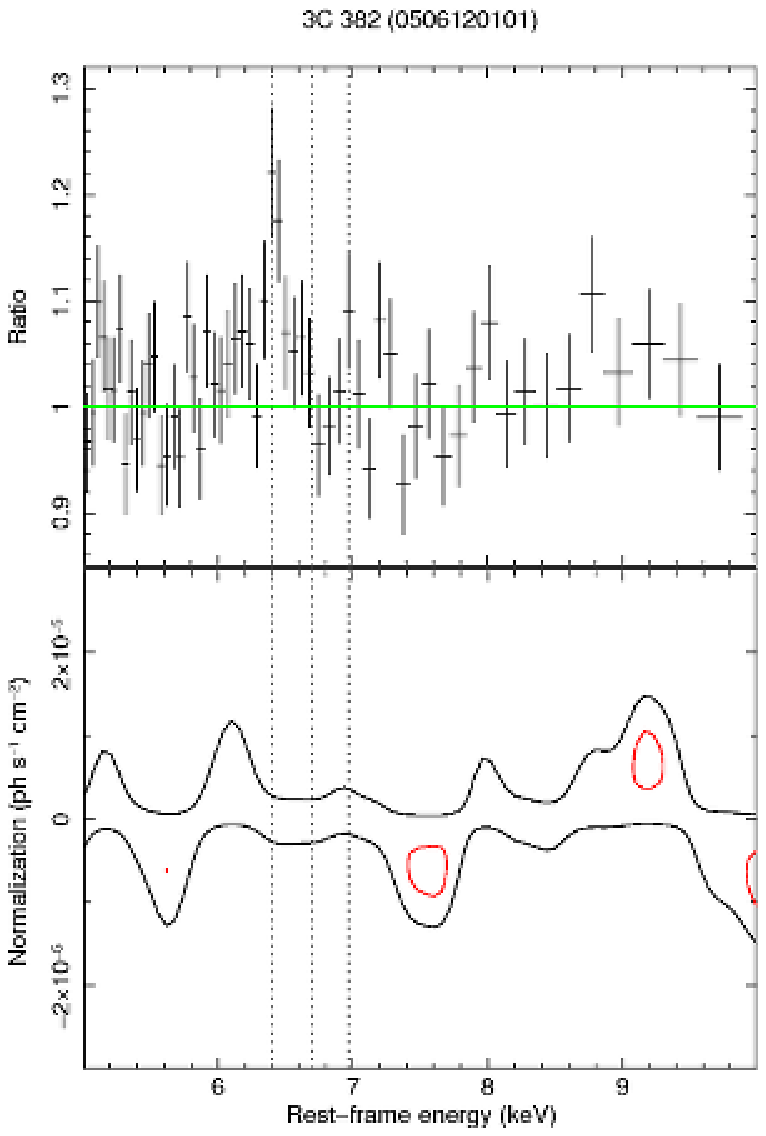}
\hspace{0.3cm}
   \includegraphics[width=4.8cm,height=6.7cm,angle=0]{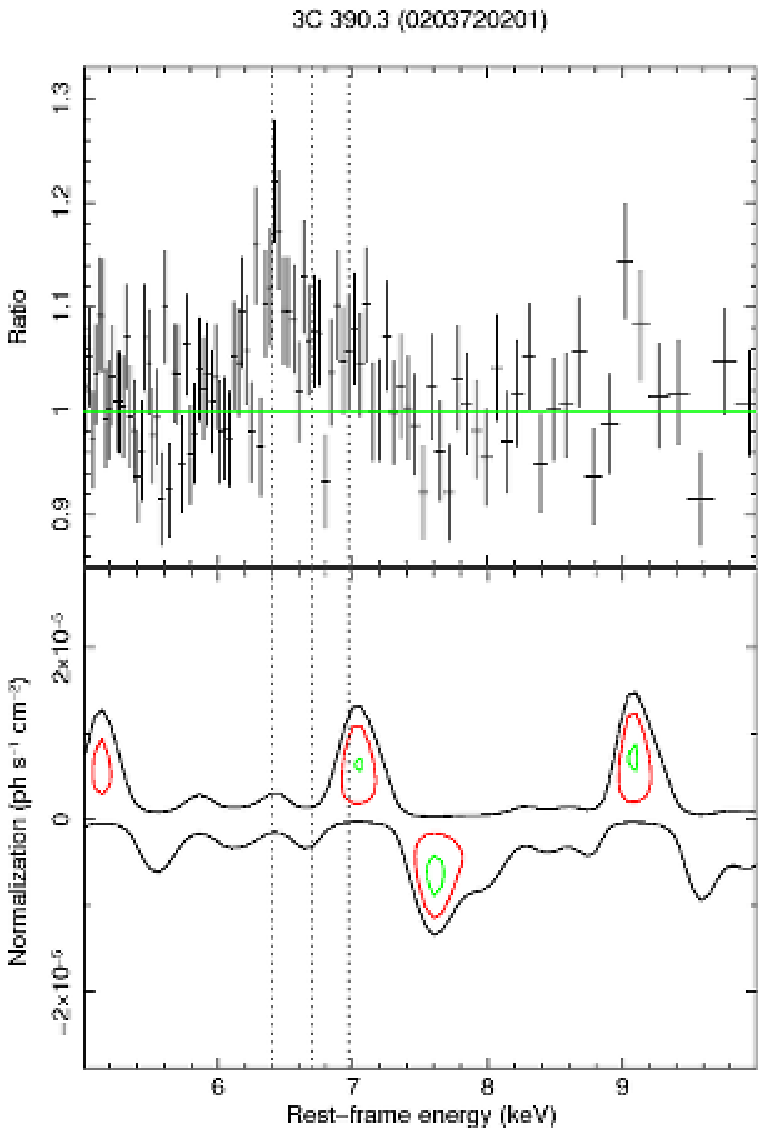}

\vspace{0.2cm}
    \includegraphics[width=4.8cm,height=6.7cm,angle=0]{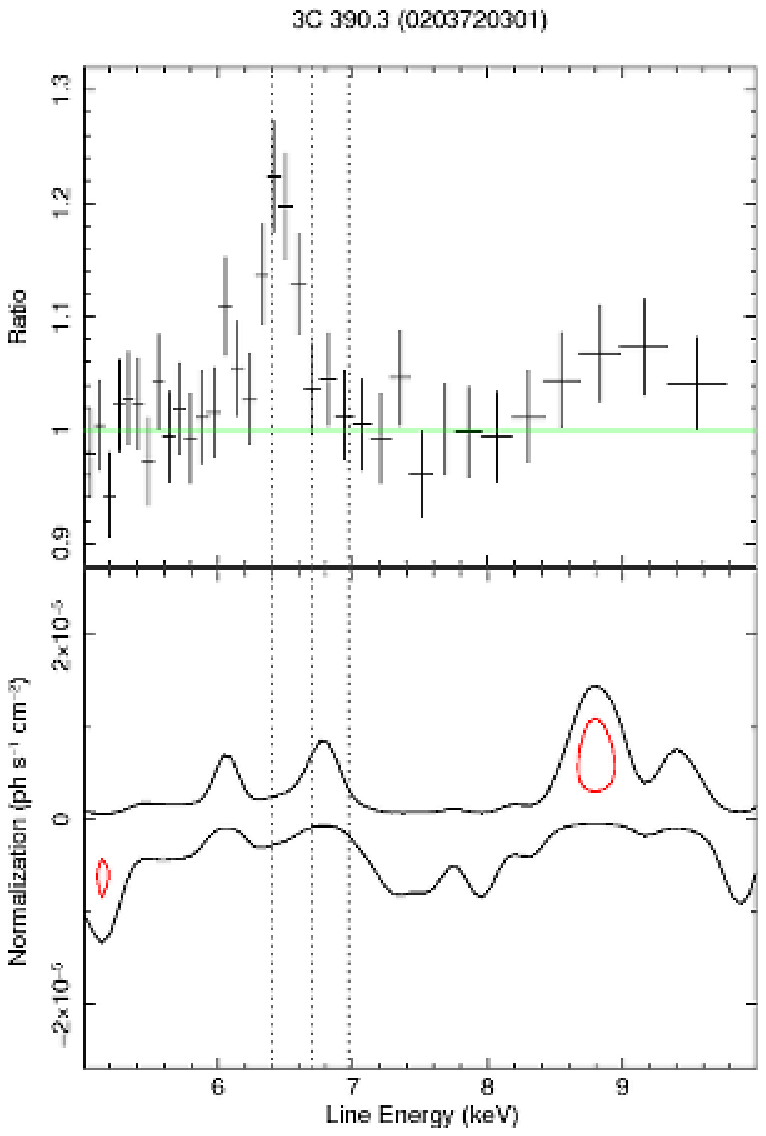}
\hspace{0.3cm}
    \includegraphics[width=4.8cm,height=6.7cm,angle=0]{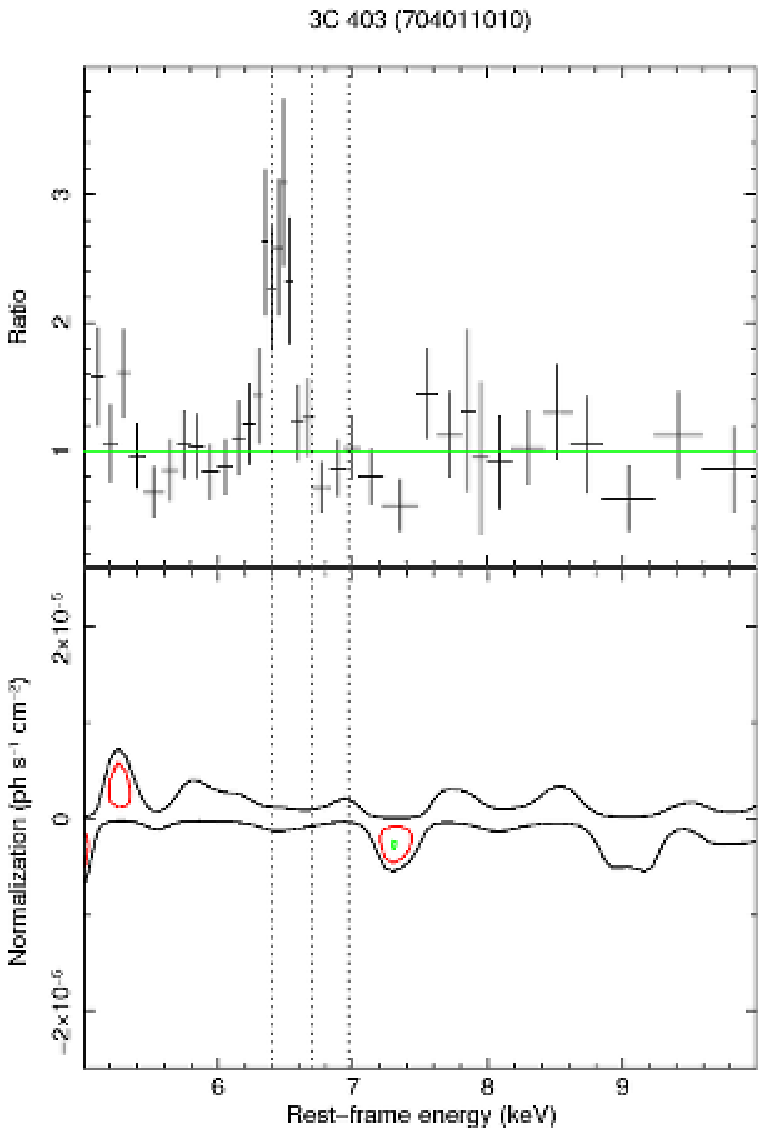}
\hspace{0.3cm}
    \includegraphics[width=4.8cm,height=6.7cm,angle=0]{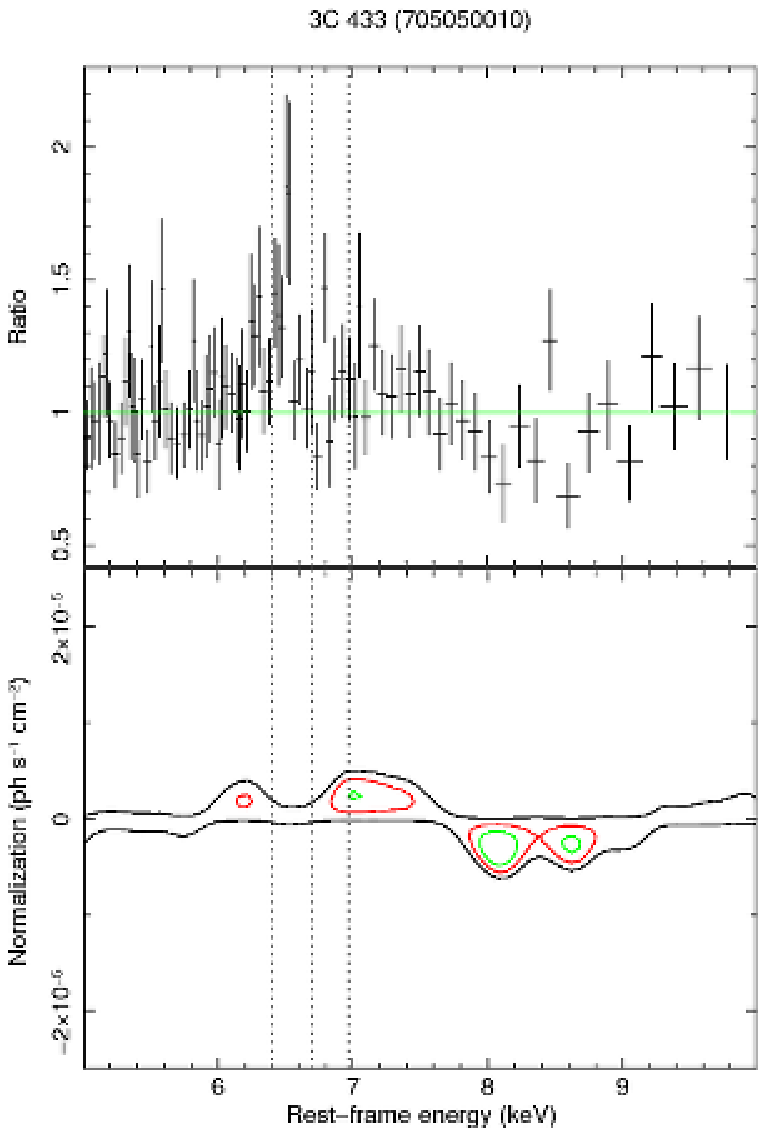}

\contcaption{-- Ratio against the continuum (\emph{upper panel}) and contour plots with respect to the best-fit baseline model (\emph{lower panel}).}
    \end{figure}


   \begin{figure}
   \centering

    \includegraphics[width=4.8cm,height=6.7cm,angle=0]{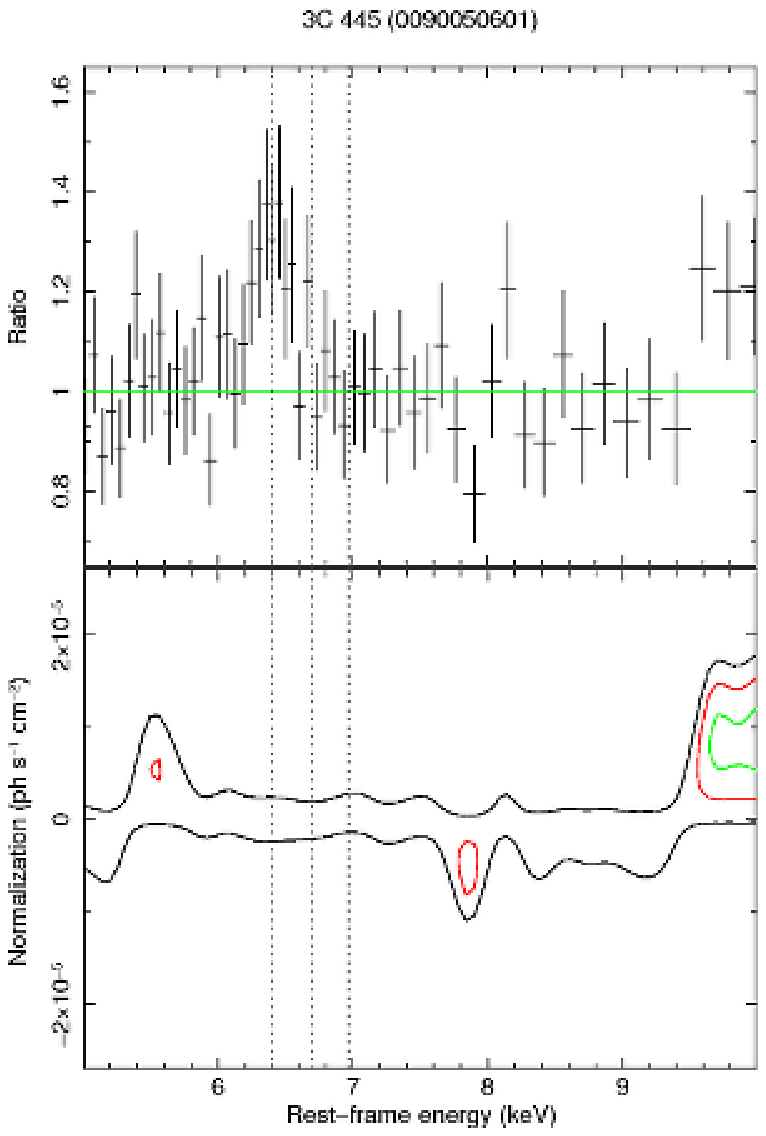}
\hspace{0.3cm}
    \includegraphics[width=4.8cm,height=6.7cm,angle=0]{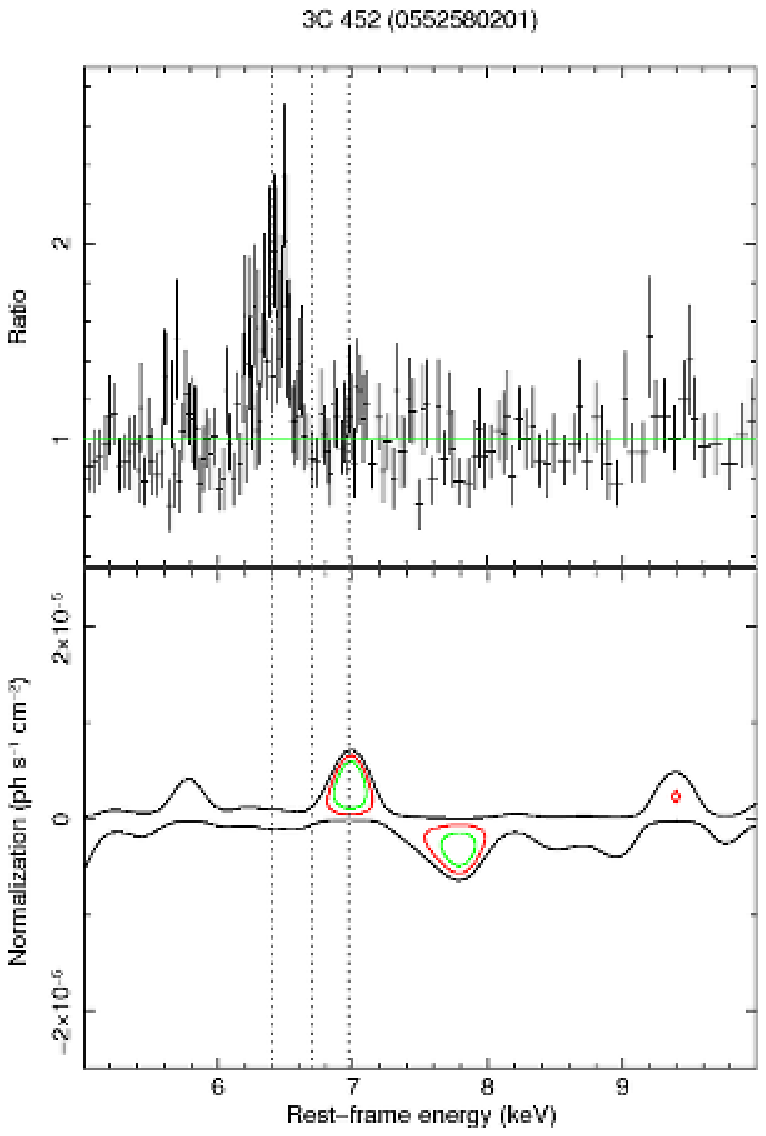}
\hspace{0.3cm}
    \includegraphics[width=4.8cm,height=6.7cm,angle=0]{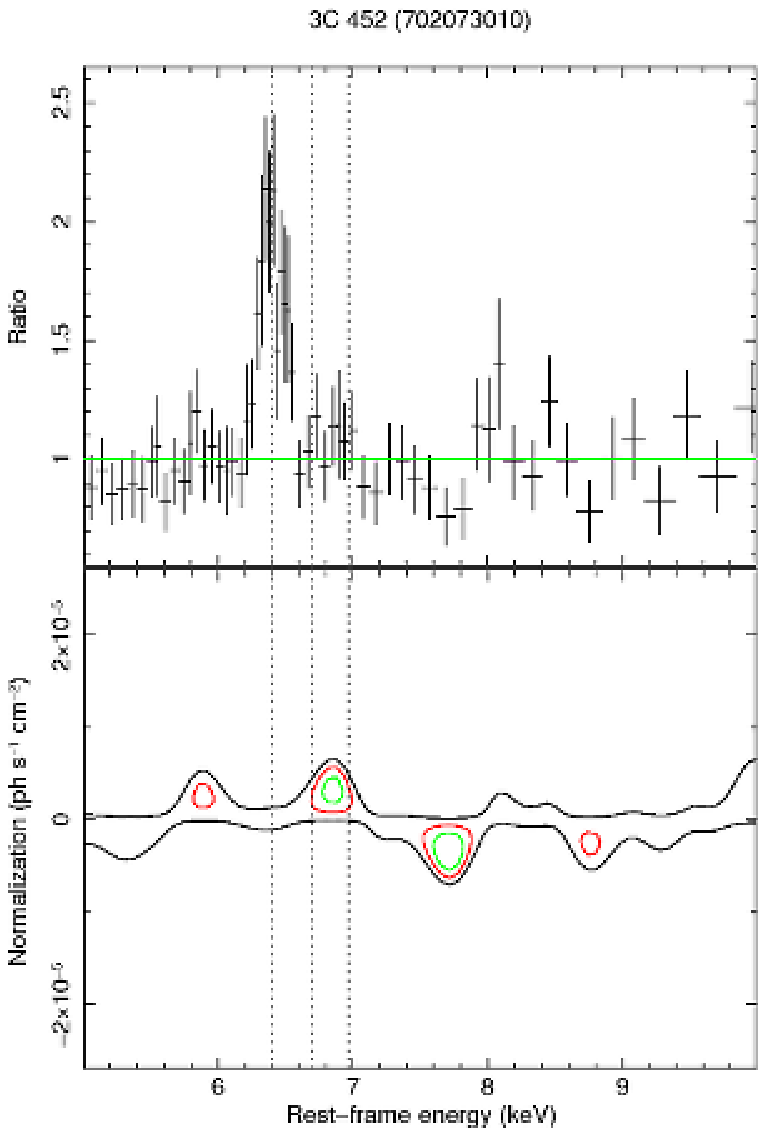}

\vspace{0.2cm}
    \includegraphics[width=4.8cm,height=6.7cm,angle=0]{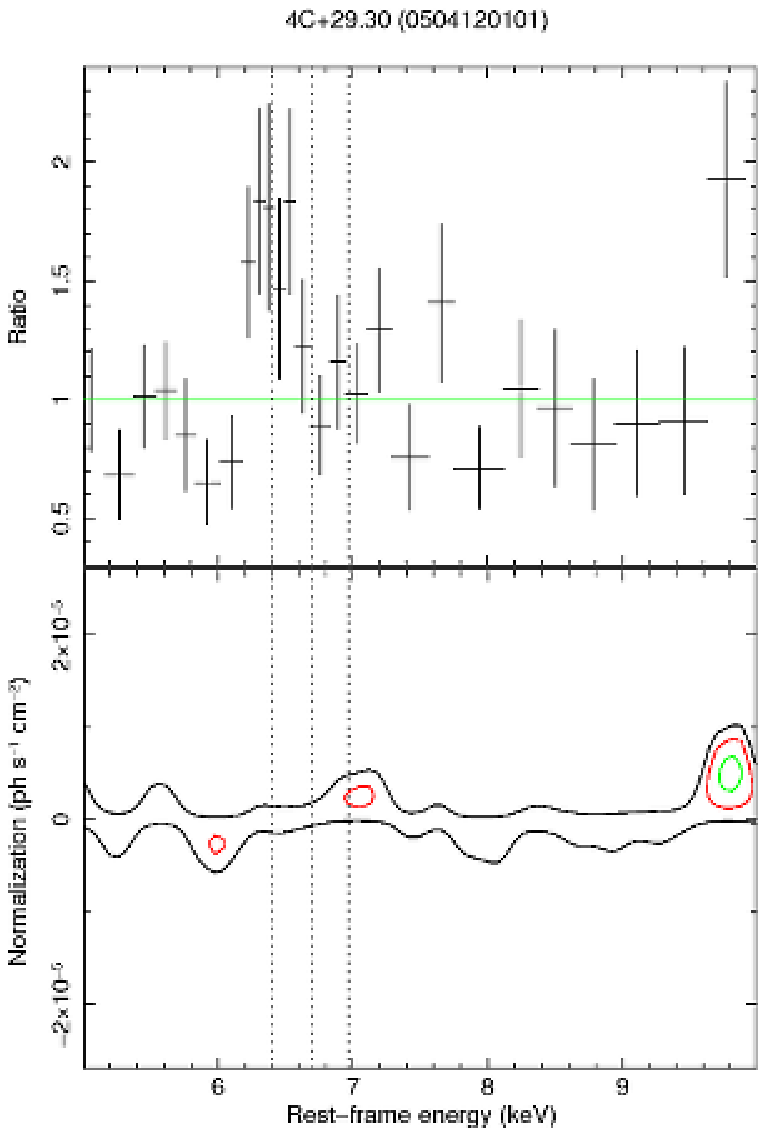}
\hspace{0.3cm} 
    \includegraphics[width=4.8cm,height=6.7cm,angle=0]{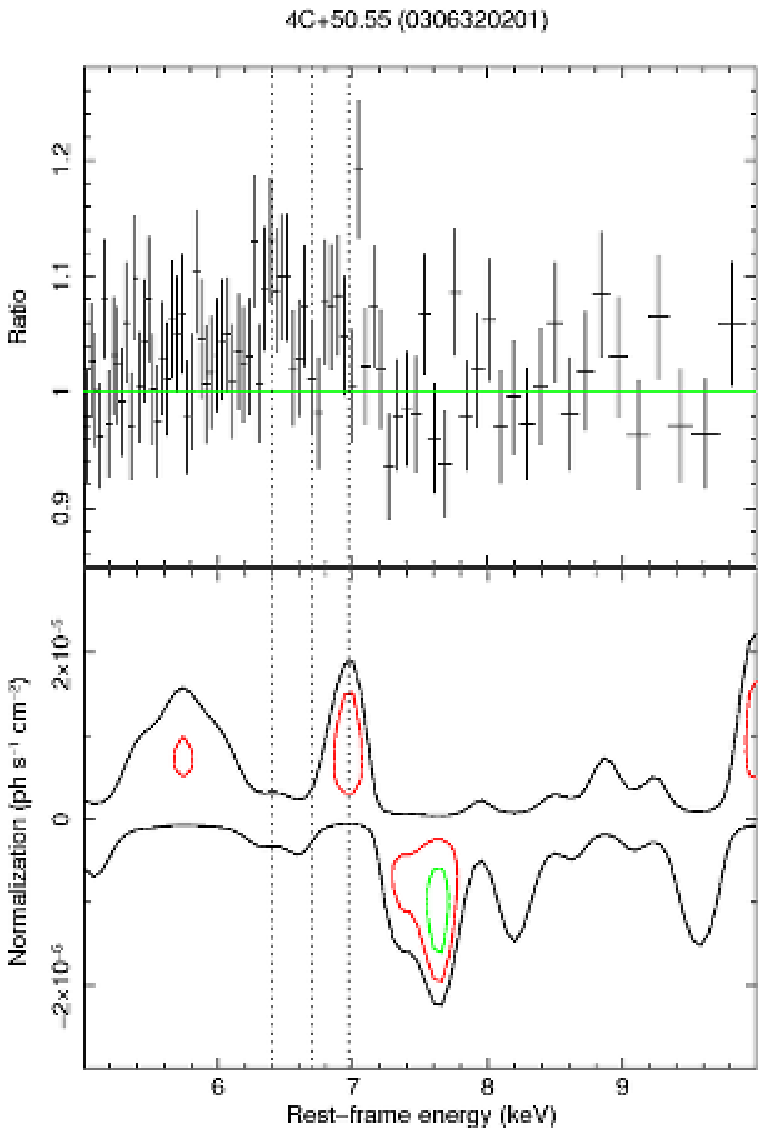}
\hspace{0.3cm}
    \includegraphics[width=4.8cm,height=6.7cm,angle=0]{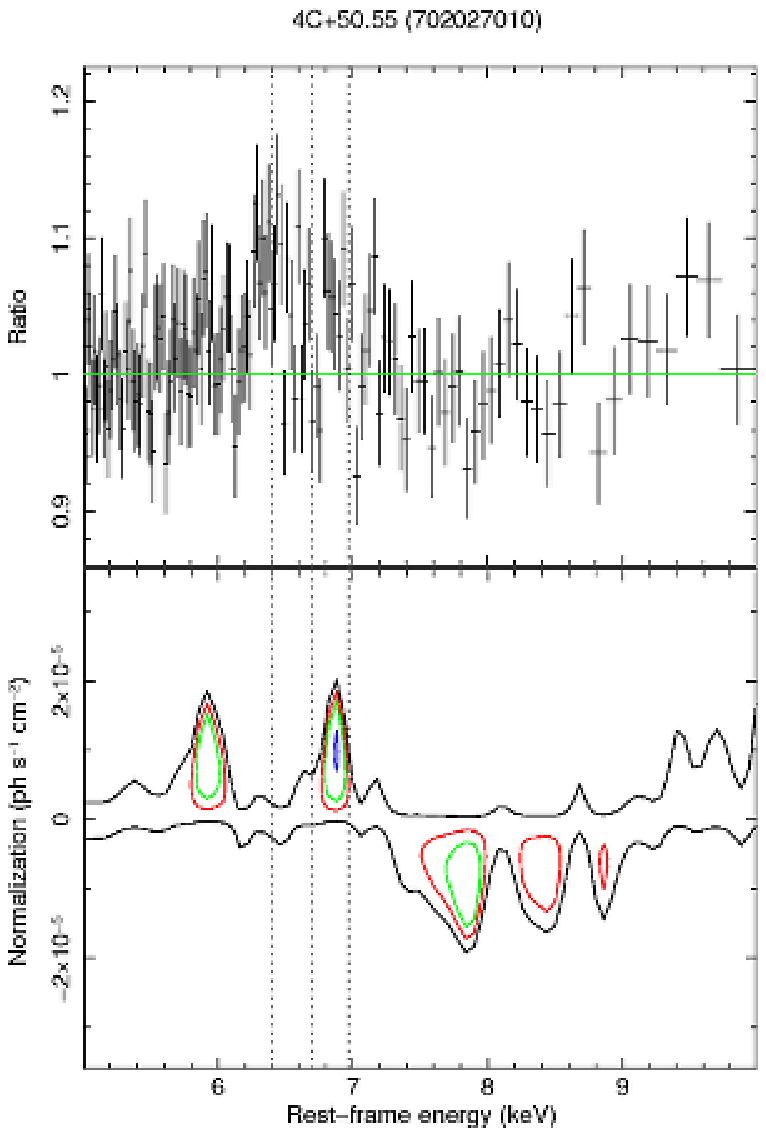}

\vspace{0.2cm}
    \includegraphics[width=4.8cm,height=6.7cm,angle=0]{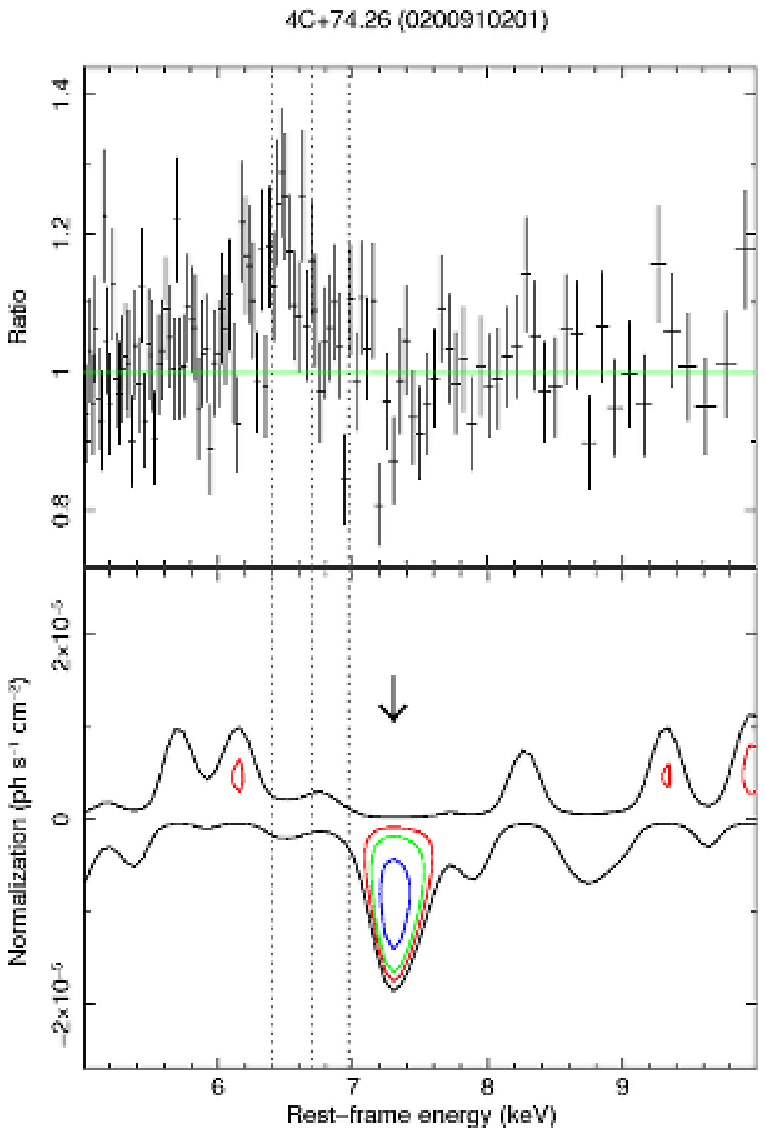}
\hspace{0.3cm}
    \includegraphics[width=4.8cm,height=6.7cm,angle=0]{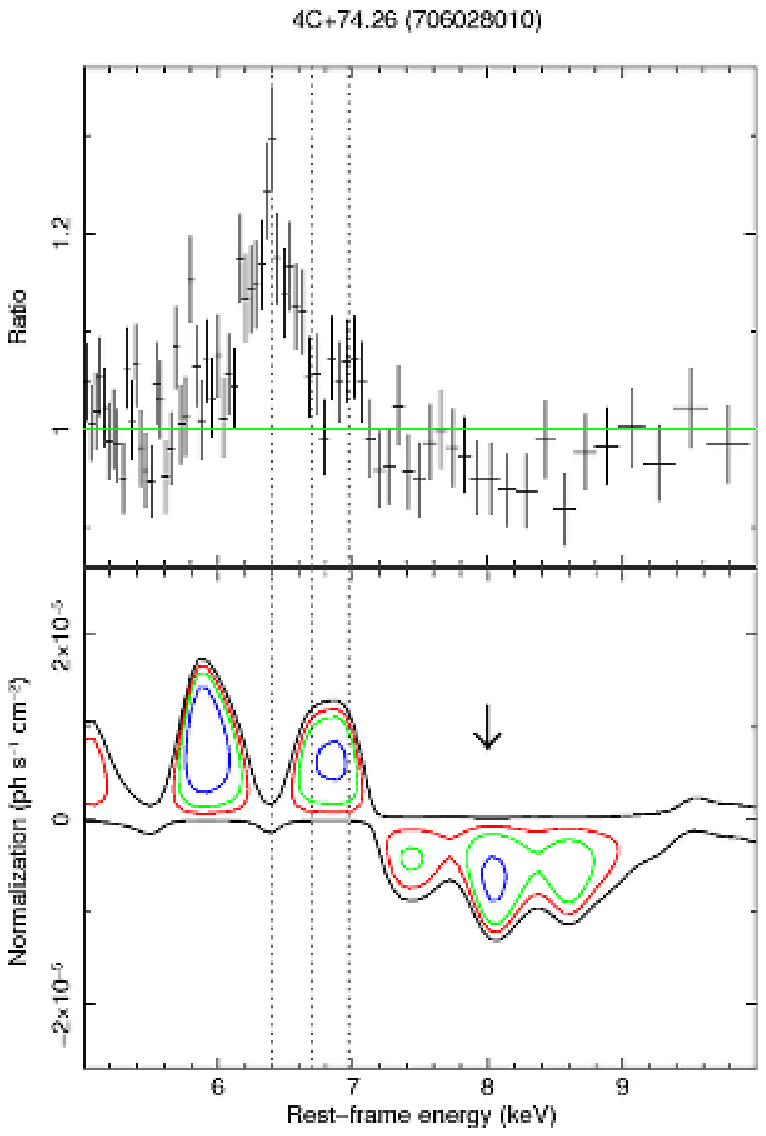}
\hspace{0.3cm}
    \includegraphics[width=4.8cm,height=6.7cm,angle=0]{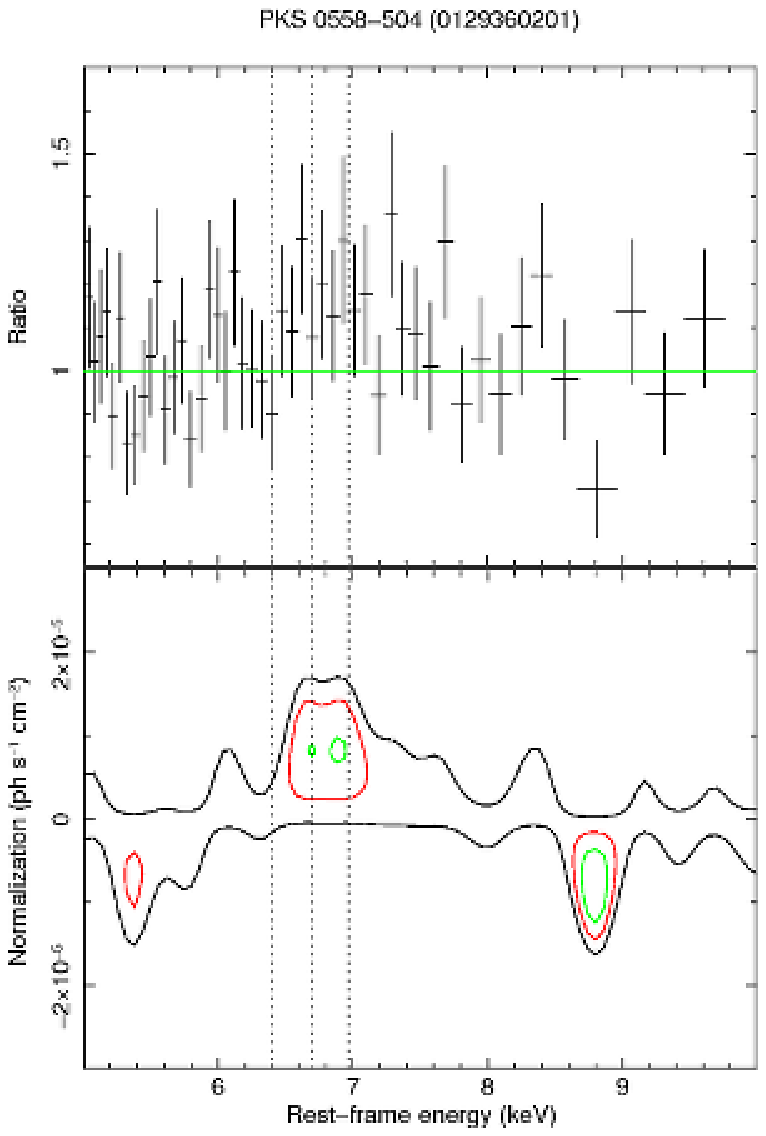}

\contcaption{-- Ratio against the continuum (\emph{upper panel}) and contour plots with respect to the best-fit baseline model (\emph{lower panel}).}
    \end{figure}


   \begin{figure}
   \centering

    \includegraphics[width=4.8cm,height=6.7cm,angle=0]{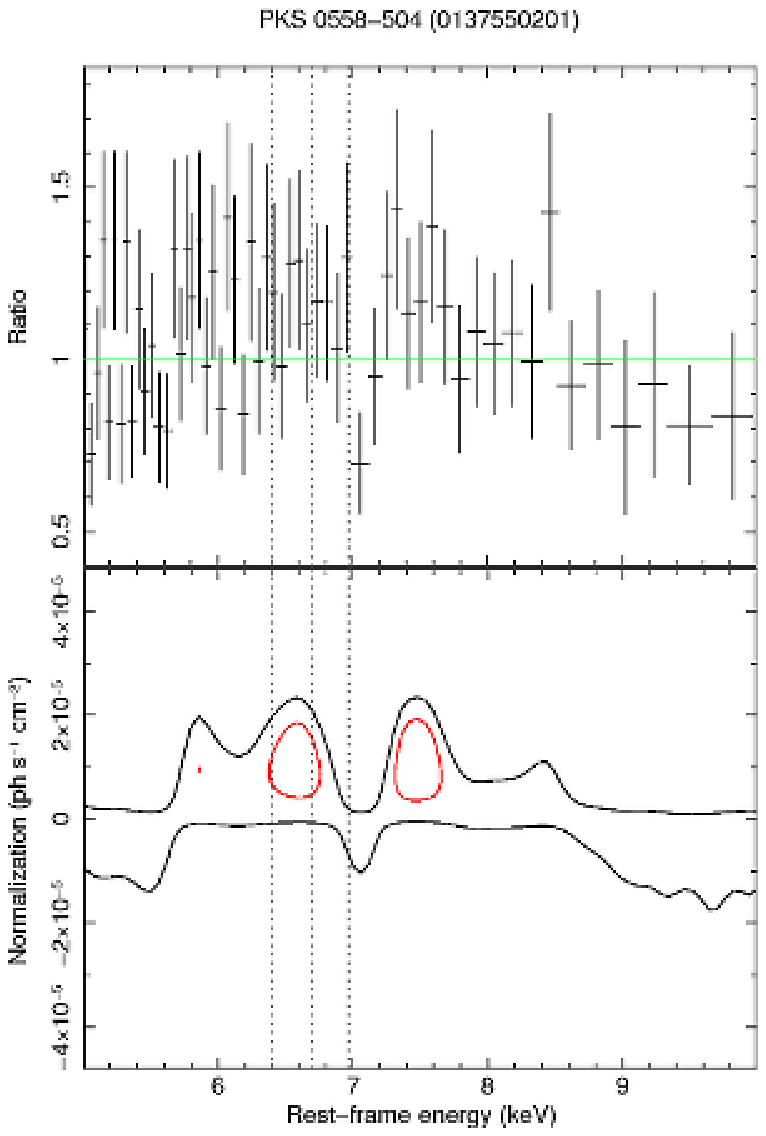}
\hspace{0.3cm}
    \includegraphics[width=4.8cm,height=6.7cm,angle=0]{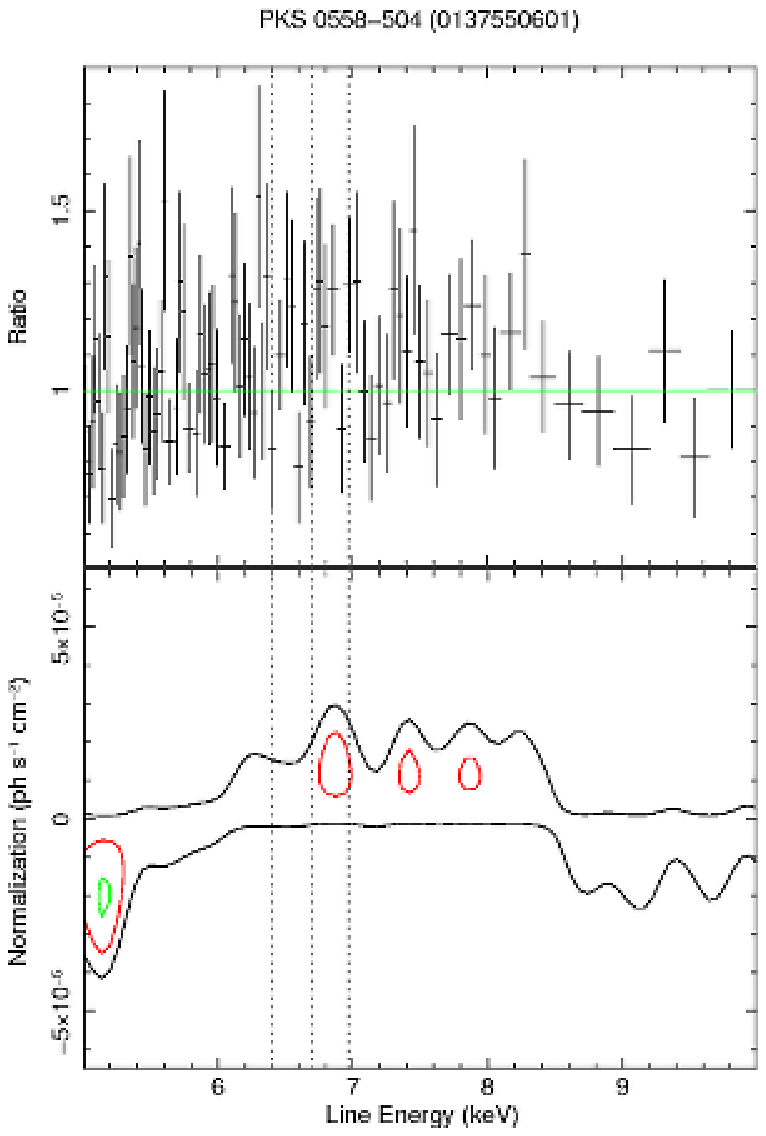}
\hspace{0.3cm}
    \includegraphics[width=4.8cm,height=6.7cm,angle=0]{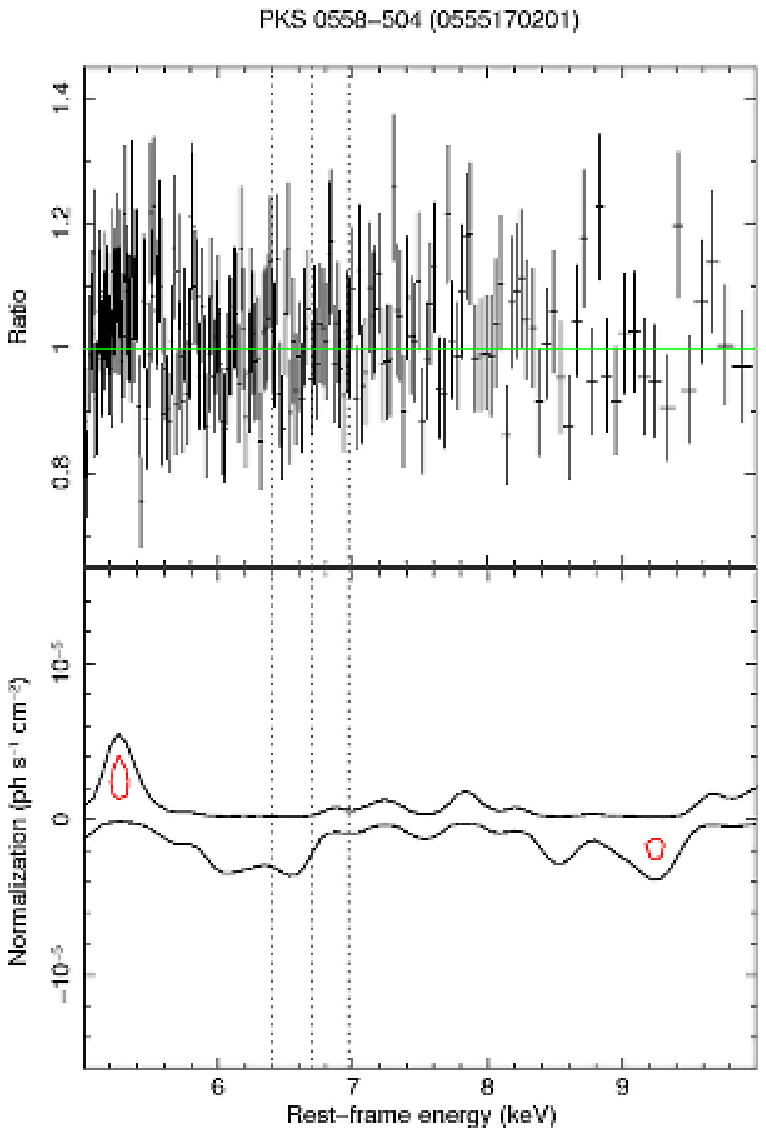}

\vspace{0.2cm}
    \includegraphics[width=4.8cm,height=6.7cm,angle=0]{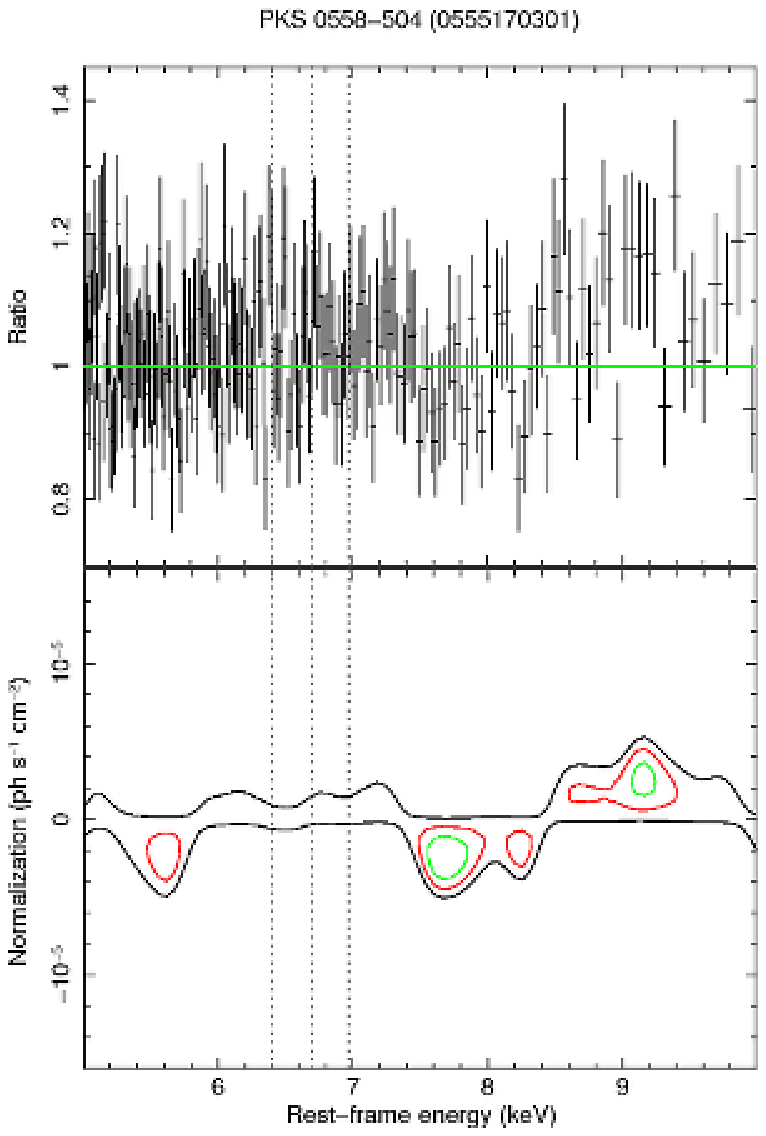}
\hspace{0.3cm} 
    \includegraphics[width=4.8cm,height=6.7cm,angle=0]{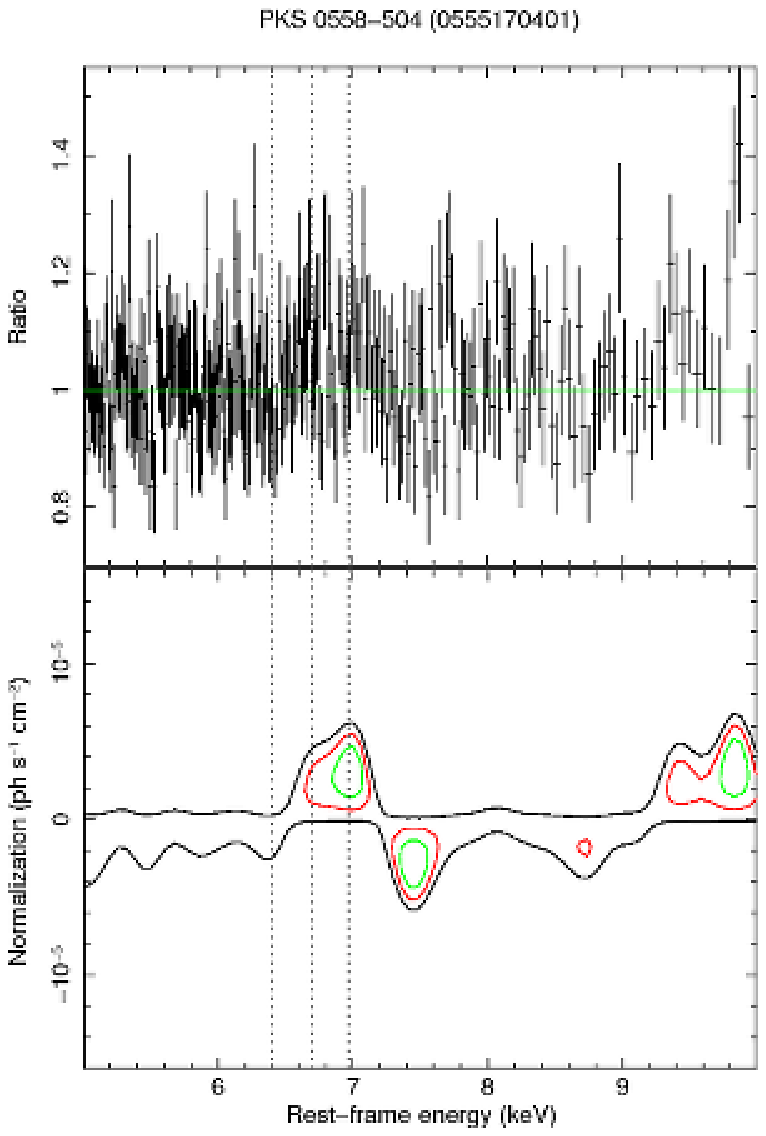}
\hspace{0.3cm}
    \includegraphics[width=4.8cm,height=6.7cm,angle=0]{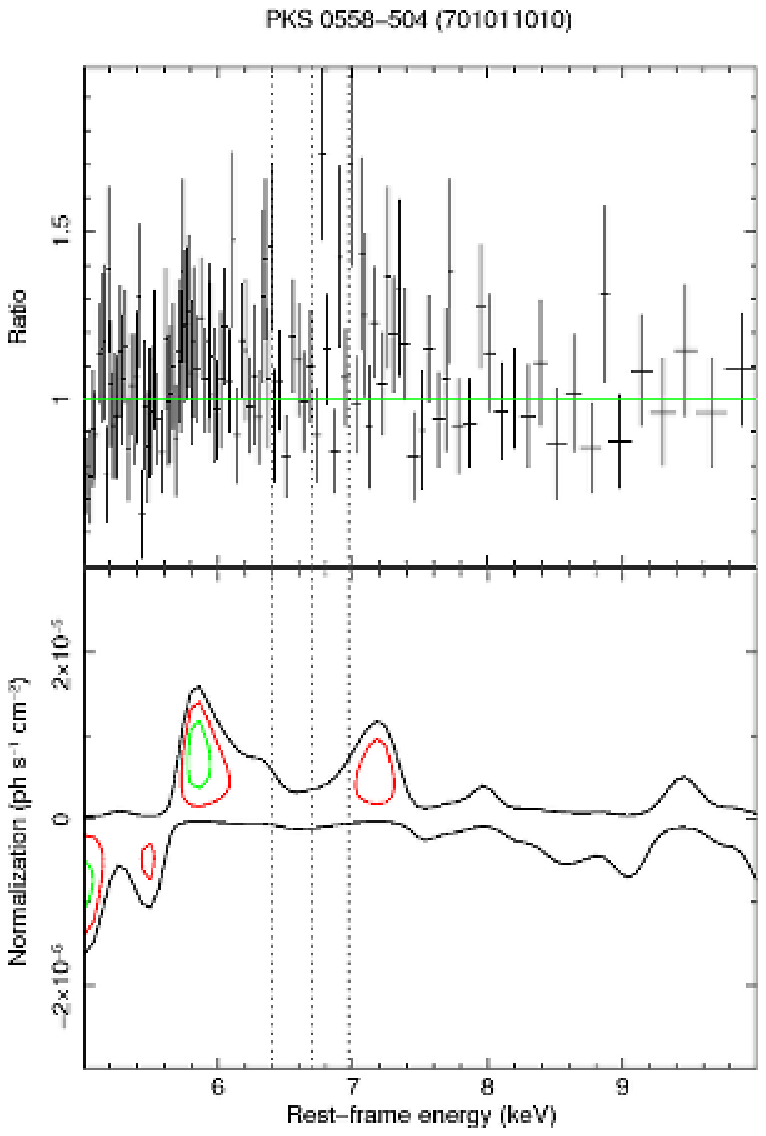}

\vspace{0.2cm}
    \includegraphics[width=4.8cm,height=6.7cm,angle=0]{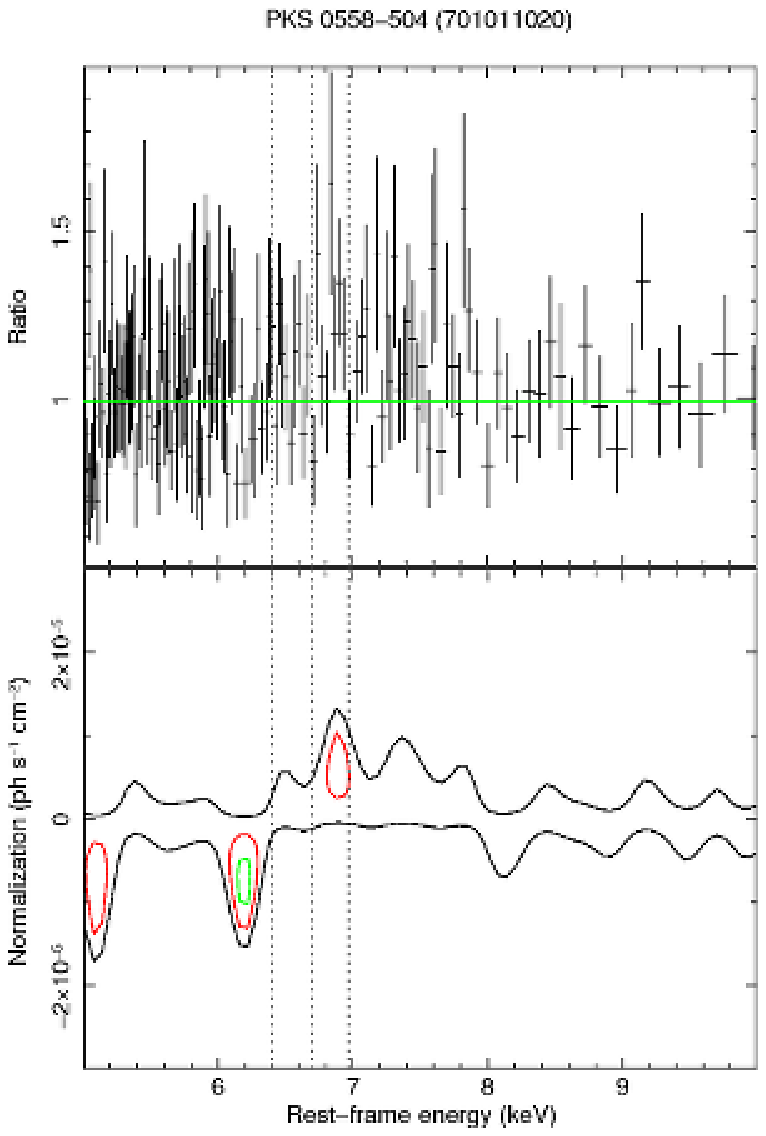}
\hspace{0.3cm}
    \includegraphics[width=4.8cm,height=6.7cm,angle=0]{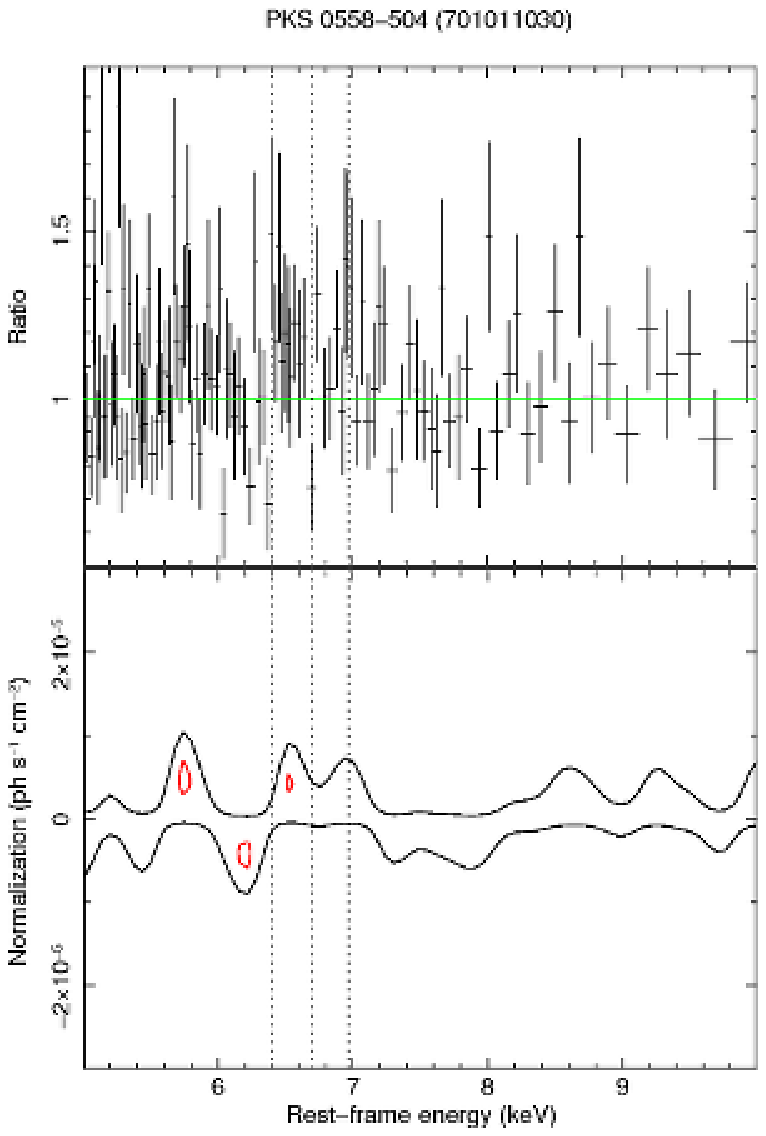}
\hspace{0.3cm}
    \includegraphics[width=4.8cm,height=6.7cm,angle=0]{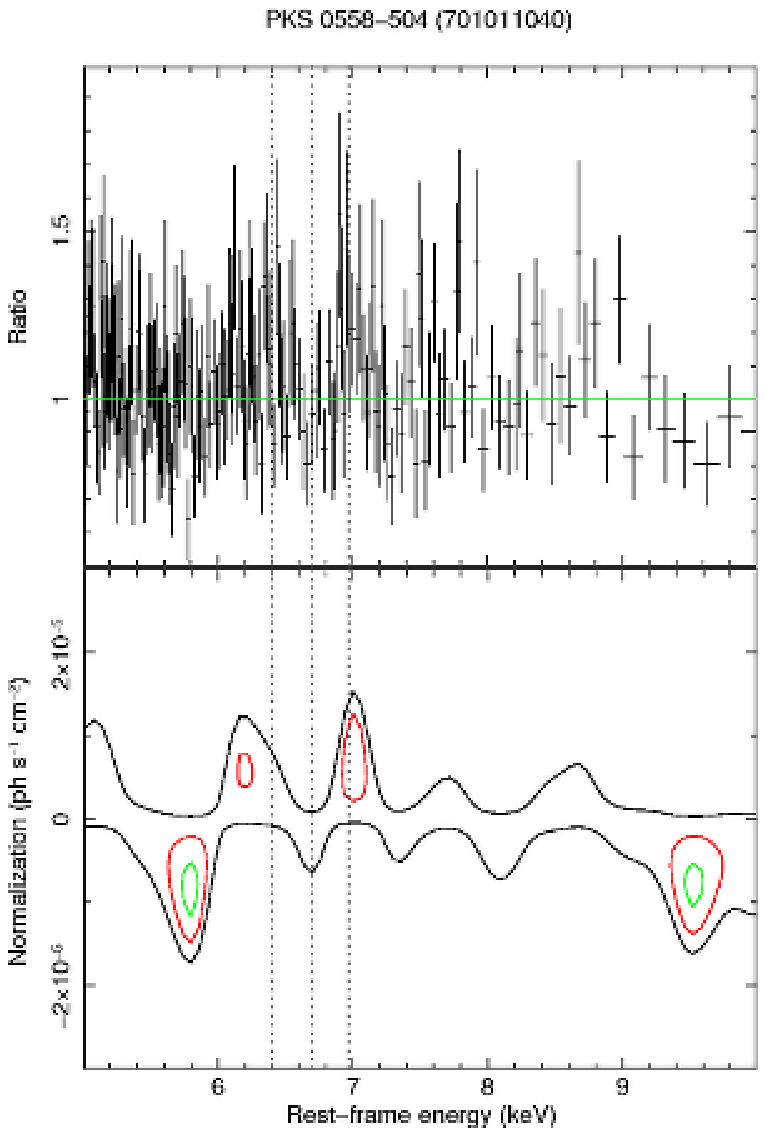}

\contcaption{-- Ratio against the continuum (\emph{upper panel}) and contour plots with respect to the best-fit baseline model (\emph{lower panel}).}
    \end{figure}


   \begin{figure}
   \centering

    \includegraphics[width=4.8cm,height=6.7cm,angle=0]{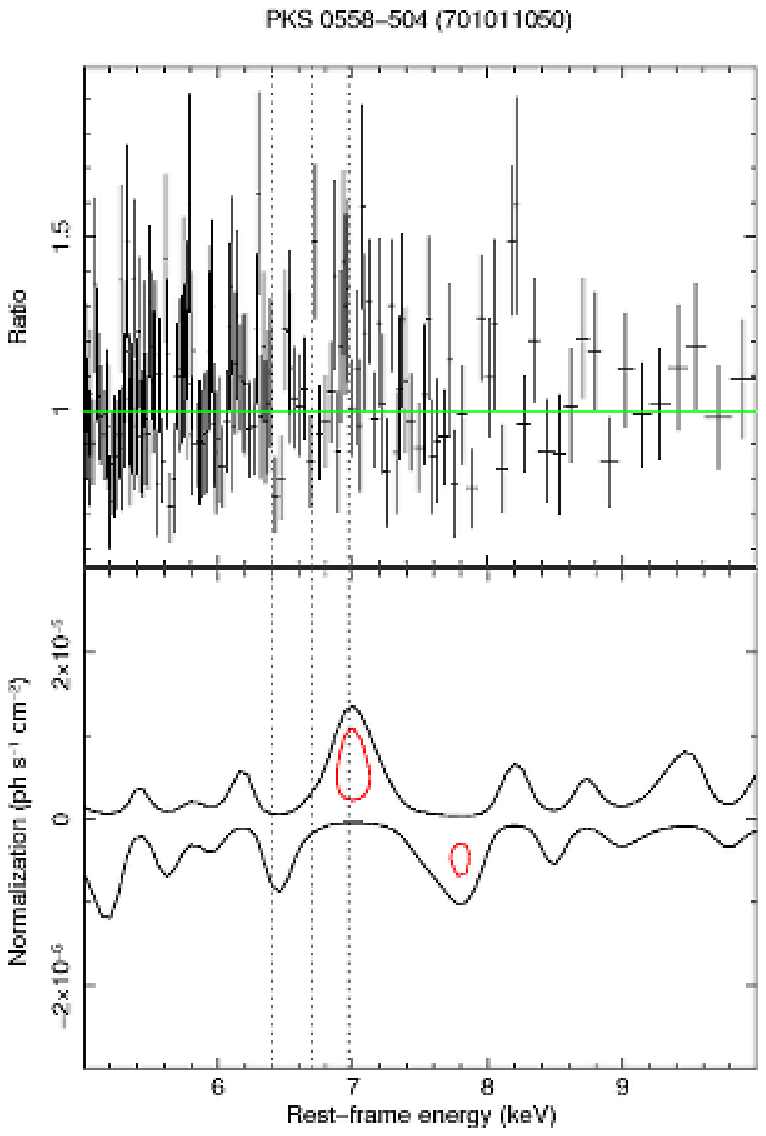}
\hspace{0.3cm}
    \includegraphics[width=4.8cm,height=6.7cm,angle=0]{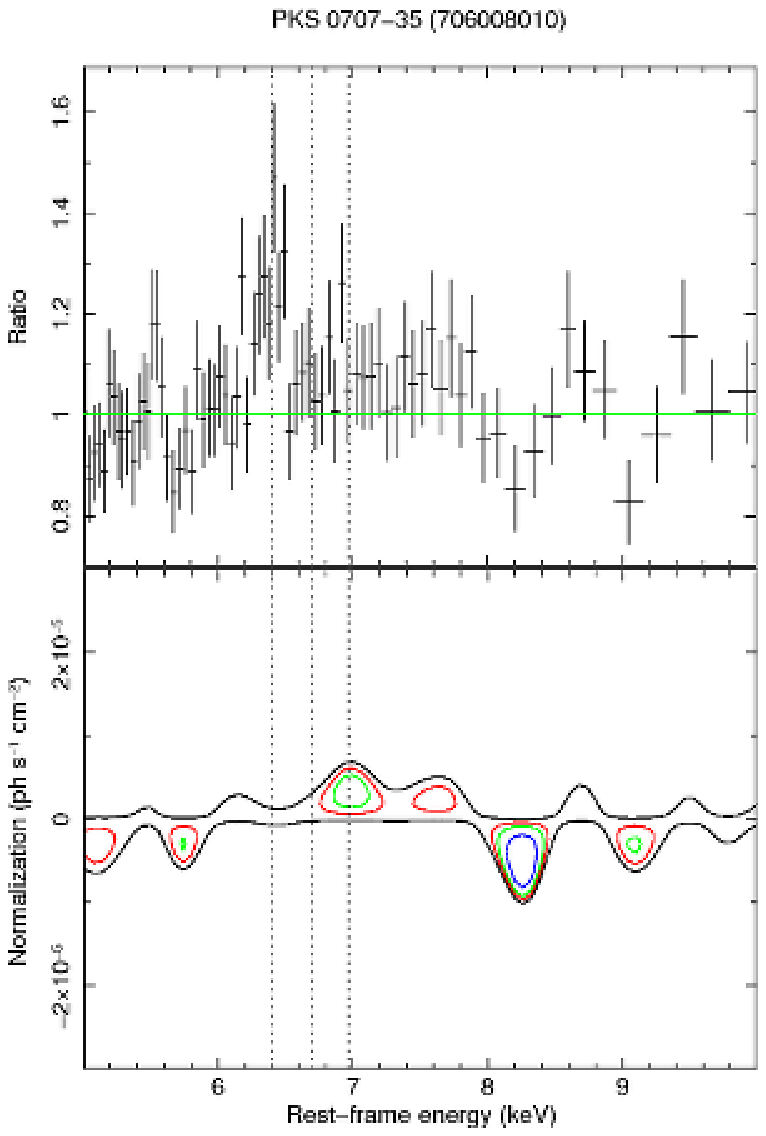}
\hspace{0.3cm}
    \includegraphics[width=4.8cm,height=6.7cm,angle=0]{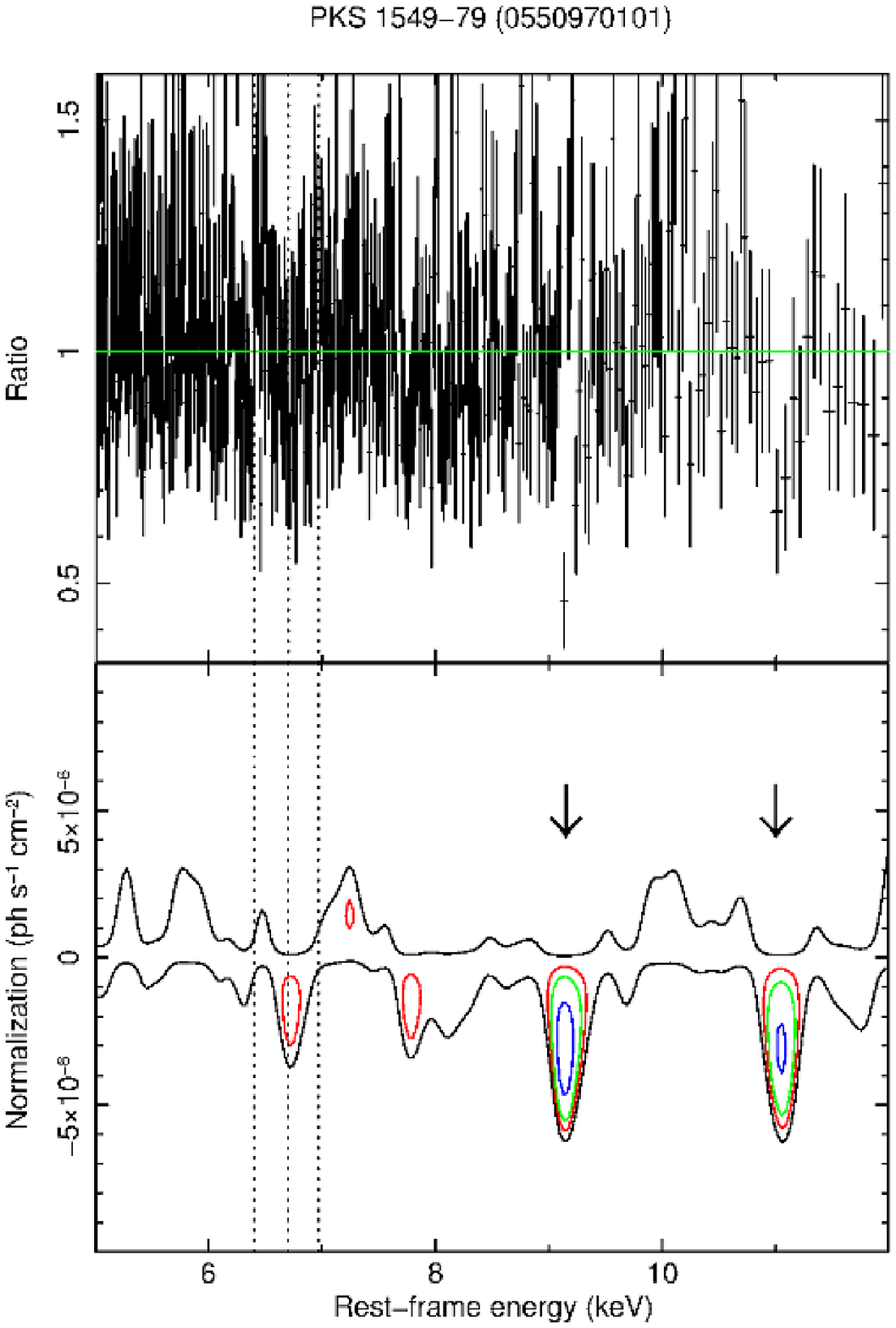}

\vspace{0.2cm}
    \includegraphics[width=4.8cm,height=6.7cm,angle=0]{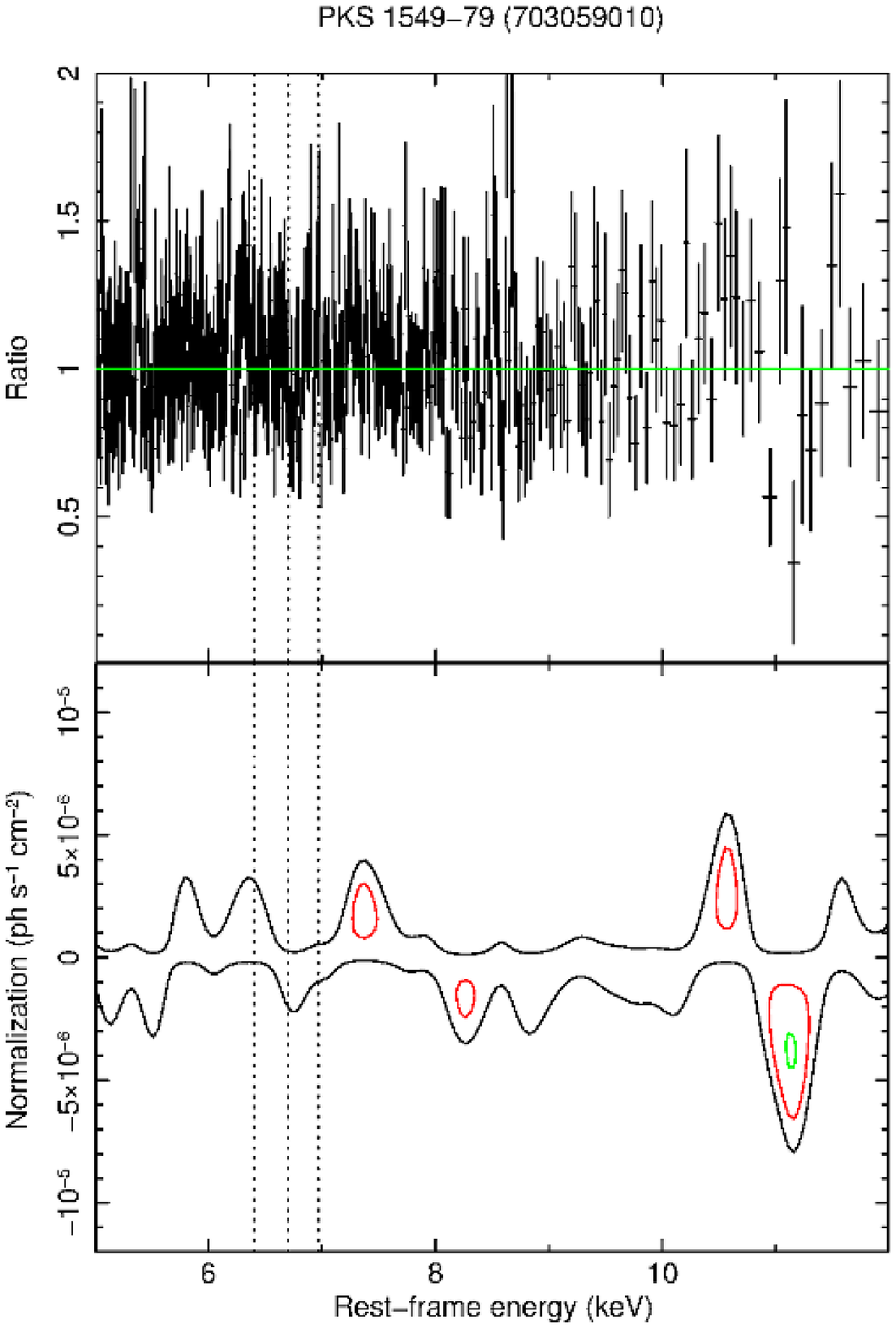}
\hspace{0.3cm} 
    \includegraphics[width=4.8cm,height=6.7cm,angle=0]{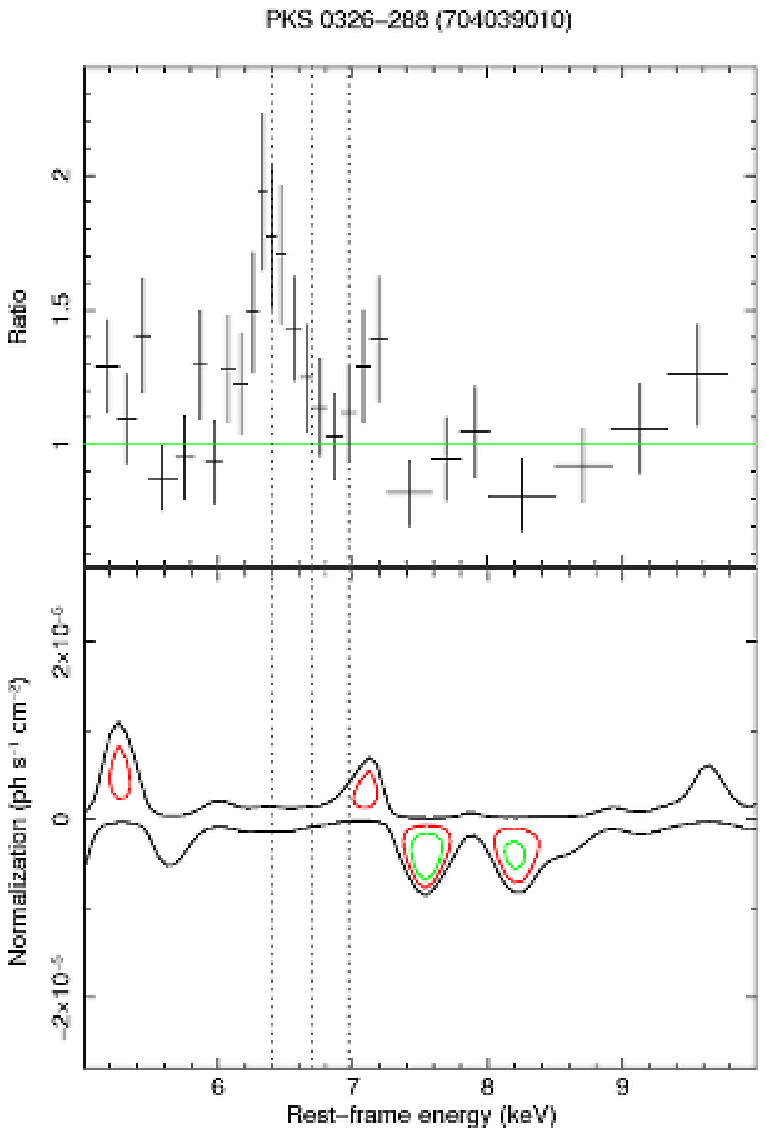}
\hspace{0.3cm}
    \includegraphics[width=4.8cm,height=6.7cm,angle=0]{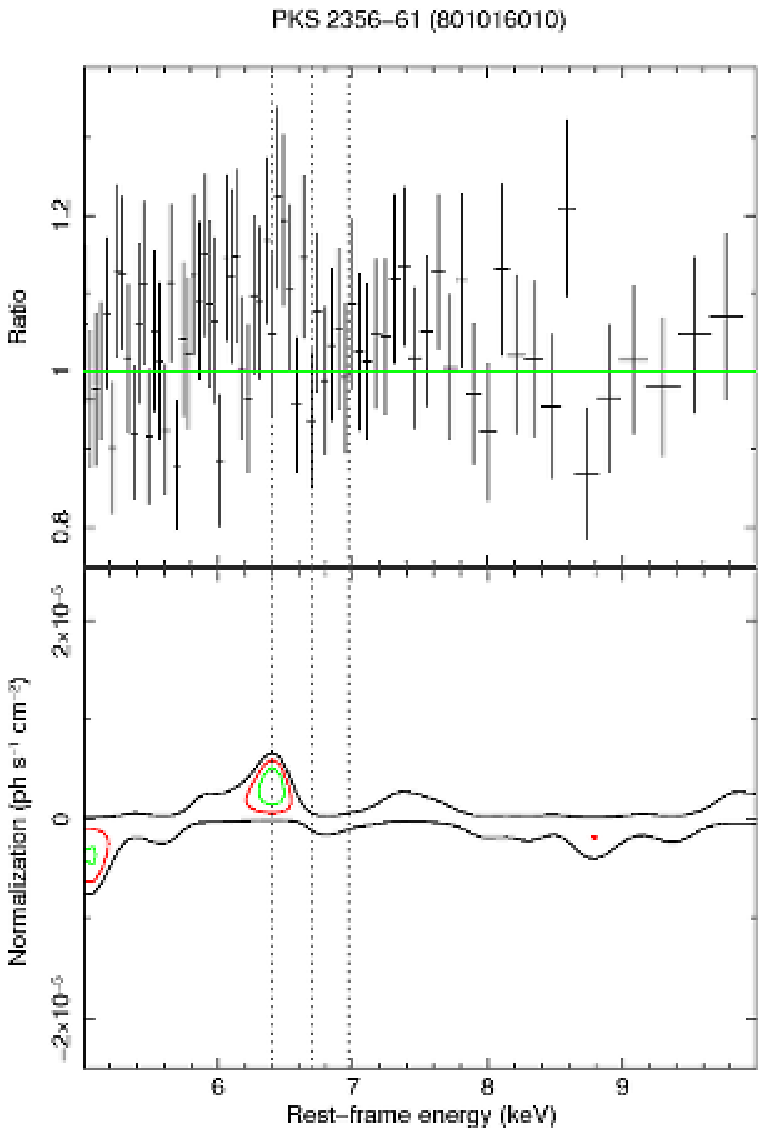}

\vspace{0.2cm}
    \includegraphics[width=4.8cm,height=6.7cm,angle=0]{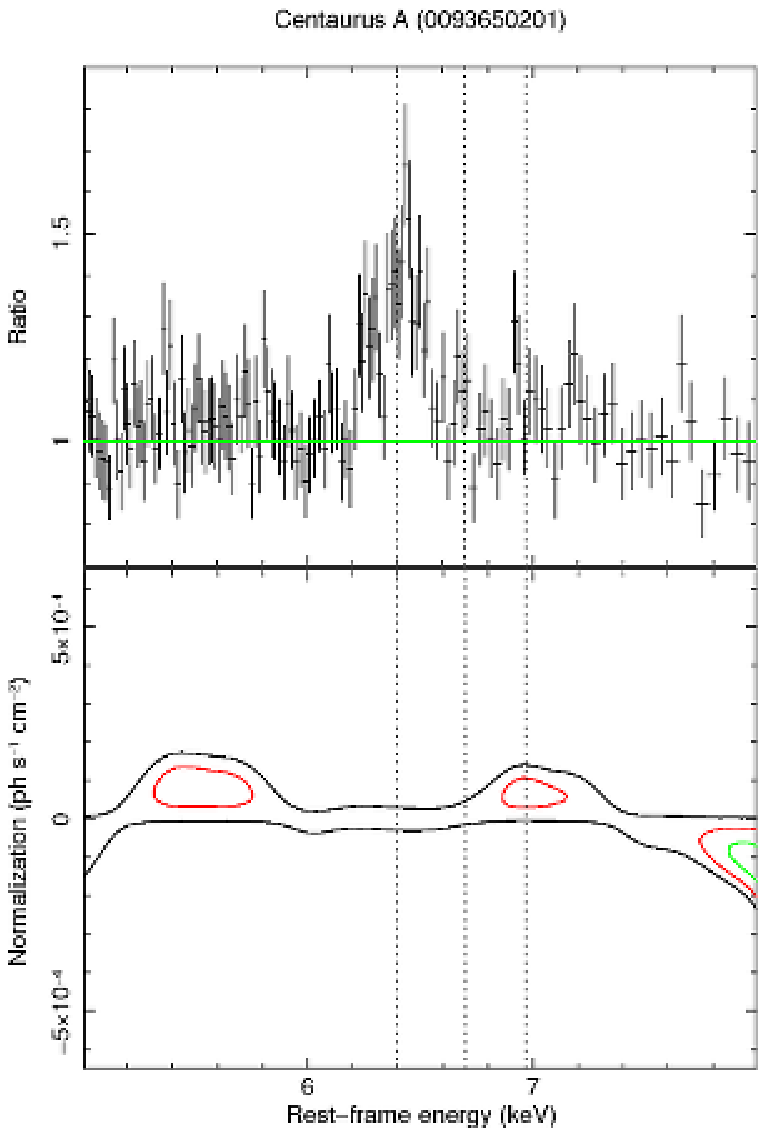}
\hspace{0.3cm}
    \includegraphics[width=4.8cm,height=6.7cm,angle=0]{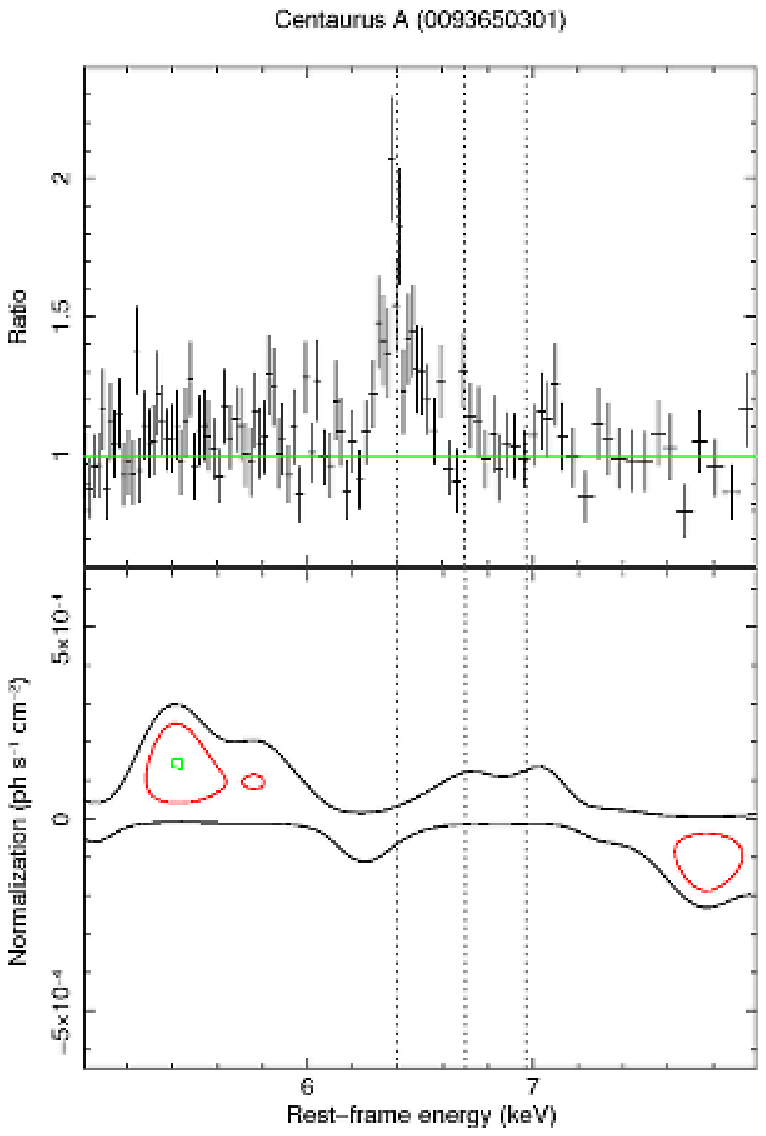}
\hspace{0.3cm}
    \includegraphics[width=4.8cm,height=6.7cm,angle=0]{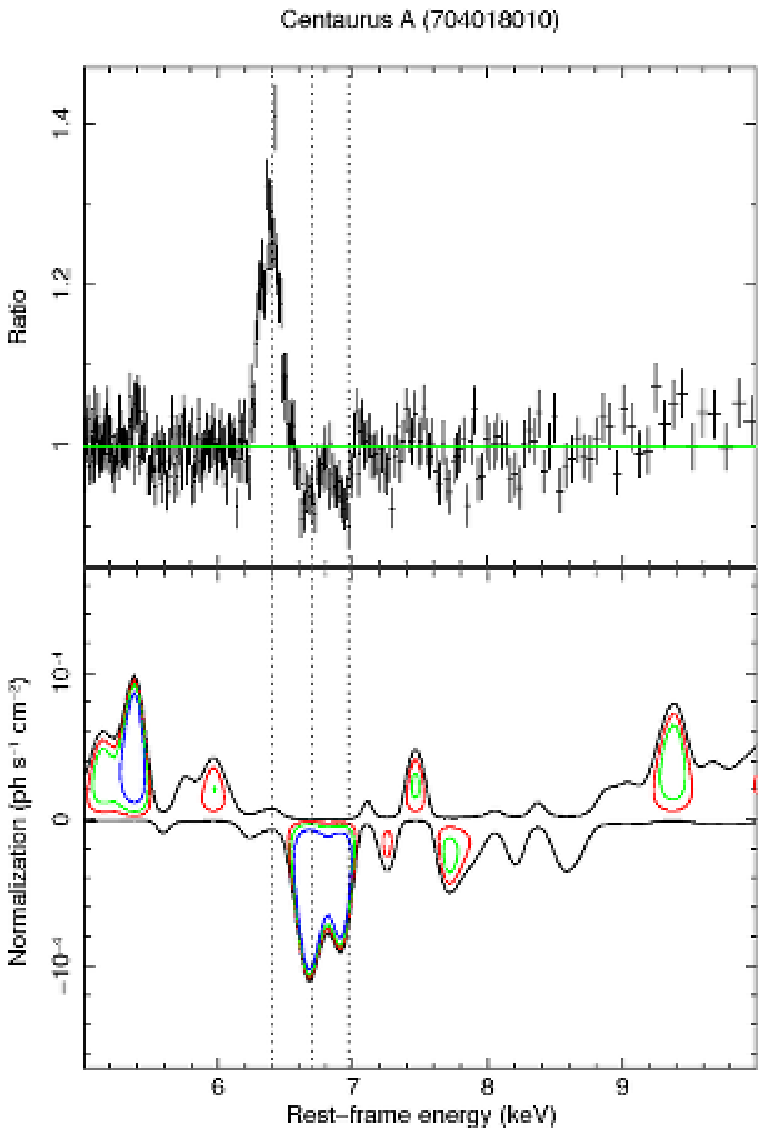}

\contcaption{-- Ratio against the continuum (\emph{upper panel}) and contour plots with respect to the best-fit baseline model (\emph{lower panel}).}
    \end{figure}


   \begin{figure}
   \centering

    \includegraphics[width=4.8cm,height=6.7cm,angle=0]{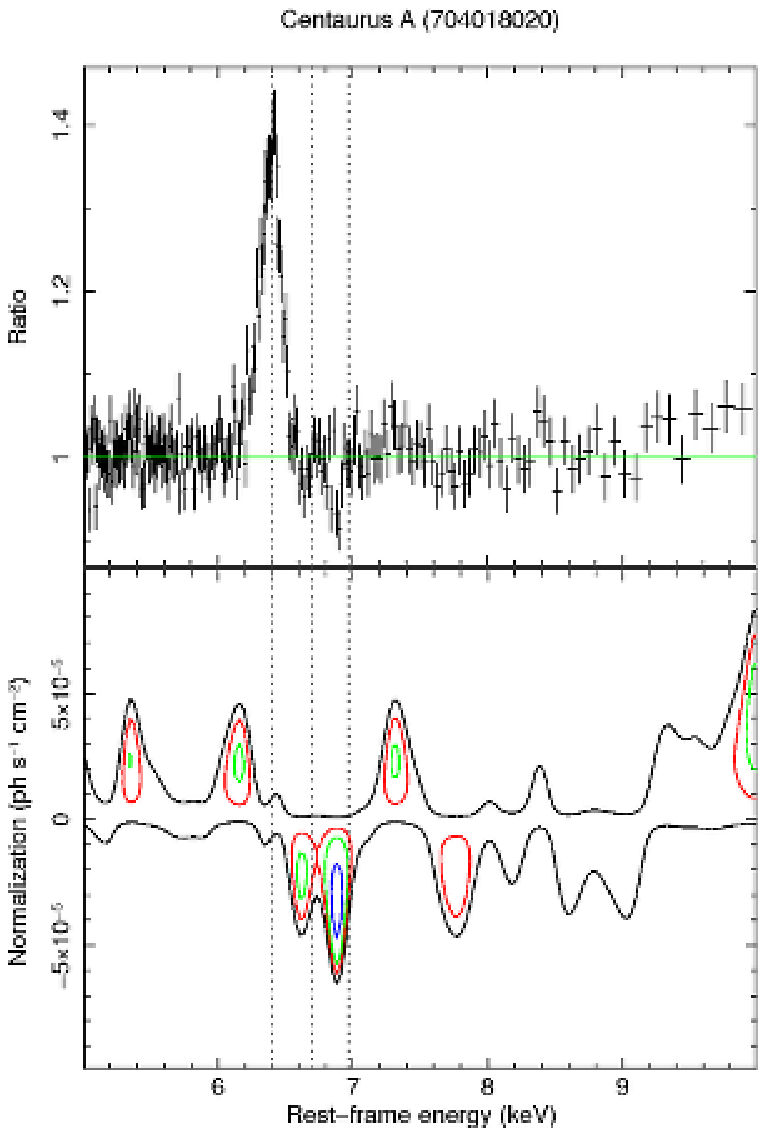}
\hspace{0.3cm}
    \includegraphics[width=4.8cm,height=6.7cm,angle=0]{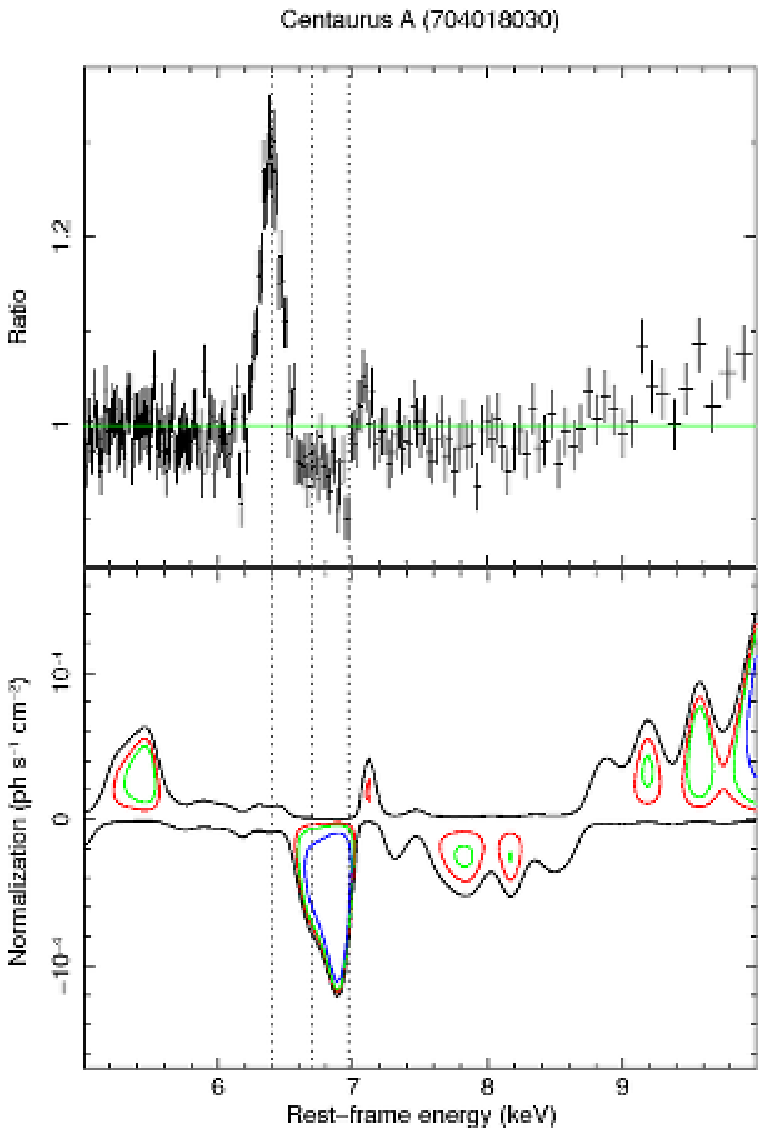}
\hspace{0.3cm}
    \includegraphics[width=4.8cm,height=6.7cm,angle=0]{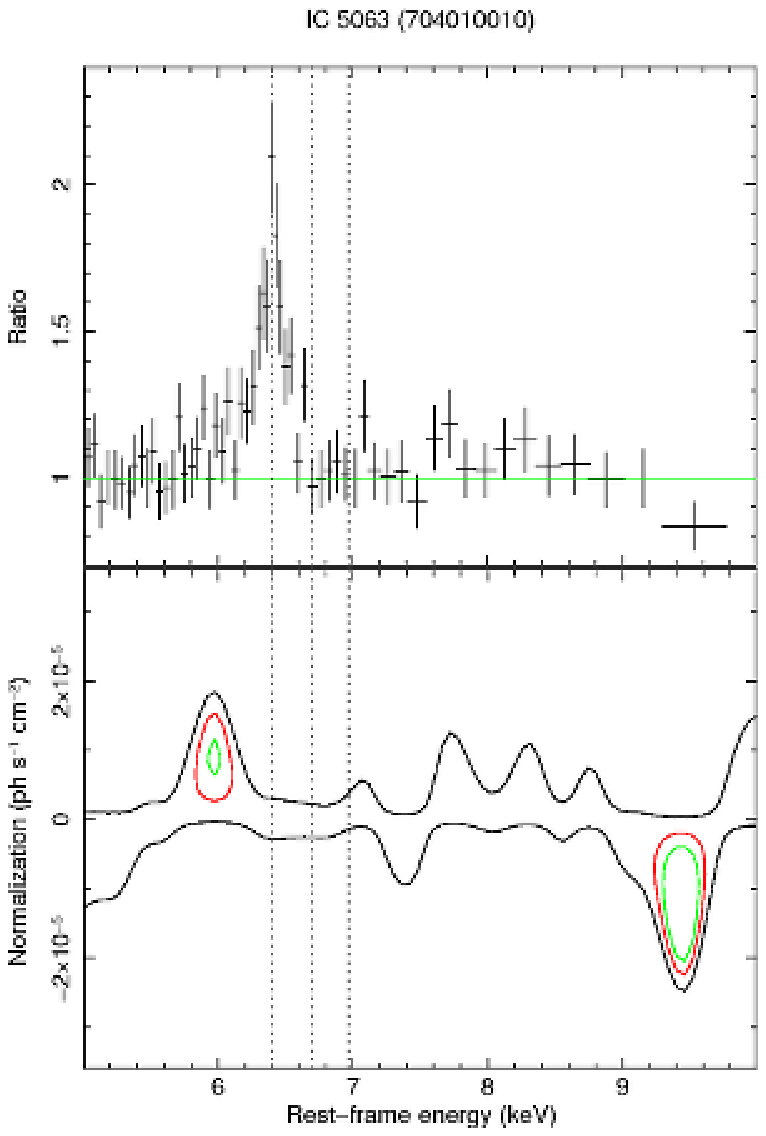}

\vspace{0.2cm}
    \includegraphics[width=4.8cm,height=6.7cm,angle=0]{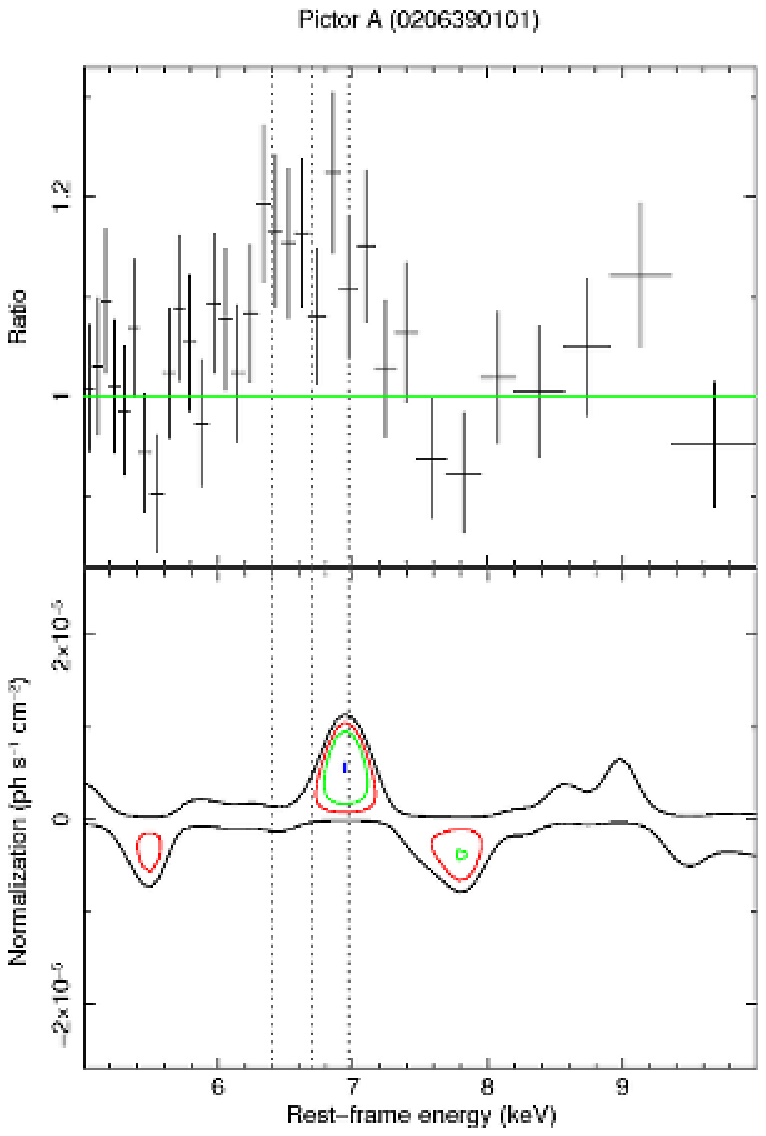}
\hspace{0.3cm} 
    \includegraphics[width=4.8cm,height=6.7cm,angle=0]{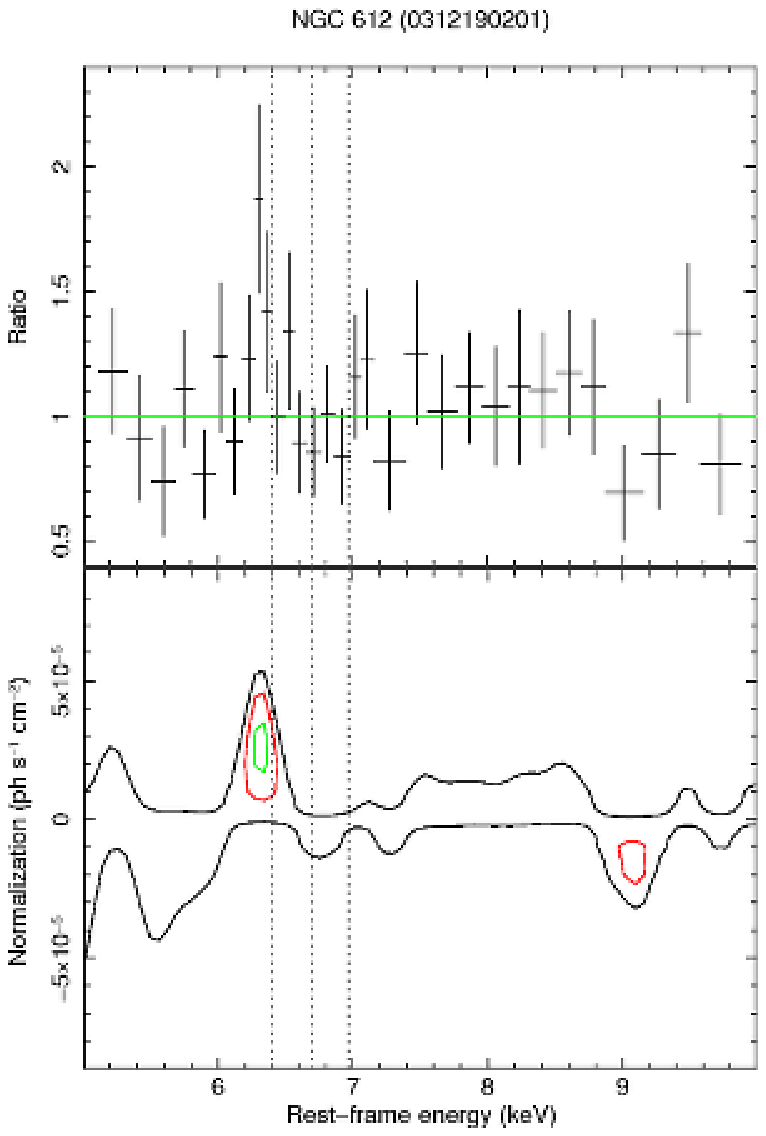}
\hspace{0.3cm}
    \includegraphics[width=4.8cm,height=6.7cm,angle=0]{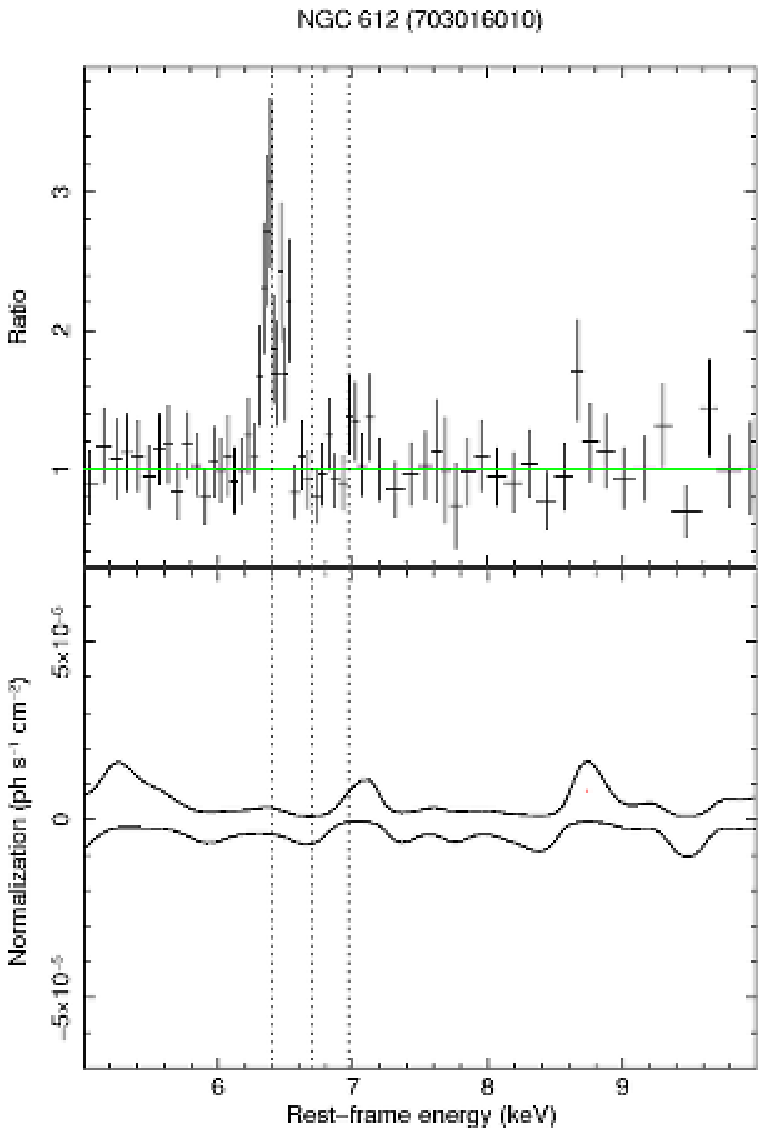}

\contcaption{-- Ratio against the continuum (\emph{upper panel}) and contour plots with respect to the best-fit baseline model (\emph{lower panel}).}
    \end{figure}

}

\end{document}